\documentclass[a4paper,USenglish,cleveref, autoref, thm-restate]{lipics-v2021}

\usepackage{amssymb}
\usepackage{tikz}
\usepackage{longtable}
\usepackage{pdflscape}
\usepackage{array}  
\usepackage{booktabs}
\usepackage{placeins}

\usepackage{caption}

\usepackage{listings}
\usepackage{inconsolata} 
\usepackage{float}
\nolinenumbers

\newcommand{\singleArbitrage}{Market Rebalancing Arbitrage}
\newcommand{\multArbitrage}{Combinatorial Arbitrage}

\hideLIPIcs  

\title{Unravelling the Probabilistic Forest: \\ Arbitrage in Prediction Markets} 

\titlerunning{Unravelling the Probabilistic Forest: Arbitrage in Prediction Markets} 

\author{Oriol Saguillo}{ IMDEA Networks, Madrid, Spain }{oriol.saguillo@imdea.org}{https://orcid.org/0009-0000-6636-8527}{}

\author{Vahid Ghafouri}{ Oxford Internet Institute, Oxford, UK }{vahid.ghafouri@oii.ox.ac.uk}{https://orcid.org/0000-0001-9068-8854}{}

\author{Lucianna Kiffer}{ IMDEA Networks, Madrid, Spain}{lucianna.kiffer@networks.imdea.org}{https://orcid.org/0000-0003-2022-7993}{}

\author{Guillermo Suarez-Tangil}{IMDEA Networks, Madrid, Spain }{guillermo.suarez-tangil@networks.imdea.org}{https://orcid.org/0000-0002-0455-2553}{}

\authorrunning{Saguillo, Ghafouri, Kiffer, and Suarez-Tangil}

\Copyright{Jane Open Access and Joan R. Public} 



\category{} 

\acknowledgements{This work was supported by a Flashbots Research Proposal FRP-51. Additionally, we thank the anonymous AFT reviewers for the helpful feedback.}

\EventEditors{John Q. Open and Joan R. Access}
\EventNoEds{2}
\EventLongTitle{42nd Conference on Very Important Topics (CVIT 2016)}
\EventShortTitle{CVIT 2016}
\EventAcronym{CVIT}
\EventYear{2016}
\EventDate{December 24--27, 2016}
\EventLocation{Little Whinging, United Kingdom}
\EventLogo{}
\SeriesVolume{42}
\ArticleNo{23}

\begin{document}

\maketitle

\begin{abstract}
Polymarket is a prediction market platform where users can speculate on future events by trading shares tied to specific outcomes, known as \emph{conditions}. 
Each market on Polymarket is associated with a set of one or more such conditions. 
To ensure proper market resolution, the condition set must be \emph{exhaustive}---collectively accounting for all possible outcomes---and \emph{mutually exclusive}---only one condition may resolve as true. Thus, the collective prices (probabilities) of all related outcomes (whether in a condition or market) should be \$1, representing a combined probability of 1 of any outcome.
Despite this design, Polymarket exhibits cases where dependent assets are mispriced, allowing for purchasing (or selling) a certain outcome for less than (or more than) \$1, guaranteeing profit.
This phenomenon, known as arbitrage, could enable sophisticated participants to exploit such inconsistencies. 

In this paper, we conduct an empirical arbitrage analysis on Polymarket data to answer three key questions: 
(Q1) What conditions give rise to arbitrage? 
(Q2) Does arbitrage actually occur on Polymarket?, and 
(Q3) Has anyone exploited these opportunities?
A major challenge in analyzing arbitrage between related markets lies in the scalability of comparisons across a large number of markets and conditions, with a naive analysis requiring $O(2^{n+m})$ comparisons. 
To overcome this, we employ a heuristic-driven reduction strategy based on timeliness, topical similarity, and combinatorial relationships, further validated by expert input.

Our study reveals two distinct forms of arbitrage on Polymarket: \emph{\singleArbitrage{}}, which occurs within a single market or condition (intra-market), and \emph{\multArbitrage{}}, which spans across multiple markets (inter-market). 
We use on-chain historical order book data to analyze when these types of arbitrage opportunities have existed, and when they have been executed by users. We find a realized estimate of 40 million USD of profit extracted across both types of arbitrage during our measurement period.

\end{abstract}

\newpage

\section{Introduction}

Forecasting future events has long been a central problem in economics and finance, where accurate predictions can have significant informational and monetary value. Traditional approaches have often relied on expert judgment or econometric models, each with their own limitations in adaptability and responsiveness to new information \cite{tetlock2017expert, armstrong2001principles}.

Prediction markets \cite{wolfers2004prediction} offer a novel approach to generating forecasts. The general idea is to pose a question openly to the public along with its possible outcomes. Participants place bets when they believe that the probability assigned to an outcome is inaccurate and may earn money over time if their predictions are correct. An example of such a market was introduced by Tradesports.com~\cite{hanson_policy_market}, which listed a security that would pay \$100 if the head of DARPA was ousted by the end of August 2003.
The novelty of this approach lies in the speed with which the probabilities were updated as new information about the event became available.
The future can also be viewed as an incomplete information game, where each individual holds a private valuation of an outcome based on their personal perception of reality. By aggregating these valuations across participants, the system can generate a more accurate probability of an event occurring. Recent prediction markets for the 2024 U.S. election have demonstrated greater accuracy in forecasting outcomes compared to traditional media forecasting tools~\cite{frick_prediction_markets}.

Polymarket \cite{polymarket2025}, a prediction market protocol on the Polygon blockchain \cite{matic2019whitepaper}, has emerged as the leading partially-decentralized protocol for making prediction markets a reality, largely due to its high levels of user engagement during the US elections. 
For instance, during the 2024 elections, Polymarket saw explosive growth, with over {\$3.7} billion in total trading volume and tens of thousands of active users placing bets on election outcomes~\cite{polymarket_pres2024}.
(See Figure~\ref{fig:volume_overview} for an overview of data on liquidity locked in the markets over time.)
At its peak, markets on Polymarket were resolving millions of dollars in open interest, with major news outlets and online communities citing market odds as real-time indicators of public sentiment.
This surge in activity is positioning Polymarket as one of the most popular forecasting ecosystems, bridging retail speculation with informational efficiency.

Arbitrage opportunities arise when two identical assets are valued differently due to a pricing mismatch between sellers. Arbitrageurs are sophisticated individuals or agents who exploit these discrepancies by purchasing the asset at the lower price and selling it to a buyer willing to pay a higher price.
In traditional finance, arbitrage is a well-studied mechanism for enforcing price consistency and improving market efficiency. 
In prediction markets like Polymarket, arbitrage plays a similar role, acting as a corrective force that aligns asset prices with their underlying probabilistic truth values. 
Unlike centralized markets, however, arbitrage on-chain introduces unique complexities: it requires cross-market, fast execution, and the ability to parse fragmented and often noisy information across multiple interrelated conditions. 
This makes the identification and exploitation of arbitrage on platforms like Polymarket a sophisticated technical challenge, especially at scale.

\subsection{Our Contributions}

To understand how arbitrageurs may be taking advantage of pricing mismatches, we set out to answer the following research questions: 
\begin{itemize}
    \item Q1. What conditions give rise to arbitrage? 
    \item Q2. Does arbitrage actually occur on Polymarket?
    \item Q3. Has anyone exploited these opportunities?
\end{itemize}

To answer these questions, we precisely characterize the types of arbitrage that arise in outcome-dependent condition spaces. We then design a methodology that combines heuristic-driven reduction with semantic analysis to efficiently identify arbitrage opportunities across Polymarket. 
Our approach leverages temporal proximity, categorization into primary topics using textual embeddings (generated by \textit{Linq-Embed-Mistral}), and large language models (LLMs) to extract combinatorial relationships and logical dependencies from market condition descriptions. 
This enables us to reduce the arbitrage search space and systematically triage market arbitrage. 
To perform our analysis, we collect historical bid data from Polymarket. We use this bid data to characterize arbitrage opportunities within conditions in a single market, and across dependent markets. Our data collection efforts span markets that were resolved over the period of one year, from April 1, 2024, to April 1, 2025.

\subsection{Related Work}

This work presents the first large-scale analysis of arbitrage on Polymarket, one of the most widely-used prediction markets. 
A central contribution of our study is the investigation of combinatorial relationships between conditions to infer and characterize dependencies that may give rise to arbitrage opportunities. 
To contextualize our approach, we survey prior research on Maximal Extractable Value (MEV) detection and highlight recent advances in leveraging LLMs to analyze and interpret complex cognitive and decision-making processes.

\subsubsection{MEV Detection} 

Maximal Extractable Value (MEV) \cite{daian2020flash} refers to the game-theoretic problem of transaction manipulation within a block, including inserting, reordering, or censoring transactions.
MEV has become a regular feature of decentralized systems, where it is used to extract profit, often to the detriment of end users. This phenomenon has been explored across a range of contexts, including in \cite{weintraub2022flash}, \cite{ferreira2024rolling}, \cite{zhang2024no}, and \cite{oz2025pandora}. In this paper, we focus on a specific class of MEV strategies: arbitrage, particularly as it applies to a new category of decentralized applications, prediction markets.
Several mechanisms have been proposed to mitigate the negative effects of MEV, as surveyed in \cite{yang2024sok}. Among these is the creation of more optimized MEV extraction environments, such as the Proposer-Builder Separation (PBS) paradigm. The goal of PBS is to reduce the influence of validators over transaction ordering and block content. Empirical studies, including \cite{heimbach2023ethereum} and \cite{oz2024wins}, have examined the implementation and impact of PBS in production.

Currently, arbitrage is generally considered a positive-sum form of MEV, as it promotes price alignment across decentralized protocols. Nonetheless, there is ongoing debate in the community about whether arbitrageurs capture a disproportionate share of value from this rebalancing process \cite{milionis2024automated}.
A key clarification for this article is our focus on non-atomic arbitrage. Unlike atomic arbitrage—where buy and sell operations are executed simultaneously and failure is impossible—non-atomic arbitrage introduces execution risk, as one leg of the trade may succeed while the other fails. Empirical measurements of non-atomic arbitrage in blockchain systems can be found in \cite{heimbach2024non} and \cite{qin2022quantifying}.

\subsubsection{LLMs}
\label{sec:relatedworks:llm}

Recent studies have demonstrated the utility of LLMs in various social and semantic annotation tasks, making them cost-effective alternatives to human annotators. Tasks such as \textit{stance detection}~\cite{cetinkaya2025cpi}, \textit{sentiment analysis}~\cite{zhu2024llm}, \textit{toxicity detection}~\cite{zhu2024llm}, detecting sociopolitical affiliation of texts~\cite{ghafouri2023kialo}, etc. have been put to large-scale applications by previous works and have proven successful even with light-weight open-source LLMs such as ``\textit{Mistral-7B-Instruct-v0.2}''~\cite{cetinkaya2025cpi}, Llama-3.2-8B-Instruct~\cite{yahan2025llama}, and DeepSeek~\cite{guo2025deepseek}.

Moreover, previous literature has extensively leveraged LLMs to analyze and understand complex cognitive processes, including the intricate structures of reasoning and the underlying logical dependencies within various tasks, often through techniques such as Chain-of-Thought (CoT) prompting~\cite{Wei2022CoT, wang2023CoT}. These methods guide LLMs to break down complex problems into intermediate steps, making the reasoning process more explicit and verifiable~\cite{yao2023CoT}.

Our LLM annotation task of detecting logical dependencies between human-generated markets falls in a hybrid domain between social annotation tasks and the analysis of logical dependencies. The perfect execution of all the mentioned tasks, including ours, relies not only on the power of the chosen LLM but also on the quality of a well-curated prompt, namely \textit{prompt engineering} (see Section~\ref{sec:background:llm}).

\section{Background}
\label{sec:background}

We present the background necessary to understand prediction markets and how LLMs can aid in understanding when markets are dependent, a first step needed to assess arbitrage. 

\subsection{Polymarket}

Polymarket is a prediction market platform that allows users to speculate on the outcomes of future events by trading shares tied to specific outcomes. Polymarket is built on top of the Polygon blockchain \cite{matic2019whitepaper}, providing some decentralized properties to the platform. Each \textbf{condition}  poses a question about a future event, such as "Will team A defeat team B in the big game?". Users can buy shares (or \textit{tokens}) in "\textit{YES}" (the condition will become true) or "\textit{NO}" (the condition will not become true) outcomes, with share prices fluctuating based on market demand and reflecting the collective belief about the likelihood of each outcome. 

A \textbf{market} is thus a future event associated with one or more conditions. For example, consider the question, "Who will win the Team A vs. Team B match?". In this case, the market would include three conditions: (1) Team A wins, (2) the match ends in a tie, or (3) Team B wins. Each condition is represented by a binary token indicating whether the condition is true or false. To ensure proper resolution, the set of conditions must be \textit{exhaustive}, collectively covering all possible outcomes of the event, and \textit{independent}, where only one condition can resolve to true. When a market contains multiple conditions, they share a market ID and are labelled as NegRisk (neg risk market).\footnote{As users can hedge against specific risks by placing bets that the event won't happen.}

\subsubsection{Market Creation}
The creation of markets \cite{polymarket_markets_created} and their associated conditions is controlled by Polymarket and must be conducted via their Discord server or by tagging the official Polymarket account on Twitter (@polymarket). 
To suggest a new market, users provide (i) a title for the market, (ii) the designated resolution source,\footnote{One or more sources that should be referred to for information on the actual outcome of the event.}, and (iii) evidence that there is demand for trading in this market. Markets are then registered on-chain, and the corresponding tokens are created for each condition and outcome. Tokens representing a condition follow the ERC-1155 standard under the Gnosis Conditional Tokens Framework \cite{gnosis2020ctf}. 

\subsubsection{Buying Positions}

Positions (tokens) in Polymarket are bought/sold based on a hybrid-decentralized Central Limit Order Book (CLOB) system \cite{polymarket_clob_intro}. Users make a bid to buy/sell some token and submit these directly to the Polymarket API. Polymarket then matches bids, and those matched bids are executed on-chain. The on-chain element ensures that users own the tokens corresponding to their positions in each market, adding an element of decentralization while the matching is done entirely in a centralized manner.

Users place limit orders to buy or sell "\textit{YES}" or "\textit{NO}" outcome shares at specified prices. The simplest type of match occurs when two users submit a sell and buy bid for a token at the same price. In this case, the token and USDC values are directly traded, and the conditional token contract emits a \textbf{OrderFilled} event logging the trade on-chain.\footnote{We use events emitted by the conditional token contract to log all buy/sells that take place on Polymarket when looking at on-chain data from contract \texttt{0x4D97DCd97eC945f40cF65F87097ACe5EA0476045}.}
Another type of match is made when the prices of opposing orders sum to \$1.00. For example, a "\textit{YES}" bid at \$0.60 matches with a "\textit{NO}" bid at \$0.40. Upon matching, \$1.00 is converted into one "\textit{YES}" and one "\textit{NO}" share, each allocated to the respective buyer. On-chain, this is done by users locking USDC tokens\footnote{The token representation of the US Dollar.} with the Polymarket exchange contract, while one new conditional token is minted corresponding to one "\textit{YES}" token and one "\textit{NO}" token of the condition. The generation of the new token emits a \textbf{PositionSplit} event and two \textbf{OrderFilled} events for each side of the buys. Similarly, two sells of opposing orders can sum to \$1.00, resulting in the two tokens being burned and each user withdrawing the corresponding USDC from the exchange contract. This action emits a \textbf{OrdersMerged} event capturing the tokens being burned, and an \textbf{OrderFilled} event for each side of the sell. 

Lastly, a user can buy both sides of a condition by locking in 1 USDC and generating both one "\textit{YES}" and one "\textit{NO}" token, which they may trade later. This emits a \textbf{PositionSplit} event. Similarly, a user may sell both positions for a condition at a cumulative price of 1 USDC. This withdraws the USDC from the exchange and burns the two tokens representing it, emitting a \textbf{OrdersMerged} event.

\subsubsection{Market Resolution}

Polymarket utilizes UMA's Optimistic Oracle to determine market outcomes \cite{umaDVM2025}. When a market concludes, a resolution is proposed. If undisputed during a designated challenge period, it is accepted as final. However, if challenged, the resolution is escalated, here UMA token holders vote to determine the outcome. In the end, each condition resolves to "\textit{True}" or "\textit{False}", and a single condition in each market resolves to "\textit{True}". This resolution is registered on-chain, and users can then claim their corresponding USDC.
This system can lead to discrepancies between the oracle's resolution and the actual event, especially in complex scenarios.\footnote{For instance, in the 2024 FIDE World Blitz Championship market \cite{polymarket2024blitz}, both Carlsen and Nepomniachtchi were declared winners of different sections, but only one outcome could be selected as "\textit{True}" due to the framework's constraints.}
Additionally, the concentration of voting power among large UMA token holders can influence resolutions, as observed in \cite{primo2024uma}, and the oracle is susceptible to potential governance attacks as discussed in \cite{feichtinger2024sok}.

\subsection{LLMs \& Prompt Engineering}
\label{sec:background:llm}

\textbf{Prompt engineering} refers to the process of crafting input queries (called \textit{prompts}) to guide an LLM toward producing useful or accurate outputs. Since these models do not inherently ``understand'' in the human sense, the way a question is phrased can significantly impact the quality of the response. Prompt engineering has emerged as a practical technique to control and refine LLM behavior without retraining the model. For instance, adding a few examples to a prompt (called \textit{few-shot prompting}) or specifying desired formats can help steer the model more effectively.

In this paper, we leverage prompt engineering practices to extract and interpret logical dependencies within the Polymarket markets to detect potential cases of arbitrage. Our prompt engineering involves explaining the task coherently, defining a set of rules for the task, and restricting the output to our specific desired JSON format while providing a sample desired response to the model (see Listing~\ref{lst:prompt-sample}). 

\section{Definitions}

We now introduce some notation to define the set of possible outcomes for a given market, and for pairs of markets. With these, we can precisely define what it means for the resolution of markets to be dependent such that there can exist arbitrage opportunities (i.e., when the resolution of some conditions implies the resolution of some other conditions, either within the same market or across markets). We then define two types of arbitrage: 1. \textbf{\singleArbitrage}, which occurs within a single market, and 2. \textbf{\multArbitrage}, which occurs between multiple markets.

\subsection{Market Dependence Taxonomy}
\label{sec:defmarketdepency}
\subsubsection{Single Market}
For a given market, which is designed to forecast the outcome of a real-world event, we define the set of all possible \textit{resolutions} of the market as the set of all possible combinations of True/False labels assigned to each condition associated with that market.

\begin{definition}[Single Market Outcome] \label{def:singlemarket}
    Let $M = \{ C_1, C_2, \dots, C_n\}$ be a market modelling an event $E$, where each $C_i$ is a boolean variable representing a possible outcome of $E$ (i.e., the $n$ conditions of the market) and so will be resolved as either \texttt{True} or \texttt{False}. The set of vectors $V = \{ v_i\}$ represents all possible unique resolutions of $M$ with each $v_i = <c_1,c_2, ..., c_n>$ where $c_j \in \{0,1\}$. Market conditions are \textbf{exhaustive}, such that $|V|=n$ and $\forall v\in V$, the following holds
        $$\sum_{c_i\in v} {c_i} = 1$$
\end{definition}

We define a market as \textit{exhaustive} as it includes all mutually exclusive conditions required to fully determine the outcome of an event. In such a market, exactly one condition must be true at resolution time, ensuring the outcome space is complete. As such, conditions within a market are dependent by definition: if one resolves to true, all others must resolve to false.

Consider the following example: let $M_1$ denote a prediction market for an election in New York with the following three mutually defined conditions:
\begin{enumerate}
    \item The Republican candidate wins in New York.
    \item The Democratic candidate wins in New York.
    \item A third-party candidate wins in New York.
\end{enumerate}

In this market, the set of conditions is \textit{exhaustive}, as it covers all possible outcomes of the election, and \textit{exclusive}, since at most one of the conditions can be true simultaneously.

\subsubsection{Multiple Markets}

Depending on how markets are created, it is possible to have two markets with a subset of conditions whose resolutions are semantically dependent. A common example in betting markets is to define both the outcome of an event and the margin by which that outcome occurs across two semantically related markets. To illustrate this, consider two markets: \textbf{Market X}, which contains a set of conditions \( C \) representing the possible outcomes of an event (e.g., which team wins), and \textbf{Market Y}, which includes a different set of conditions \( C' \) that expresses the \emph{margin} or \emph{degree} of the outcome (e.g., the score difference).

For example, consider a football match. Market X might include conditions indicating whether Team A or Team B wins, while Market Y could contain more granular conditions, such as specific scorelines or a minimum winning margin for Team A. Suppose Market Y includes a condition stating that ``\textit{Team A wins by at least 2 goals}''. Then any context which has a resolution of market Y with ``\textit{Team A wins by at least 2 goals}'' as True must also have a resolution of market X with Team A also winning. Thus, the state space of possible resolutions of the two markets combined is more limited. 

\begin{definition}[Market Outcome for Two Markets] \label{def:multimarkets}
    Let $M_1$ and $M_2$ be two markets with condition sets  $\{ C_1, \dots, C_n\}$ and $\{ C'_1, \dots, C'_m\}$ respectively. We define the set of possible unique resolutions for each market as $V_1$ and $V_2$ respectively, and the combined set of possible \textbf{joint} unique resolutions as $V_1 \times V_2 = \{v_i\}$ with each $v_i = <c_1,\dots, c_n,c'_1, \dots, c'_m>$ representing a possible resolution of both markets with subsets $<c_1,\dots, c_n>\in V_1$ and $<c'_1,\dots, c'_m>\in V_2$. 
    
    While each market is exhaustive, i.e., $|V_1|=n$ and $|V_2|=m$, the set of possible joint resolutions have two cases: \\
    (1) If $|V_1 \times V_2| = n \cdot m$, the two markets are \textbf{independent} of each other.\\
    (2) If $|V_1 \times V_2| < n \cdot m$, the two markets are \textbf{dependent}, and $\exists S \subset M_1$ and $S' \subset M_2$ where $S$ and $S'$ are \textbf{dependent subsets} s.t. $\forall v\in V_1 \times V_2$ the following holds: 
    $$ \sum_{c_i \in S} c_i = \sum_{c'_j\in S'} c'_j$$

\end{definition}

In other words, two markets being independent means any resolution of one market leaves all possible resolutions of the other market as possibilities. When two markets are dependent, however, there is one or more conditions in one market where if any of those conditions resolve to true (or all resolve to false), there is one or more conditions in the other market where one of them must resolve to true (or respectively all to false).

In this work, we focus on single market and two market dependencies as an initial study of arbitrage on prediction markets built atop blockchain protocols. One can imagine extending the above definition to a set of $n$ markets $\{M_1,M_2,\dots,M_n\}$, where $M = M_1\times M_2\times \dots \times M_n$ is the set of possible unique resolutions of the join $n$ markets. If $|M|=\Pi_{i\in[1,n]}|M_i|$, the outcomes of all markets are independent of each other. If $|M|<\Pi_{i\in[1,n]}|M_i|$, there exists some dependency between some subset of the markets. Characterizing these dependencies quickly becomes more complicated and is out of the scope of this work.

\subsection{Arbitrage Labels Taxonomy}

In this section, we present a taxonomy of the two primary forms of arbitrage that can arise in order-book-based prediction markets.

\subsubsection{\singleArbitrage{} Label}

In prediction markets, the price of a "\textit{YES}" token is interpreted as the market-implied probability of that outcome occurring. For events with multiple mutually exclusive outcomes—each represented as a distinct condition—the sum of the "\textit{YES}" token prices should, in theory, equal 1. When this condition is violated (e.g., if the total is less than 1), arbitrage opportunities arise. Traders can take long positions on all outcomes (or short positions if the sum exceeds 1), securing a risk-free profit when the market resolves. We refer to this as \textbf{\singleArbitrage}, where arbitrageurs restore market consistency by adjusting positions or submitting orders until probabilities realign with logical constraints.

\begin{definition}[\singleArbitrage] \label{def:singlearbitrage}
Let $M$ be a market with a set of conditions $\{ C_1, \dots, C_n\}$ where $\mathbf{val}(Y_i,t)$ is the price of the \text{Yes} position for condition $C_i$ at time $t$. We say that a \textbf{Long \singleArbitrage} opportunity exists at time $t$ if:

$$ \sum_{i} \mathbf{val}(Y_i,t) < 1 ,$$

\noindent and a \textbf{Short \singleArbitrage} opportunity exists at time $t$ if:

$$ \sum_{i} \mathbf{val}(Y_i,t) > 1 .$$

\end{definition}

In a Long \singleArbitrage, the total cost of acquiring one unit of each of the "\textit{YES}" tokens is less than 1. 
Since one of the conditions must resolve to true (i.e., the final token value will be 1), this position allows for a \textit{guaranteed} profit of $1- \sum_{i} \mathbf{val}(Y_i,t)$.

In a Short \singleArbitrage, since the total cost of acquiring one unit of each "\textit{YES}" token is more than 1, this implies the "\textit{NO}" tokens are undervalued. One could buy one unit of all "\textit{NO}" positions, and when the market resolves, the \textit{guaranteed} profit is 
$$n-\sum_{i} \mathbf{val}(N_i,t)=\sum_{i} \mathbf{val}(Y_i,t)-1$$
where $N_i$ is the "\textit{NO}" position, as the sum value of one token of all "\textit{NO}" resolutions is $n-1$. 

An alternative shorting strategy is for an arbitrageur to buy a unit of both positions for each condition (i.e., create a \textbf{Split} for each condition for 1USDC) and then sell the "\textit{YES}" position right away, immediately taking advantage of the overvaluation and gaining the profit of $\sum_{i} \mathbf{val}(Y_i,t)-1$.\footnote{We can use the same definition to define the much simpler \textbf{arbitrage in a single condition} if we think of the "\textit{YES}" and "\textit{NO}" positions as effectively two conditions: If they sum to $<1$, an arbitrageur buys both positions; If they sum to $>1$, an arbitrageur \texttt{SPLIT}s the condition and sells both positions.}

With both long and short \singleArbitrage, the profit gained is the difference between the sum probabilities of all the "\textit{YES}" conditions and 1 (i.e., $|\mathbf{val}(Y_i,t)-1|$). We note that due to the order book nature of Polymarket, each of the above trades is non-atomic; thus, there is always some risk in attempting the arbitrage.

\subsubsection{\multArbitrage}

Given two dependent markets as defined above, we can express the precise market conditions that allow for arbitrage and the resulting arbitrageur strategy. A \multArbitrage{} opportunity arises between two markets when it is possible to construct a portfolio of bets across conditions in both markets such that at least one bet is guaranteed to win. 

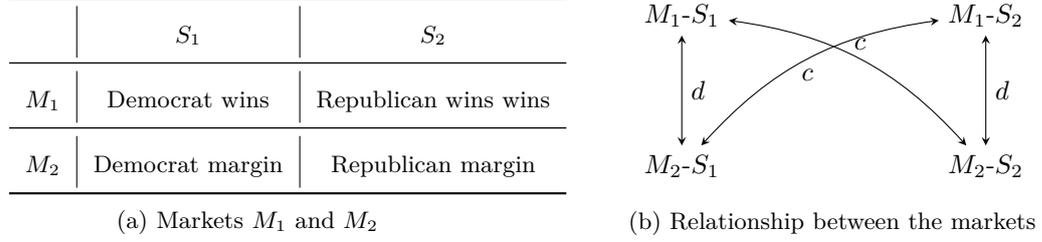
\begin{figure}[h!]
\centering

\begin{minipage}[c]{0.45\textwidth}
\centering
\renewcommand{\arraystretch}{1.5}
\begin{tabular}{c|c|c}
\toprule
 & $S_1$ & $S_2$ \\
\midrule
$M_1$ & Democrat wins & Republican wins wins \\
\midrule
$M_2$ & Democrat margin & Republican margin \\
\bottomrule
\end{tabular}

\vspace{0.5em}
\small (a) Markets $M_1$ and $M_2$
\end{minipage}
\hfill
\begin{minipage}[c]{0.45\textwidth}
\centering
\begin{tikzpicture}[>=stealth, node distance=2cm]
\node (m1s1) at (0,2) {$M_1$-$S_1$};
\node (m1s2) at (0,0) {$M_2$-$S_1$};
\node (m2s1) at (4,2) {$M_1$-$S_2$};
\node (m2s2) at (4,0) {$M_2$-$S_2$};

\draw[<->] (m1s1) -- node[right] {$d$} (m1s2);
\draw[<->] (m2s1) -- node[right] {$d$} (m2s2);

\draw[<->,bend left=20] (m1s1) to node[above] {$c$} (m2s2);
\draw[<->,bend left=20] (m1s2) to node[below] {$c$} (m2s1);

\end{tikzpicture}

\vspace{0.5em}
\small (b) Relationship between the markets
\end{minipage}

\caption{
        \textbf{Markets and state relationships.} 
        This figure illustrates two dependent prediction markets: \(M_1\), which declares the winner of a state election, 
        and \(M_2\), which specifies the winning margin. For each market, we define two mutually exclusive and exhaustive 
        states, \(S_1\) and \(S_2\). An assignment of a state in one market can imply the corresponding outcome in the other. 
        In the relationship diagram, edges labelled \(d\) indicate that the connected states are \textit{dependent} , and edges labelled \(c\) indicate the states are (share opposite token valueations -- YES in one corresponds to NO in the other) \textit{complementary} (both must evaluate to TRUE or both to FALSE).
    }
    \label{fig:market_dep}
\end{figure}

\begin{definition}[\multArbitrage] \label{def:multiarbitrage}
    Let $M_1$ and $M_2$ be dependent markets with some dependent subsets $S \subset M_1$ and $S' \subset M_2$. We say that a \textbf{\multArbitrage} opportunity exists at time $t$ if either of the following holds:\\
    (1) If $\sum_{c\in S} \mathbf{val}(T_c,t)< \sum_{c'\in S'} \mathbf{val}(T_{c'},t)$, then an arbitrage opportunity exists holding "\textit{YES}" positions for conditions in $S$ and "\textit{YES}" positions for conditions in the complement of $S'$.\\
    (2) If $\sum_{c\in S} \mathbf{val}(T_c,t)> \sum_{c'\in S'} \mathbf{val}(T_{c'},t)$, then an arbitrage opportunity exists holding "\textit{YES}" positions for conditions in the complement of $S$ and "\textit{YES}" positions for conditions in $S'$.
\end{definition}

Note that holding all "\textit{YES}" positions for the complement of a set $S\subset M$, $\overline{S} = M-S$, is equivalent to holding all "\textit{NO}" positions for the subset $S$. Figure~\ref{fig:market_dep} shows the dependency graph for an example pair of markets. 
As with \singleArbitrage, for holding a "\textit{YES}" position for some condition $C$, one can either buy a unit of the "\textit{YES}" token or buy 1 USDC of the condition in a Split and sell the "\textit{NO}".
With \multArbitrage{}, however, we consider arbitrage strategies only in holding the "\textit{YES}" positions of complementary subsets across two markets (e.g., "\textit{YES}"  of $M_1-S_1$ and $M_2-S2$). 
The profit of the \multArbitrage{} is the difference in the market values of the dependent conditions, i.e., $|\sum_{c\in S_1} \mathbf{val}(T_c,t) - \sum_{c'\in S_1} \mathbf{val}(T_{c'},t)|$\footnote{We don't consider strategies holding the "\textit{NO}" position as multiple "\textit{NO}"s can be true at once (by design), so there is not a symmetric profit between the positions of the two markets.}.

\subsection{Arbitrage Analysis}

Given the above definitions, we perform the following arbitrage analysis on Polymarket data (cf. Section~\ref{sec:data_colection} for our data collection methodology). 

\begin{itemize}
    \item[(i)] In Section~\ref{sec:results_dependent_markets}, we employ an instance of the LLM \textit{DeepSeek-R1-Distill-Qwen-32B}~\cite{deepseek_qwen32b} together with market description data to capture the state space of possible market resolutions to then infer semantic dependency between pairs of markets and their conditions. 

    \item[(ii)] In Section~\ref{sec:results_detected_arbitrage}, we analyze historical orderbook data to detect when \singleArbitrage{} opportunities existed for each market, and when \multArbitrage{} opportunities have existed for pairs of dependent markets in our measurement window.
    
    \item[(iii)] In Section~\ref{sec:results_captured_arbitrage}, we further examine the orderbook data to determine whether any participants exploited these arbitrage opportunities, and explore the behavior of key actors involved.

\end{itemize}

\section{Data Collection}
\label{sec:data_colection}

For this study, we analyzed Polymarket data spanning markets that resolved over the period of one year, from 1st April 2024 to 1st April 2025. Our analysis required both the textual descriptions of each market queried from the Polymarket API \cite{polymarket_pyclob_2024} and the on-chain historical record of executed bids.

\subsection{Market Descriptors}

We retrieved market metadata directly from the Polymarket API using the official Python client \cite{polymarket_pyclob_2024}.\footnote{A tutorial on accessing this data is available in \cite{whittaker2024polymarket}} 
In this section, we present only the relevant fields needed to understand the market and condition structures, as well as their connections with the underlying smart contracts. In Appendix~\ref{appendix:descriptors} we provide a full overview of all fields.

To detect arbitrage opportunities, it is essential that dependent markets share the same end date and describe the same underlying event in different ways. However, we observed inconsistencies in the \texttt{end\_date\_iso} field among markets with the same market ID, which should not occur, as all associated conditions are expected to resolve simultaneously. To address this issue, we computed the most frequent end date among conditions within the same market ID. In cases where multiple dates shared the highest frequency, we selected the latest date as the canonical \texttt{end\_date\_iso}.

\textbf{Our data set consists of two types of markets: 8659 single-condition markets, and 1578 multiple-condition markets (i.e., NegRisk markets) made up of 8559 conditions, totalling 17218 conditions across all markets.}

\subsubsection{Topic Analysis}
\begin{figure}[t]
    \centering
    \resizebox{0.85\linewidth}{!}{ 
        \begin{minipage}{\linewidth}
    \begin{subfigure}{\linewidth}
        \centering
        \includegraphics[width=\linewidth]{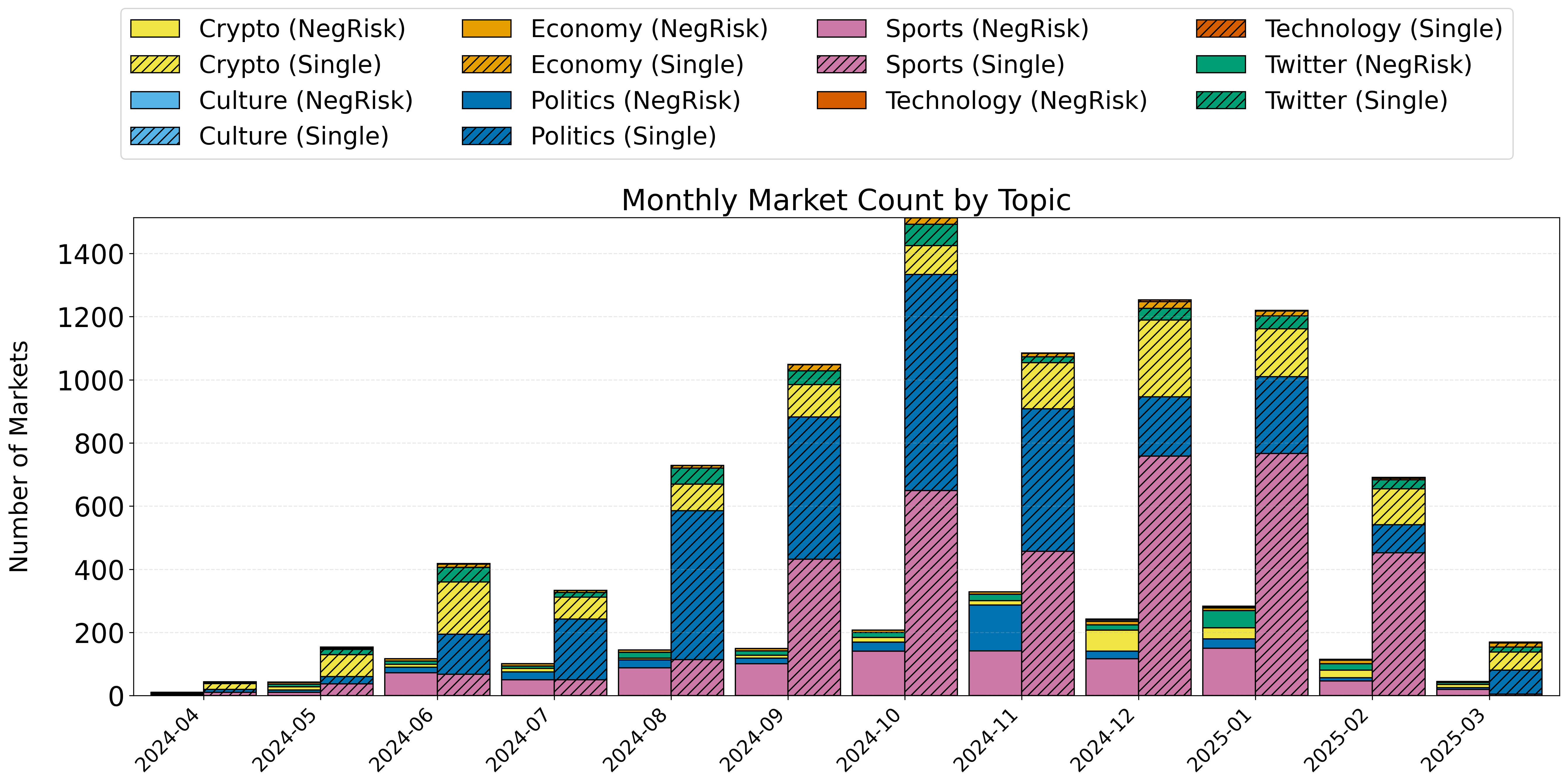}
    \end{subfigure}
    \begin{subfigure}{\linewidth}
    \centering
        \includegraphics[width=\linewidth]{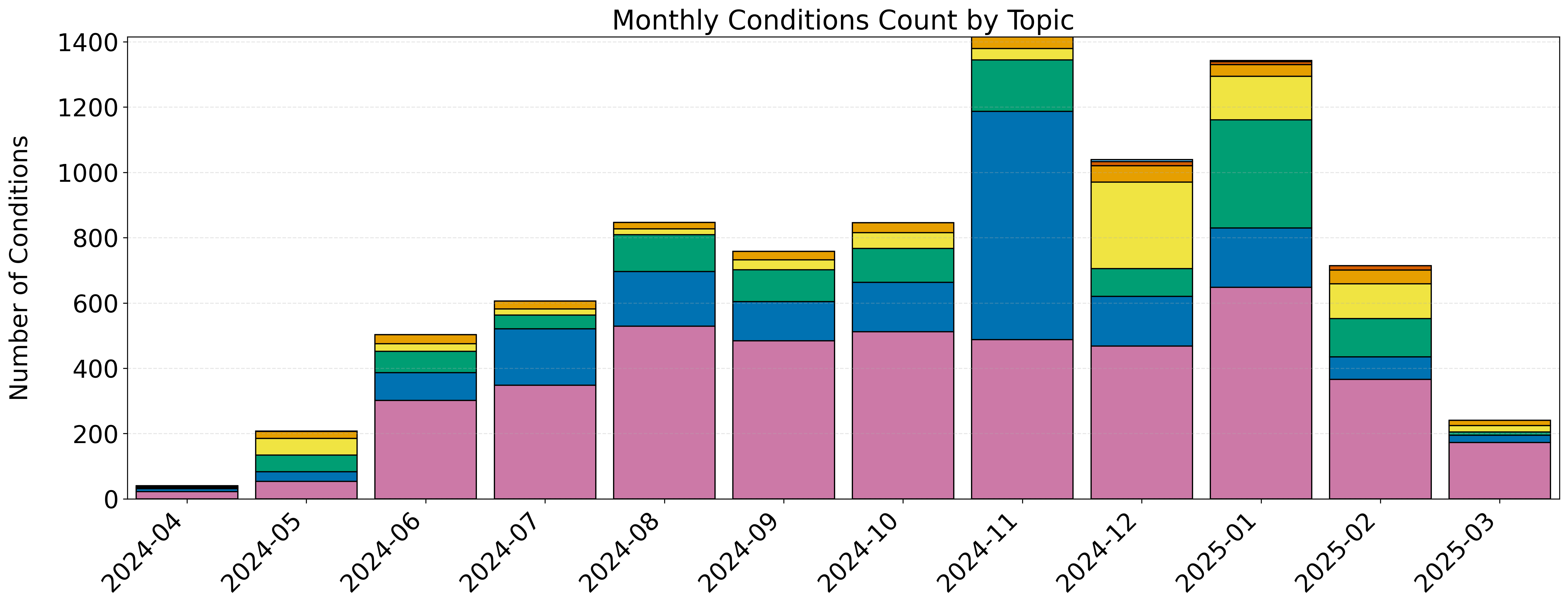}
    \end{subfigure}
    \end{minipage}
    }
    \caption{Top shows the total number of markets by topic and end-date, differentiating between the single-condition markets and the neg risk markets. Bottom shows the same value but for total conditions in the neg risk markets. Overall, politics and sports dominate in popularity.}
    \label{fig:markets_overview}
\end{figure}

We group markets into primary topics using the topic categories listed on the Polymarket website \cite{polymarket2025}: \texttt{["Politics", "Economy", "Technology", "Crypto", "Twitter", "Culture", "Sports"]}. We first use \textit{Linq-Embed-Mistral} model~\cite{linq_embed_mistral} to generate vector embeddings for both market questions and topics. The model was the best-performing \textit{open-source} text embedder~\cite{mteb_leaderboard} at the time of conducting the experiments. We then compute the cosine similarity between each question's embedding and all topics' embeddings, and assign the question to the topic with the highest similarity. 
Figure~\ref{fig:markets_overview} plots the number of markets per topic per end date for our measurement period. We see that the top categories are Politics and Sports, with the U.S. election falling in our measurement period, which clearly illustrates a rise in Polymarket activity.

To validate the classification performance, we randomly sampled 100 instances from the dataset and manually labeled them which yielded an accuracy of 92\%. In several cases, markets framed as ``\textit{If politician X says Y}'' were classified under \textit{Politics}, which aligns with our interpretation of the market context. However, certain markets revealed limitations in the labeling scheme; for example, markets such as ``\textit{If the weather will be X}'' were sometimes categorized under \textit{Twitter}, despite lacking a clearly appropriate label. Since the primary goal of this labeling was to reduce the search space, we aligned the categories with those used by Polymarket. 

In Section~\ref{sec:results_detected_arbitrage}, when analyzing dependencies between pairs of markets, we consider only pairs of markets within a given topic and end date to limit our search space to markets more likely to relate to the same event (e.g., markets about a match between team A and team B should both resolve on the day of the match when the outcome is known. This is due to the controlled/centralized nature of market creation). 

\subsection{Historical Bid Data} 
To get Polygon on-chain data for Polymarket, we use the Alchemy public node API~\cite{alchemy}. While users submit bids directly to Polymarket, those bids which are matched are registered on-chain by Polymarket operators. All conditional tokens (a "\textit{YES}" or "\textit{NO}" token for each market condition) are an instance of an ERC-1155 Conditional Token, which is managed by the Polymarket Conditional Token Contract at address \texttt{0x4D97DCd97eC945f40cF65F87097ACe5EA0476045}. While there are several contracts that make up the logic of Polymarket, all trades and liquidity operations are eventually registered as an \texttt{EVENT} by this contract. In particular, we care about the \texttt{OrderFilled}, \texttt{PositionSplit}, and \texttt{PositionsMerge} events. 

Section~\ref{sec:background} describes the various bids that take place, which can be a simple match of a buy and sell order, or a match of two buys for each side of a condition, such that new tokens are created. Ultimately, the \texttt{OrderFilled} event captures any time a token is traded for USDC and who is the buyer/seller (the seller can be the Polymarket exchange contract if new tokens are minted). Whenever new tokens are minted (or conversely destroyed), the \texttt{PositionSplit} event registers the USDC being locked (or conversely \texttt{PositionsMerge} and withdrawn) and by whom (again, this can be the Polymarket exchange). 

\begin{figure}[t]
    \centering
    \resizebox{0.85\linewidth}{!}{ 
        \begin{minipage}{\linewidth}
    \begin{subfigure}{\linewidth}
        \centering
        \includegraphics[width=\linewidth]{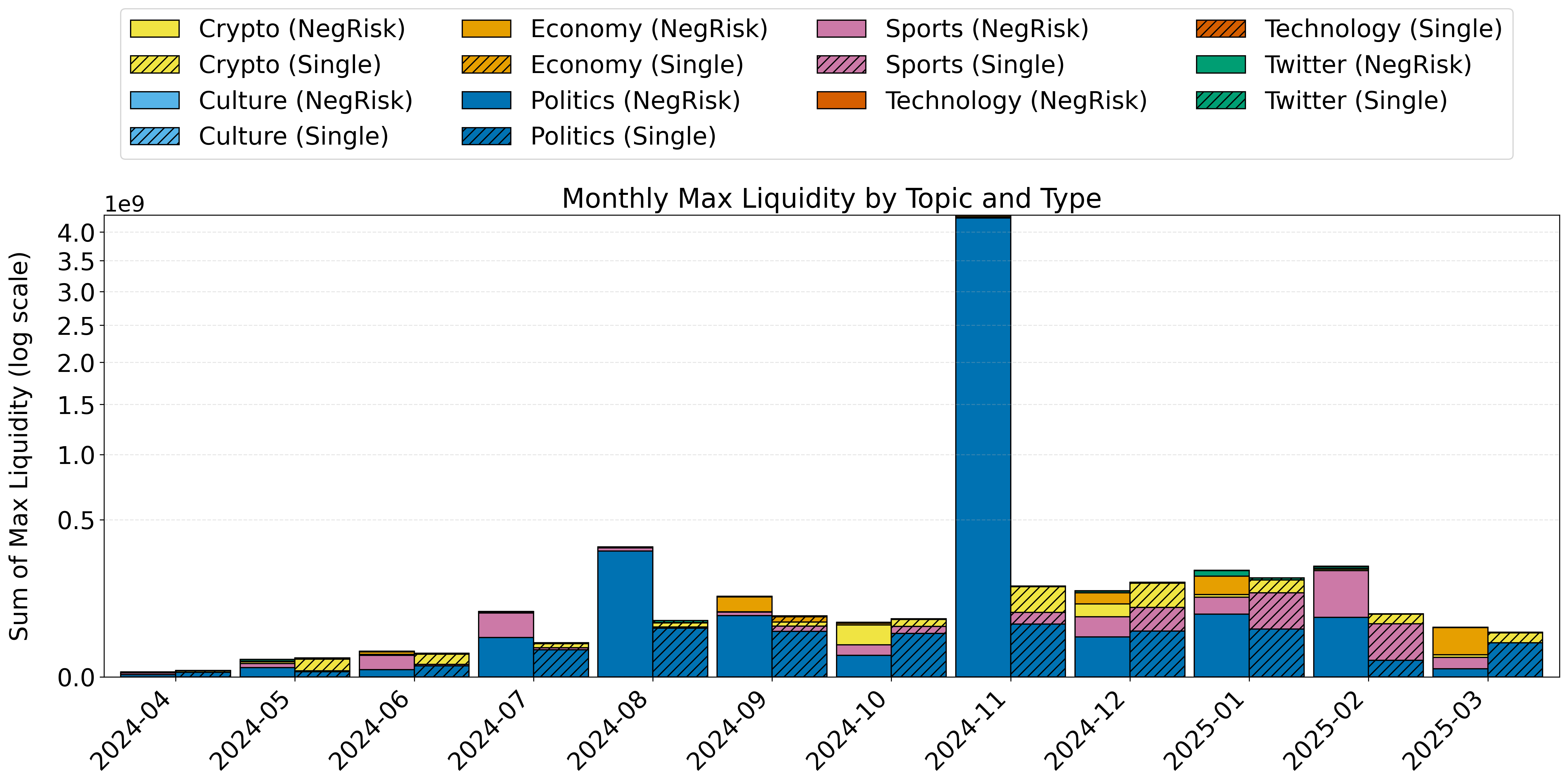}
    \end{subfigure}
    \begin{subfigure}{\linewidth}
    \centering
        \includegraphics[width=\linewidth]{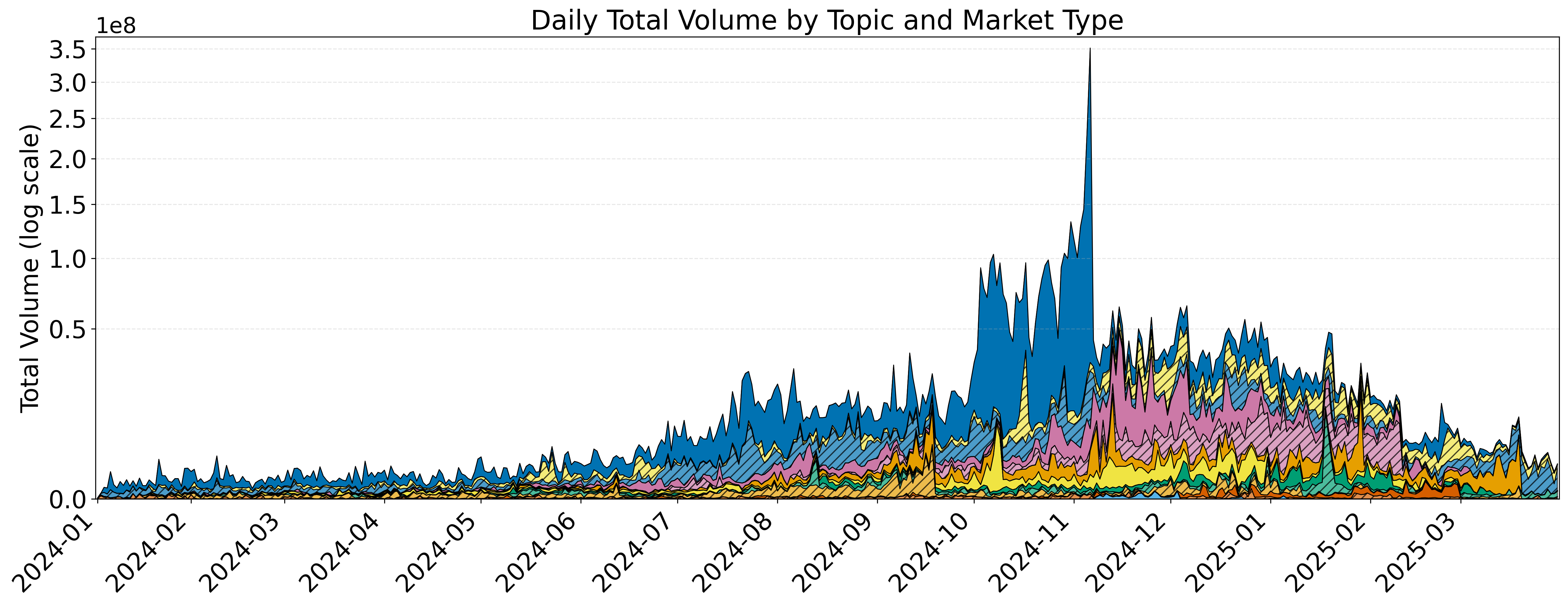}
    \end{subfigure}
    \end{minipage}
    }
    \caption{(Top) Total liquidity per market by end date and topic. (Bottom) Total volume of executed bids by market type over time. We see the U.S. election markets were the primary driver of activity in Polymarket at the time.
    }
    \label{fig:volume_overview}
\end{figure}

We query the transaction traces for the Polymarket Conditional Token Contract for blocks from 1st January 2024 to 1st April 2025 and filter for the three events and only conditional tokens for the markets in our measurement period. Figure~\ref{fig:volume_overview} shows the total locked volume in markets ending in each month, with the U.S. election surpassing all other markets, and the total volume over time by market cap.

\section{Markets Analysis: Detecting Market Dependencies} 
\label{sec:results_dependent_markets}

\begin{figure}[t]
    \centering
    \includegraphics[trim=0 125 75 0, clip,width=0.9\linewidth]{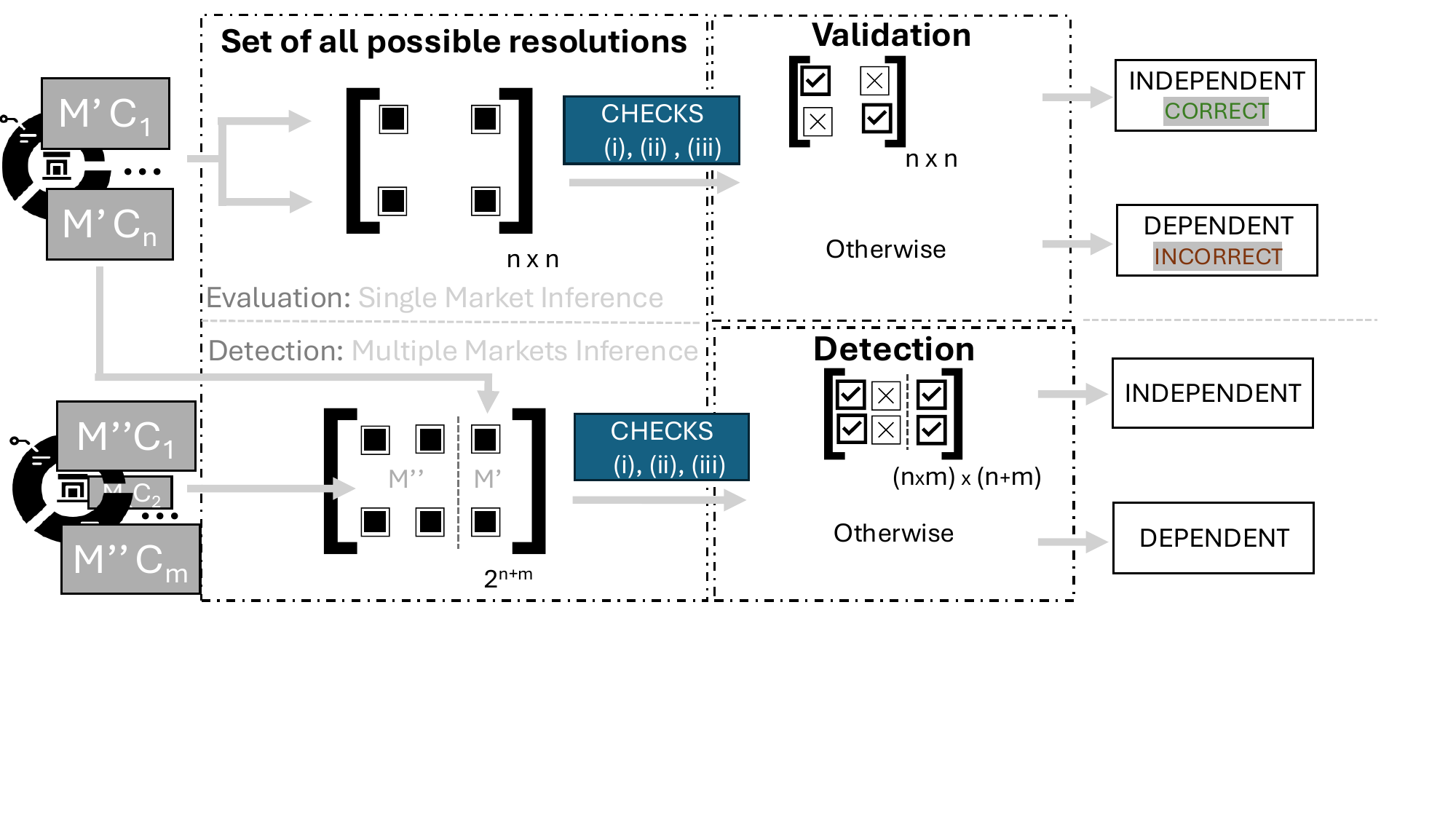}
    \caption{Our approach to detect market dependencies in a nutshell.}
    \label{fig:architecture}
\end{figure}

We begin by developing a methodology that leverages LLMs to automatically detect semantic dependencies between markets as defined in Section~\ref{sec:defmarketdepency} and represented in Figure~\ref{fig:architecture}. 
While our goal is to determine whether conditions in two separate markets are dependent, we begin with the single market case, where we know all conditions are dependent, to validate the capabilities of the LLM for reasoning.

\subsection{Single Market Inference}

Given a collection of conditions, our goal is to output the set of all possible resolutions of the markets. Our approach uses the reasoning capabilities of an LLM to evaluate the logical consistency between a given assertion and a set of statements, using structured inference. We abstract away the prediction market context of this task, and instead focus on \textit{semantic} dependency between statements.

Our first approach iterates exhaustively through the conditions. We take the set of conditions of a market (structured as questions in the question variable), and assert to the LLM that one condition is true and ask it whether the remaining conditions can be true. An example of a market \( M \) is defined by the following condition questions: "Will team A win the Turtle Bowl?", "Will team B win the Turtle Bowl?", and "Will team A and team B tie in the Turtle Bowl?". In this setup, the general structure for the LLM prompt is as follows: the \texttt{assertion\_statement} corresponds to the specific condition being evaluated (e.g., ``Team A will win the Turtle Bowl"), while the remaining condition questions are assigned indices to define the mapping between column indices and condition indices.

Again, for conditions in the same market, the LLM should return that the remaining questions must resolve to \textit{False}. While this approach works, it would involve $n$ calls to the LLM for a market with $n$ conditions. We thus compose queries such that the space of all possible outcomes is computed in a single call, and the LLM returns a JSON representation of the output space (see Appendix~\ref{appendix:prompt} for the full prompt).

We run our prompt for all NegRisk markets on election day 2024-11-05 with assigned\_topic as Politics and test the following conditions: (i) The LLM returns a valid JSON, as sometimes the LLM gets stuck in a logical loop. (ii) The returned JSON is correct, meaning there are $n$ unique vectors, and (iii) each vector has exactly one true condition. \textbf{We get that out of the 128 markets tested, only 4 do not return a valid JSON, with 101 (81.45\% ) fulfilling all conditions.}

Looking at the failed prompts, we observe that the LLM cannot handle too many conditions at a time -- a known limitation of LLMs failing to handle large prompts well~\cite{nec2024,microsoft2024}. 
 
This would be further problematic when we check the dependency between two markets, as we will concatenate conditions between markets (further increasing the prompt size). 

We thus pre-process markets with more than 4 conditions and reduce them to 4 or fewer conditions with the most total traded volume (in \textit{YES}, and \textit{NO} tokens), and a fifth condition to catch all other outcomes. In Appendix~\ref{appendix:liqui-4} we show that over 90\% of all liquidity in a market resides in the top 4 conditions.
Note that the fifth condition is a logical "OR" of all remaining conditions, thus it preserves logical dependencies. 

\subsection{Multiple Markets Inference}

We extend the LLM analysis to pairs of markets. Recall that due to the centralized nature of how markets are created, markets relating to the same event should share an end date. We thus check pairwise dependencies across markets within the same topic group in the same day. Given a pair of markets, we take the union of all conditions (in their reduced format to max 5 conditions), and pass them to the LLM prompt as one set. We perform the following consistency check: (i) the LLM returns a correct JSON, (ii) for each vector of assignments of conditions, there is exactly one true value in each set and (iii) the set of vectors is of size at most $n$+$m$ for pairs of reduced markets of size $n$ and $m$. Conditions (ii) and (iii) check that the set of vectors returned by the LLM are a valid assignment for a market (i.e., exactly one condition in a market must resolve to "True").\footnote{We see a pattern in the pairs that fail this check, where there is a loop in the LLM's reasoning, resulting in it eventually returning the exhaustive set of "True"/"False" assignments of a vector of size $2^{n+m}$.}

\noindent{\bf Non-US Elections.} Of the {2267 pairs of markets checked outside the primary U.S. election group ("Politics" group with end date November 5th, 2024), 30 did not return any JSON, and 203 failed the other two checks. Of the remainder, 2033 pairs were classified as independent, and 1 pair in this set was classified as dependent. Table~\ref{tab:dependent_no_ele} in App~\ref{appendix:dependent} shows the details of this pair. While the resolution of some conditions impact the state space of outcomes for others, it does not strictly satisfy our \multArbitrage definition of Section~\ref{sec:defmarketdepency}. Such dependencies are left for future work.

\noindent{\bf US Elections.} 
Taking the Politics group with end date November 5th, 2024, we have 128 NegRisk and 177 Single markets. We run all 46360 pairs of markets, we get that 353 do not return a JSON, and 4374 returned an output which does not satisfy (ii) or (iii). In the end, we get a total of \textbf{40057 independent markets and 1576 markets characterized as dependent}. Of the pairs characterized as dependent, 129 are between two Single markets, 1353 are between a NegRisk and a Single market, and 94 are between two NegRisk markets. We run the vector of assignments through a checker that verifies whether there exist subsets in the two markets that satisfy Definition 3 \footnote{This check involves computing all possible subsets in a pair, an exponential task, but tractable for 5 conditions}. We get 4 NegRisk-NegRisk, 94 Single-Sigle, and 276 NegRisk-Sigle pairs. \textbf{We manually check these 374 pairs and get 11 NegRisk-NegRisk and 2 NegRisk-Single pairs, which satisfy our \multArbitrage~definition.} The majority of false positives corresponded to pairs of markets with some weaker notion of dependency (e.g., a market for who wins a certain swing state, and another for who wins the election) or the LLMs conflating U.S. election outcomes (e.g., the popular vs. electoral college votes, Senate vs. House elections, etc.). Eight markets contributed to 1469 invalid pairs, largely due to their own inherent ambiguity, they are listed in the Appendix~\ref{tab:invalid_markets}.  

\section{Markets Analysis: Detecting Arbitrage Opportunities}
\label{sec:results_detected_arbitrage}

We now use the history of executed bids to explore when arbitrage opportunities existed in each market. We take all executed bids for each position (this includes both the USDC amount and the token amount) and calculate a weighted average price for each position, weighted by the token amount. We compute this average over some time frame $T$. We note that arbitrage happens during periods of volatility, so the larger the time frame, the less volatility we capture; we take the average over one block, carrying forward the last known price for up to 5K blocks (2.5 hours) if a token isn’t traded; Otherwise, we set the price to 0 (a token that stops being traded effectively has no value). Additionally, we want to capture arbitrage when the outcomes of markets are not yet known (i.e., there is sufficient uncertainty such that the market is liquid -- there is a market to buy/sell the different outcomes), so we only look at times when no position (no token) is worth more than \$0.95 (that is, it has a predicted outcome greater than 95\% probability).
 
Lastly, we limit our analysis to opportunities with a profit of at least \$0.05 on the dollar to focus on the higher-reward opportunities given the risk.\footnote{Since placing multiple orders in an order book is non-atomic (only a subset of the attempts may succeed), there is some inherent risk to attempting arbitrage.} 

We show an example condition in Figure~\ref{fig:cond_ex} with the price calculations over time for each position and whether an arbitrage opportunity existed ($|1-\text{VWAP Sum} |>0.02$). We see how market uncertainty creates opportunity for arbitrage, and show that players actually capitalized on these opportunities (and at a larger profit margin than our averaged estimates). 

\begin{figure}[t]
  \centering
    \includegraphics[width=\linewidth]{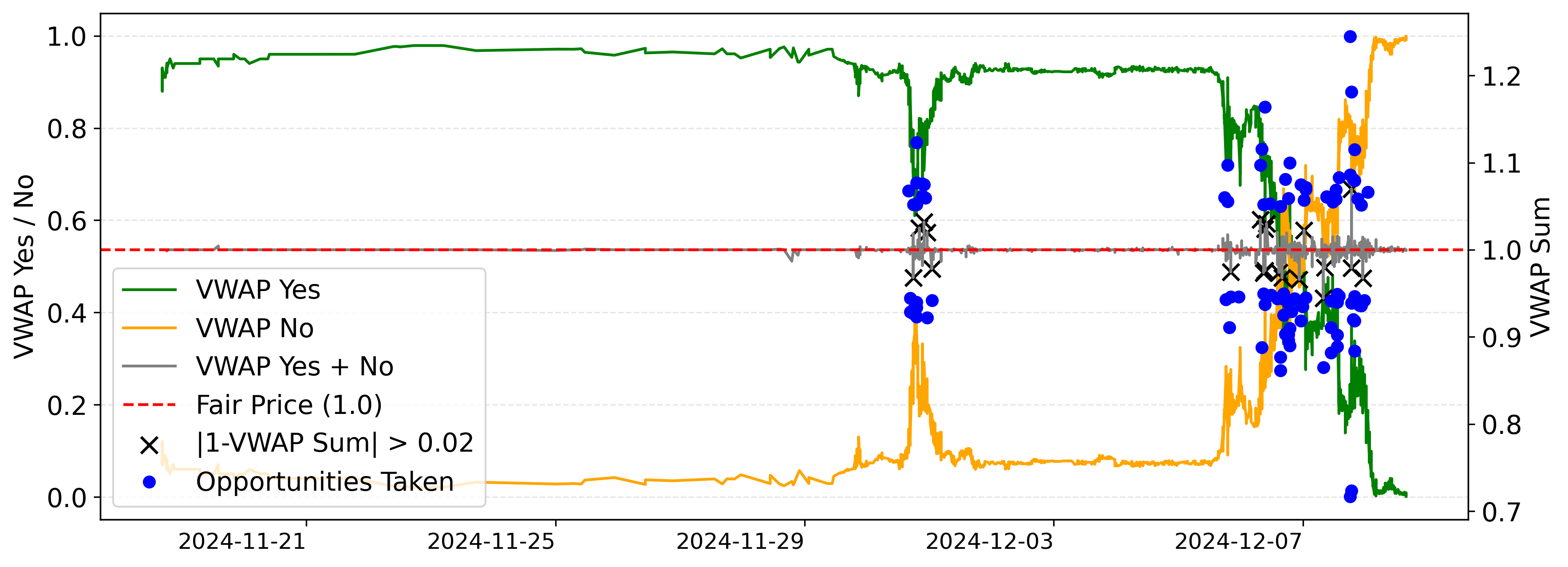}
  \caption{A view of the market behavior of the "Will Assad remain President of Syria through 2024?" condition. We plot the VWAP price of each position, when we detect an arbitrage opportunity ($|1-\text{VWAP Sum} |>0.02$), and the events where an arbitrageur profited from the opportunity (computed in Section~\ref{sec:results_captured_arbitrage}. Because we calculate token prices from an average of executed bids, we underestimate the margin of profit an arbitrageur is able to realize.} 
  \label{fig:cond_ex}
\end{figure}

\subsection{Arbitrage Within Single Conditions}

\FloatBarrier

\begin{figure}[t]
  \centering
  \begin{subfigure}[b]{0.4\textwidth}
    \centering
    \includegraphics[width=\linewidth]{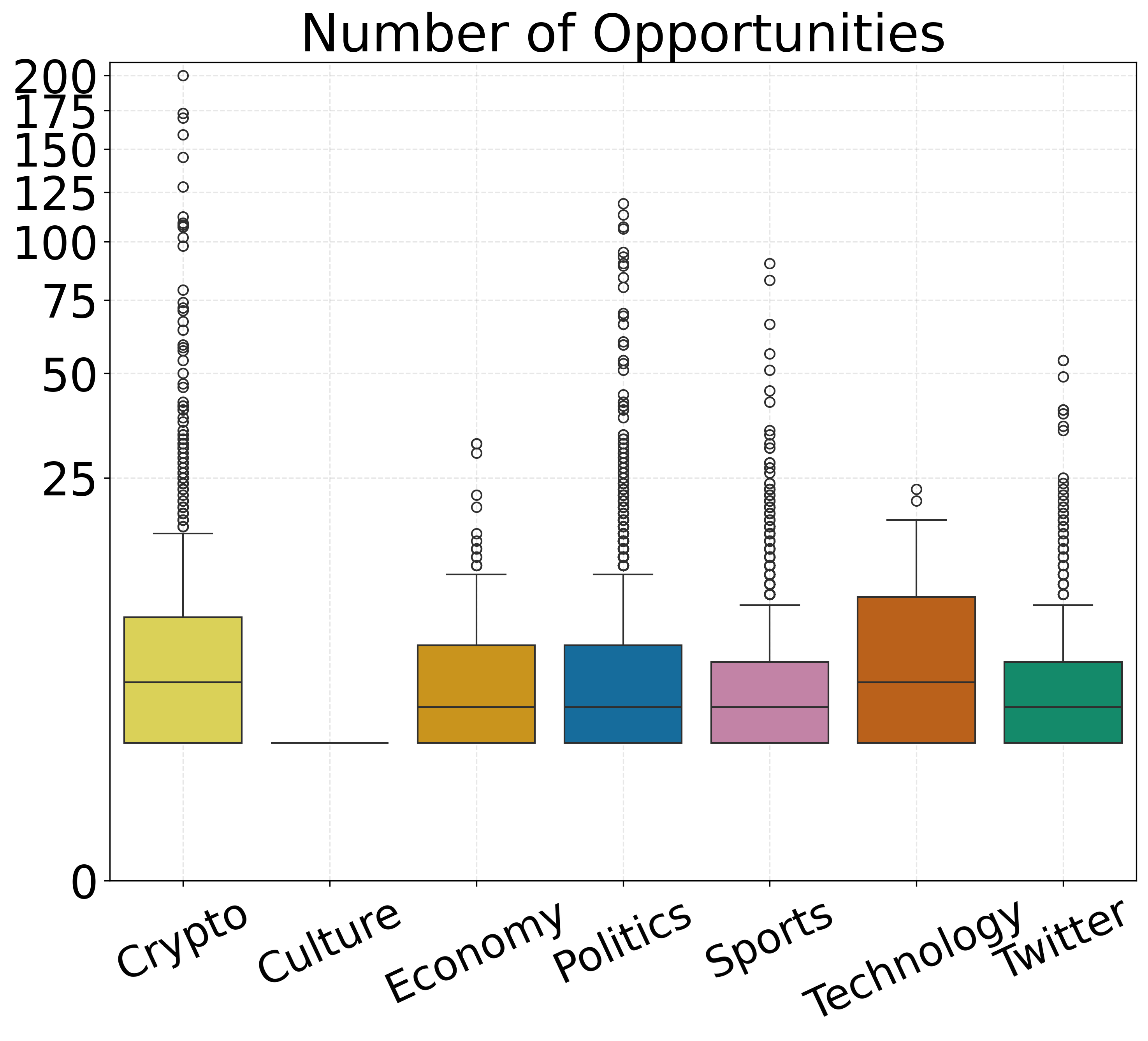}
    \vspace{-15pt} 
  \end{subfigure}
  \hspace{1.5em}
  \begin{subfigure}[b]{0.4\textwidth}
    \centering
    \includegraphics[width=\linewidth]{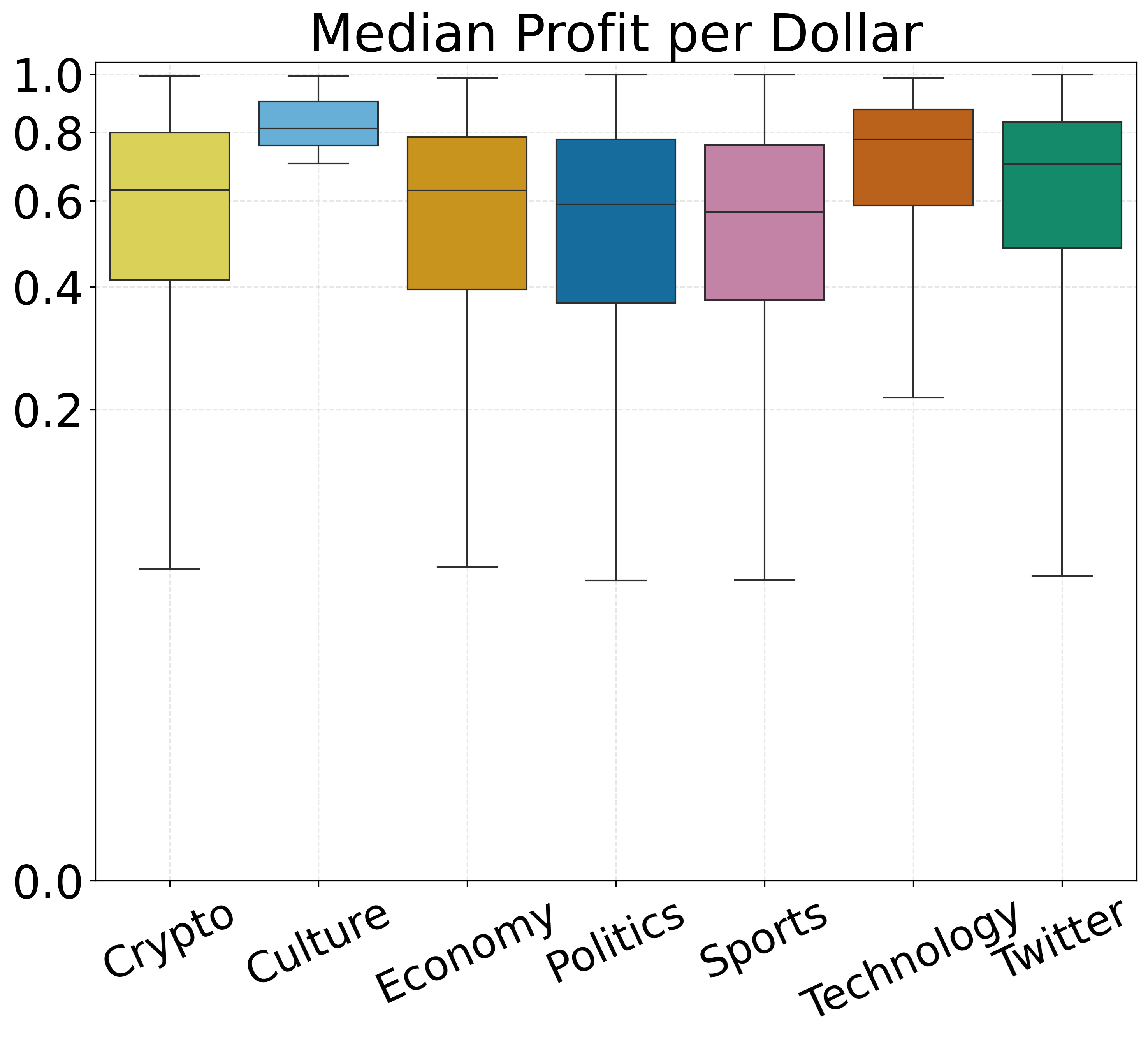}
    \vspace{-15pt} 
  \end{subfigure}
  \caption{Arbitrage opportunities detected within a single condition. In Appendix~\ref{app:detect_cond}, we show NegRisk and Single Markets separately. We see that most conditions have only a few opportunities, with Crypto having the biggest outliers. Most surprisingly, the median profit on the dollar for these opportunities is much higher than our 2 cents bound, and highlights large market inefficiency. 
  } 
  \label{fig:cond_arb_box}
\end{figure}

\begin{figure}[h!]
    \centering
    \resizebox{0.85\linewidth}{!}{ 
        \begin{minipage}{\linewidth}
            \begin{subfigure}{\linewidth}
                \centering
                \includegraphics[width=\linewidth]{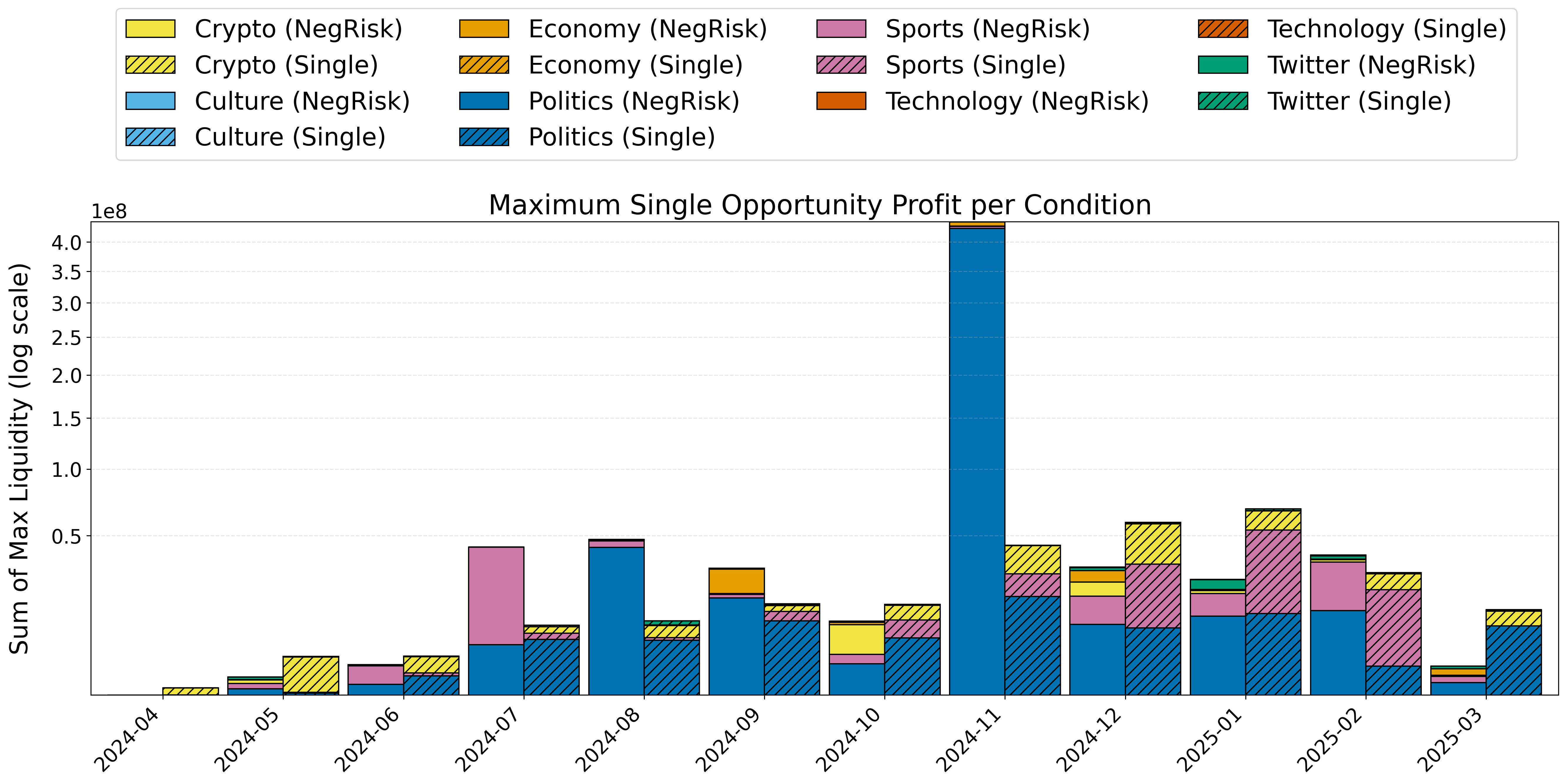}
            \end{subfigure}
            
            \begin{subfigure}{\linewidth}
                \centering
                \includegraphics[width=\linewidth]{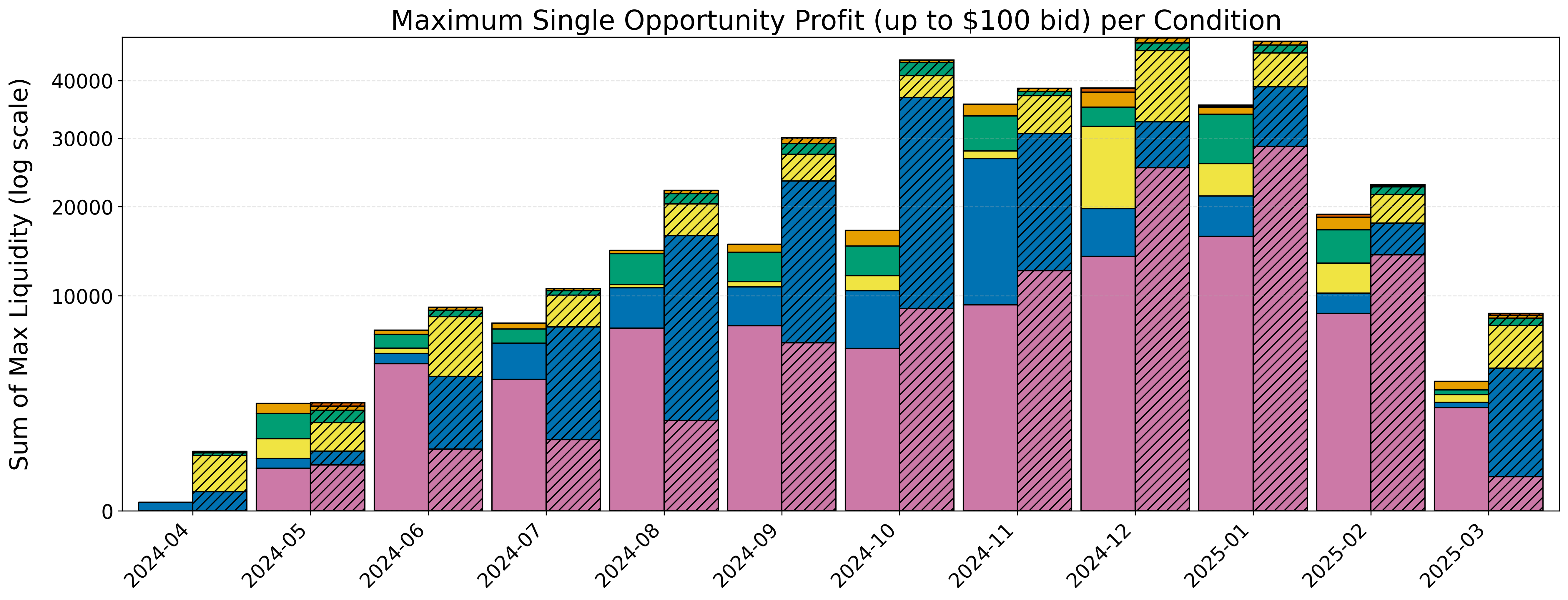}
            \end{subfigure}
        \end{minipage}
    }
    \caption{We explore the total arbitrage possible if an arbitrageur were to take advantage of the single most profitable opportunity in each condition at the maximum liquidity (top), and up to just \$100 of liquidity (bottom). The latter shows us that the existence of arbitrage is quite split among markets, and many more opportunities exist in Sports markets. Without capping profit (top), we see that the very lucrative opportunities are predominantly in Politics, with higher profits per condition (mostly in NegRisk markets). In Appendix~\ref{app:detect_cond}, we compare results from price averages over 100 blocks and we find that we capture less arbitrage generally, and particularly less in the Politics markets of November and Sports markets of July -- both of which were highly exploited cf. Fig.~\ref{fig:conditions_detected}.}
    \label{fig:con_arb_stack}
\end{figure}

We begin by exploring arbitrage within a single condition, the "\textit{YES}" and "\textit{NO}" positions of a single outcome.
In total, our data set consists of 17.2K conditions (8.56K from NegRisk and 8.66K from Single markets), of these 7,051 conditions have at least one arbitrage opportunity within our parameters (2,628 NegRisk, 4,423 Single), with most conditions having only a few opportunities (see Fig.~\ref{fig:cond_arb_box}). All arbitrage opportunities observed are \textit{long}, meaning the sum of the price of "\textit{YES}" and "\textit{NO}" is less than 1. We see the median price of the sum of conditions (i.e., profit per Dollar) is around \$0.60 for all topics of markets, showing remarkable market inefficiency.

As taking advantage of one arbitrage opportunity may impact the prices afterwards, we bound the possible arbitrage profit by considering the most profitable arbitrage opportunity for each condition. 
We define the maximum profit of an opportunity as the price for both "\textit{YES}" and "\textit{NO}" tokens times the total amount of tokens that exist (i.e., the maximum one could purchase from the market at that time). 
Figure~\ref{fig:con_arb_stack} top shows the sum of this value across all conditions, differentiating between the NegRisk and Single condition markets. We see that most potential profit comes from markets related to politics, particularly those related to the U.S. 2024 presidential election. Assuming one could take advantage of arbitrage at 1\% of the available tokens, this is still millions in potential profit. 

\subsection{Arbitrage Within Markets}

\begin{figure}[t]
    \centering
    \begin{subfigure}{.48\linewidth}
        \includegraphics[scale =0.22]{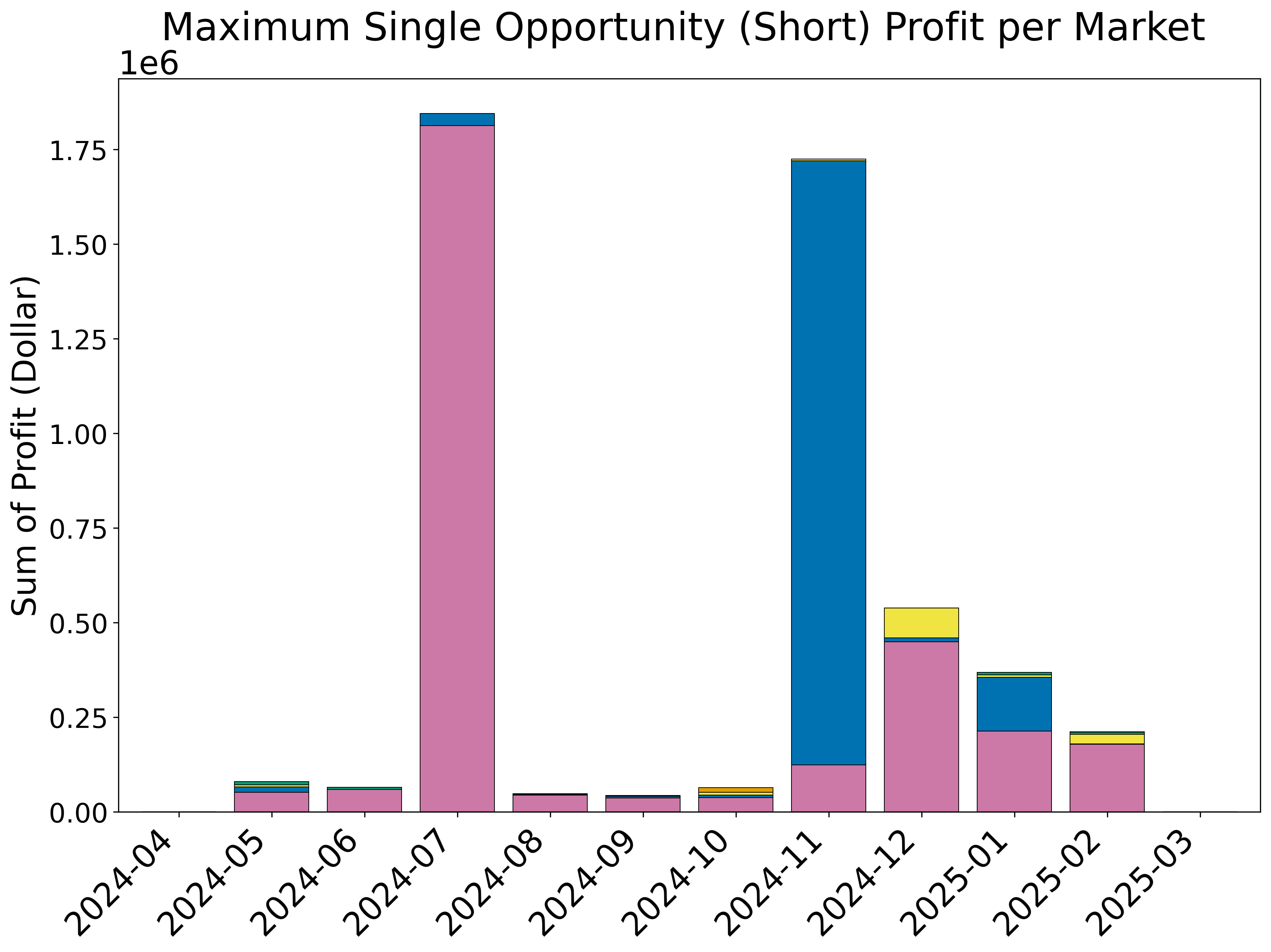}
    \end{subfigure}
    \hfill
    \begin{subfigure}{.48\linewidth}
        \includegraphics[scale =0.22]{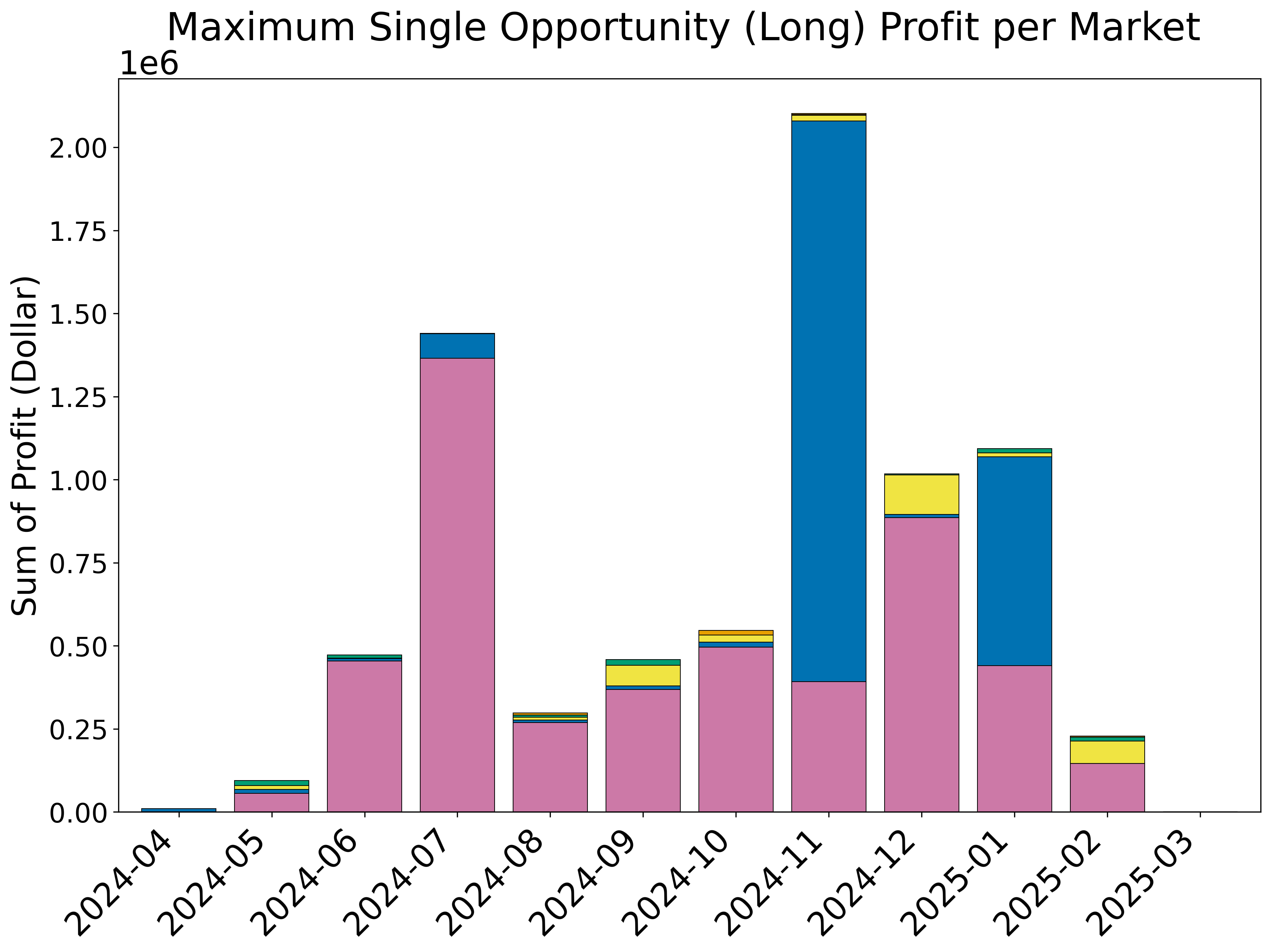}
    \end{subfigure}
    \caption{Here we explore the total arbitrage possible across each market if an arbitrageur were to take advantage of the single most profitable opportunity. Overall, Sports consistently has more profit, particularly in Long opportunities, besides a couple of outliers in Politics, suggesting Sports markets are often overvalued.}
    \label{fig:market_arb_stack}
\end{figure}

Next, we consider arbitrage that is possible between the conditions of a NegRisk market. Of the 1578 NegRisk markets, 662 had at least one arbitrage opportunity. In Appendix~\ref{app:detect_mark} we show the distribution of several characteristics of these opportunities. In general, we see many more opportunities per market, averaging around 100 per market, with quite high outliers, particularly in Sports. Within a market we also see both short (sum of ``\textit{YES}'' is more than \$1) and long (less than \$1) arbitrage, though the average maximum profit on the dollar is more for the long arbitrage, but with short having significant outliers.

We again look at the maximum profit possible for each market from a single opportunity. Since each condition in a market may have a different volume, we take the minimum volume across all conditions that have a probability more than 2\% (when performing arbitrage, it may be low risk to ignore low probability events). Figure~\ref{fig:market_arb_stack} shows the sum of these opportunities for \textit{long} and \textit{short} positions separately. While the average maximum opportunity for both short and long arbitrage across topics is approximately the same (40 cents on the Dollar), we see in the cumulative of these opportunities that Sports dominates across all months except for the U.S. election period. In Appendix~\ref{app:detect_mark}, we again consider a modest arbitrageur with a \$100 budget; there is twice as much profit in long arbitrage, with Sports dominating both, suggesting arbitrage opportunities are generally common within Sports markets.

\subsection{Arbitrage Across Markets}
Lastly, in this section, we look at possible arbitrage opportunities across the 13 dependent pairs of markets during the 2024 U.S. presidential election. For the pairs of markets, we again focus on arbitrage relating to the price of the "\textit{YES}" token of all conditions. In Figure~\ref{fig:pair_arb_dist}, we show the distribution of profit per USDC and maximum profit for each opportunity given the liquidity at the time. The pairs are shown in order of number of opportunities, with Pair 8 having none, and the median among the rest being 8 opportunities (pairs 2, 1, and 4 have 72, 176, and 6630 opportunities, respectively). The markets of Pair 4 are on who will win the popular vote, and if the popular vote winner will the presidency (recall the description of each pair is in Appendix~\ref{appendix:dependent_pairs}). We note that though we were able to observe arbitrage opportunities, they are largely during lower liquidity moments, and at lower profit than the previous sections (the average max profit is around \$100, suggesting total token volume in the market of less than 2K).

\begin{figure}[t]
    \centering
    \begin{subfigure}{.45\linewidth}
        \includegraphics[scale =0.27]{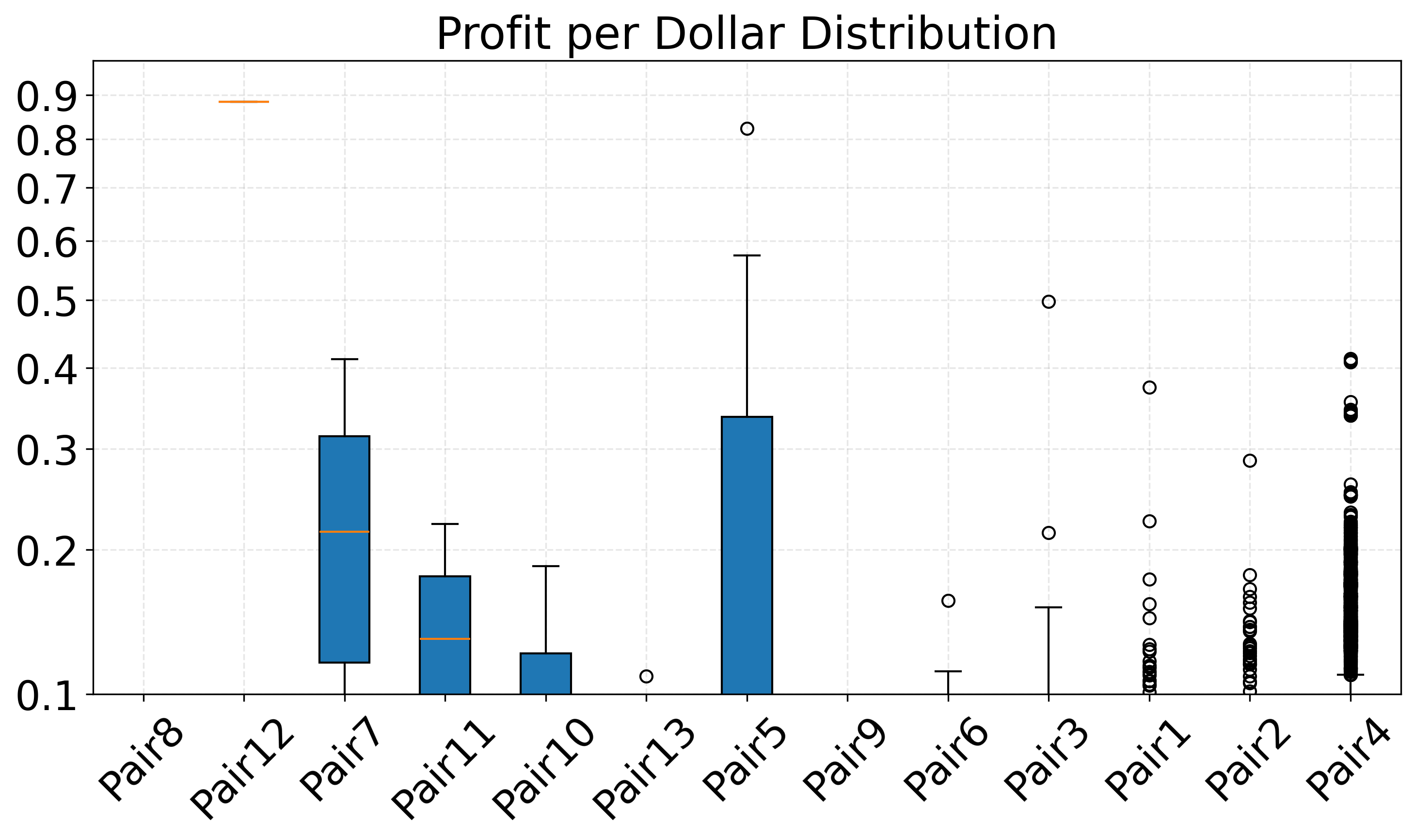}
    \end{subfigure}
    \hfill
    \begin{subfigure}{.45\linewidth}
        \includegraphics[scale =0.27]{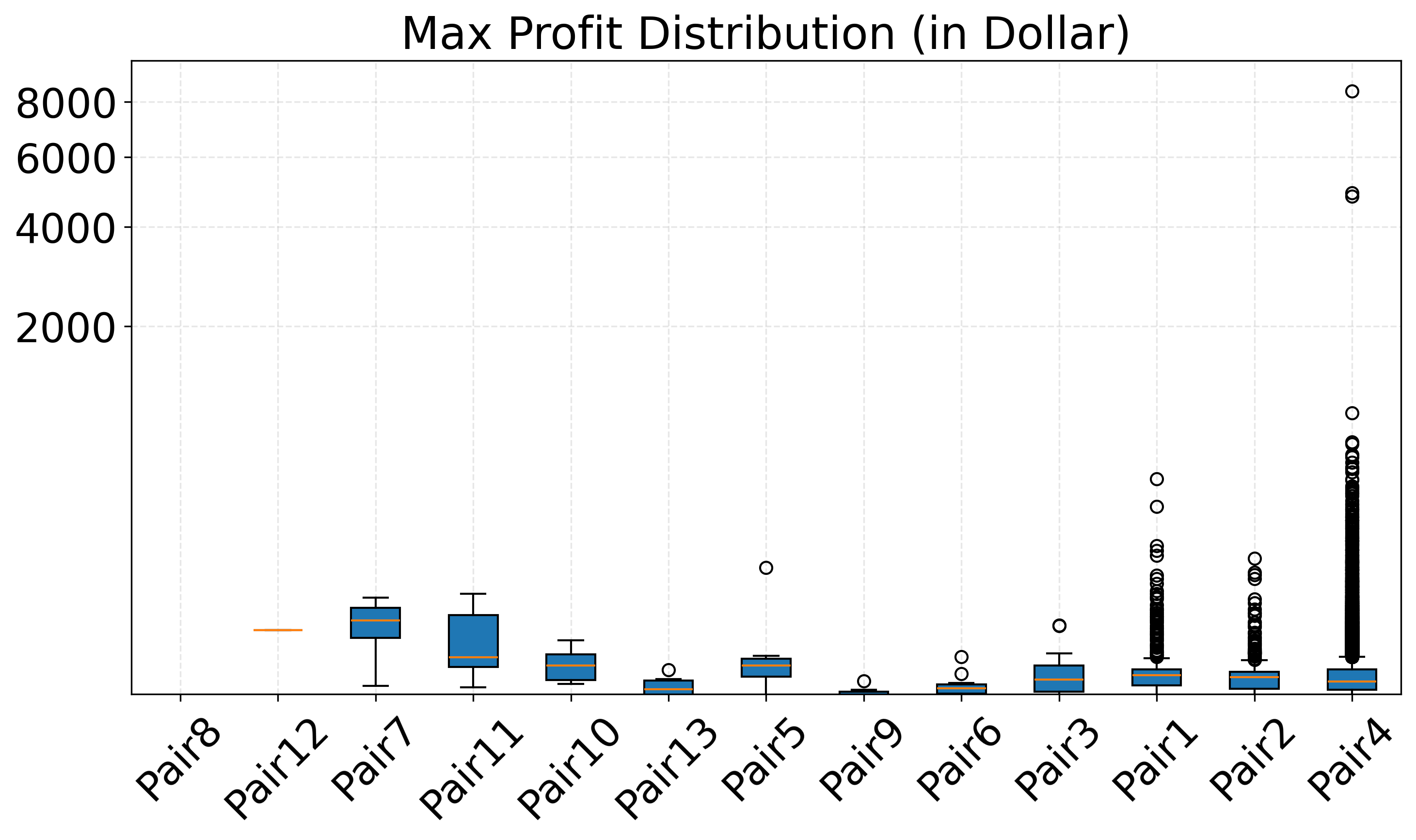}
    \end{subfigure}
    \caption{Distribution of profit per Dollar (left) and profit assuming maximum liquidity (right) for the 11 pairs of dependent markets from the U.S. election in order of number of arbitrage opportunities.}
    \label{fig:pair_arb_dist}
\end{figure}

\section{Markets Analysis: Uncovering Arbitrageurs} 
\label{sec:results_captured_arbitrage}

\FloatBarrier
\begin{figure}[t]
    \centering
    \includegraphics[width=0.9\textwidth]{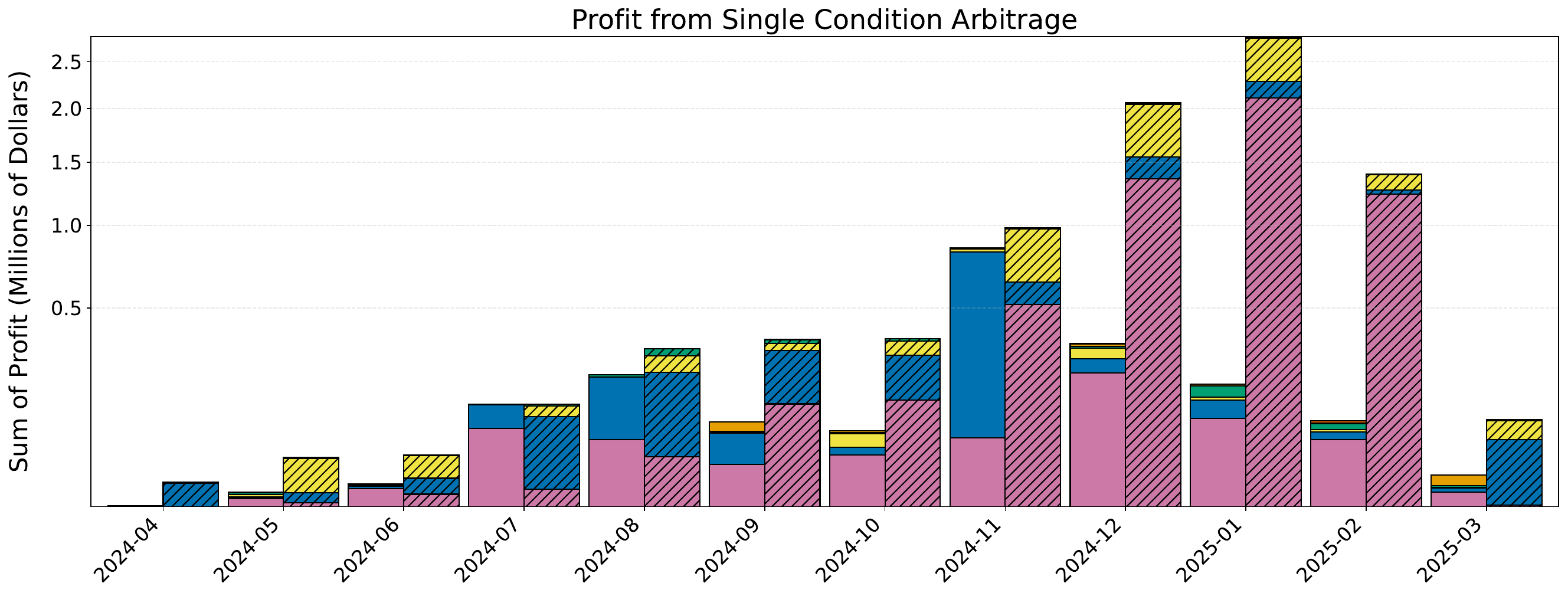}
    \caption{Total profit made by all users in single condition arbitrage. About 1\% of our estimated U.S. election opportunities were exploited by users. Interestingly, Sports Single markets dominate in exploited opportunities, surpassing the election profit.}
    \label{fig:conditions_detected}
\end{figure}

\subsection{Bids Processing and Window Size}
\begin{figure}[t]
    \centering
    \begin{minipage}{0.48\textwidth}
        \centering
        \includegraphics[width=\linewidth]{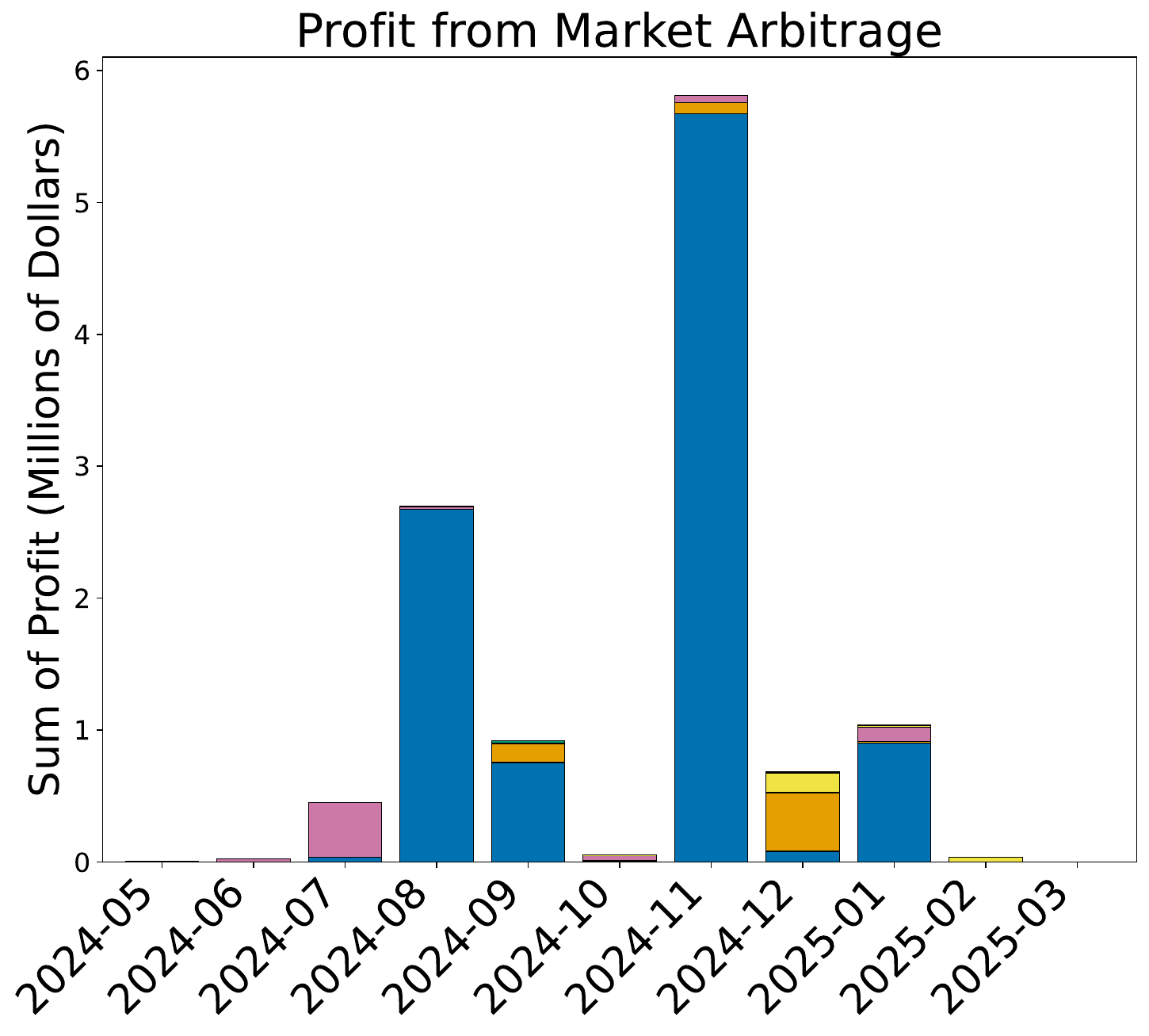}
    \end{minipage}
    \hfill
    \begin{minipage}{0.48\textwidth}
        \centering
        \includegraphics[width=.95\linewidth]{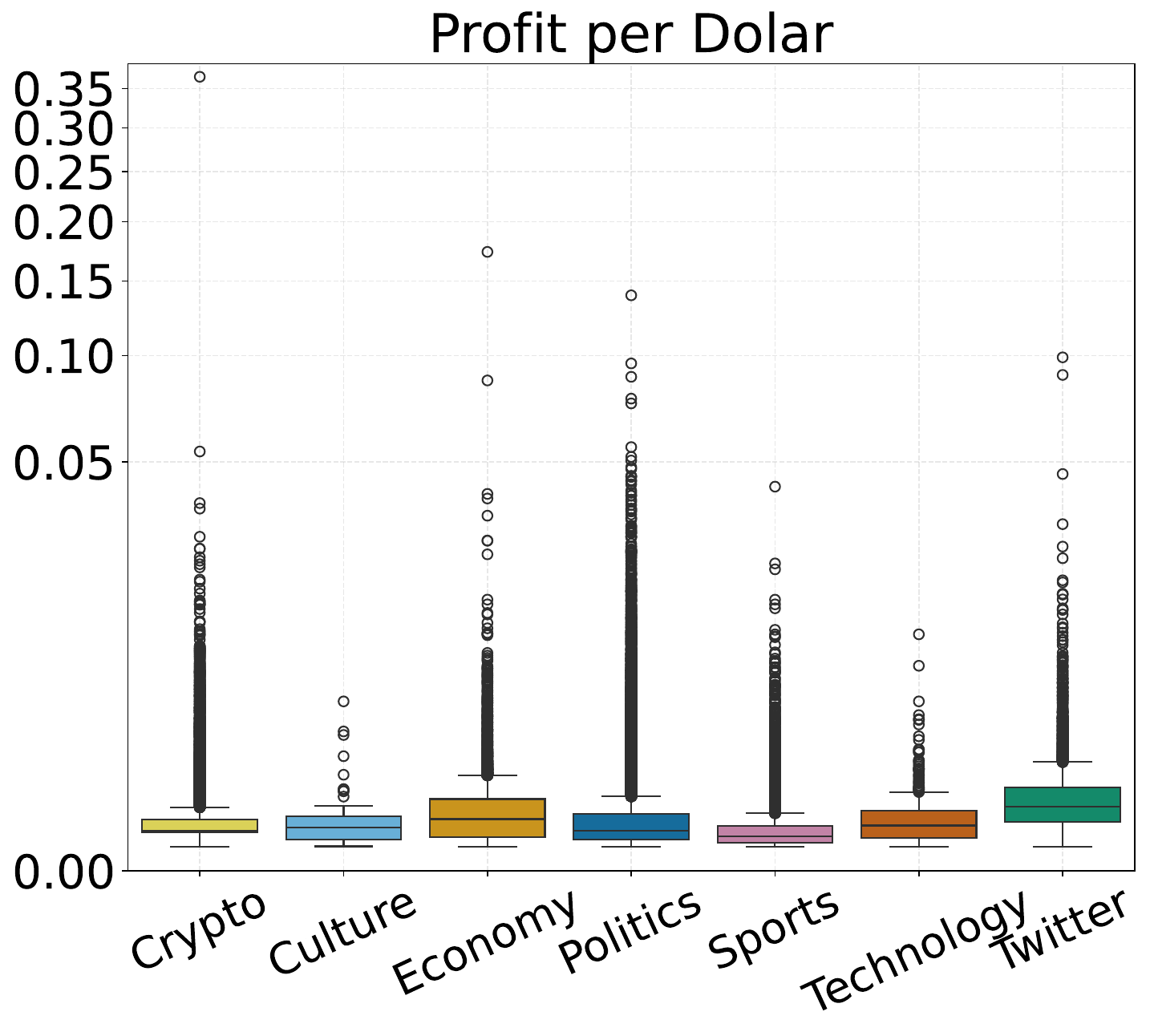}
        \label{fig:total_profit_per_dolar}
    \end{minipage}
    \caption{(Left) The total profit obtained through Rebalancing Arbitrage in NegRisk markets. We see that most profit is found in markets relating to Politics. Sports is surprisingly absent from this figure, likely due to a smaller scale of profits. (Right) The distribution of opportunities by potential yield per dollar. Most opportunities exhibit low returns, while a few outliers -- primarily in  Crypto, Politics, and Twitter markets -- offer significantly higher yields.}
    \label{fig:markets_detected}
\end{figure}

Having explored the space of arbitrage opportunities, we next detect whether any actors took advantage of these opportunities. For each user (single Polygon address), we take the history of all bids -- buys, sells, splits accompanied with sells(the latter can be used the compute the price for which the user held the opposing position, e.g., a user that creates 1 "\textit{YES}"/"\textit{NO}" position and then sells the "\textit{YES}" for \$0.70 effective holds a "\textit{NO}" at price \$0.30.).

In Appendix~\ref{appendix:bids} we summarize the set of 86 million bids. To make the data more manageable but still capture the majority of arbitrage value, we filter bids below \$2.00. We then group all executed bids from a user and consider bids within a time window $T$ as part of the same opportunity. We set $T$ to 950 blocks (approx. 1 hour) to capture some delay in bids being matched (75\% of bids fall within this window -- see App.~\ref{appendix:bids} for the distribution).
%
We then calculate a rolling window of positions a bought and at what price.
We calculate profit from the minimum amount of tokens held across all positions minus the price to acquire the positions. It is important to clarify that we do not take into account the fees because Polymarket currently does not charge per trade executed.

\subsection{Arbitrage Within Single Conditions}
We first consider strategies within a single condition -- recall from Section~\ref{sec:results_captured_arbitrage} that the largest profit was seen in this type of arbitrage. We look at the price the user acquired "\textit{YES}" and "\textit{NO}" tokens and find all instances when the sum of prices deviates from \$1. Figure~\ref{fig:conditions_detected} shows the total arbitrage captured by all users -- we see that these types of opportunities are largely captured. Of the two strategies, the total profit from \textit{buying below one dollar} amounts to \$5{,}899{,}287.427, whereas the total profit from \textit{selling above one dollar} is \$4{,}682{,}074.77. 

An interesting phenomenon we observed was the presence of opportunities with extremely high market discounts. The most striking example was executed by the user \texttt{@Tutaaa91},
who simultaneously purchased both \textit{"YES"}/\textit{"NO"} tokens for less than \$0.02 each, resulting in a single-trade profit of \$58{,}983.36 (other two trades from this account also exhibited high returns).
This phenomenon arises when prices are mismatched with real-world probabilities, creating an opportunity to exploit the discrepancy -- an interesting area for future work.

\subsection{Arbitrage Within and Between Markets}

Next, we look at strategies within a single market. Many markets include extraneous low-probability conditions that a user can safely ignore when performing arbitrage on the higher-probability events. To handle this, we use the approach from Section~\ref{sec:results_detected_arbitrage} to estimate the price of the missing positions. We include this missing price in characterizing if arbitrage exists (to not undercount the total probabilities). 

We calculate the total profit for each strategy as follows: buying "\textit{YES}" \$11{,}092{,}286.31, selling "\textit{YES}" \$612{,}188.83, selling "\textit{NO}" \$4{,}264.33, and buying "\textit{NO}" \$17{,}307{,}113.81. Compared with single condition arbitrage, selling becomes more difficult across multiple markets, and buying seems to be the predominant strategy. 

Interestingly, buying "\textit{NO}" outperforms other strategies. In fact, Polymarket announced that some accounts are making a lot of profit just by buying "\textit{NO}" positions (see related Polymarket tweets~\cite{polymarket_tweet1,polymarket_tweet2}).

We plot the total profit from single market arbitrage and the distribution of profit on the Dollar in Figure~\ref{fig:markets_detected}. Surprisingly, Sports are largely absent from the plots -- maybe a less explored venue for arbitrageurs. Additionally, Politics dominates in extracted arbitrage, with the U.S. election in November and predominantly two markets in August relating to the Democratic party pick and VP pick following Biden's drop from the election.

We also investigate arbitrage opportunities among the 13 dependent U.S. election pairs identified in Section~\ref{sec:results_dependent_markets}. No executed arbitrage was detected in the 2 NegRisk-Single pairs. Among the remaining 11 pairs, we found evidence of value extraction in 5 cases. Recall that our check focuses on the sum of \texttt{YES} outcomes across dependent subsets of the pairs. The extracted total arbitrage across market pairs is as follows: pair~2, \${60{,}236.71}; pair~4, \${18{,}472.31}; pair~1, \${15{,}818.53}; pair~3 \${629.16},  (see Appendix~\ref{appendix:dependent_pairs} for pair descriptions). Interestingly, one market on the GOP presidential win margin is in pairs 1 and 2, and one market on the balance of power among the presidency, House, and Senate, is in pairs 2 and 4.

\begin{figure}[t]
    \centering
    \includegraphics[scale =0.35]{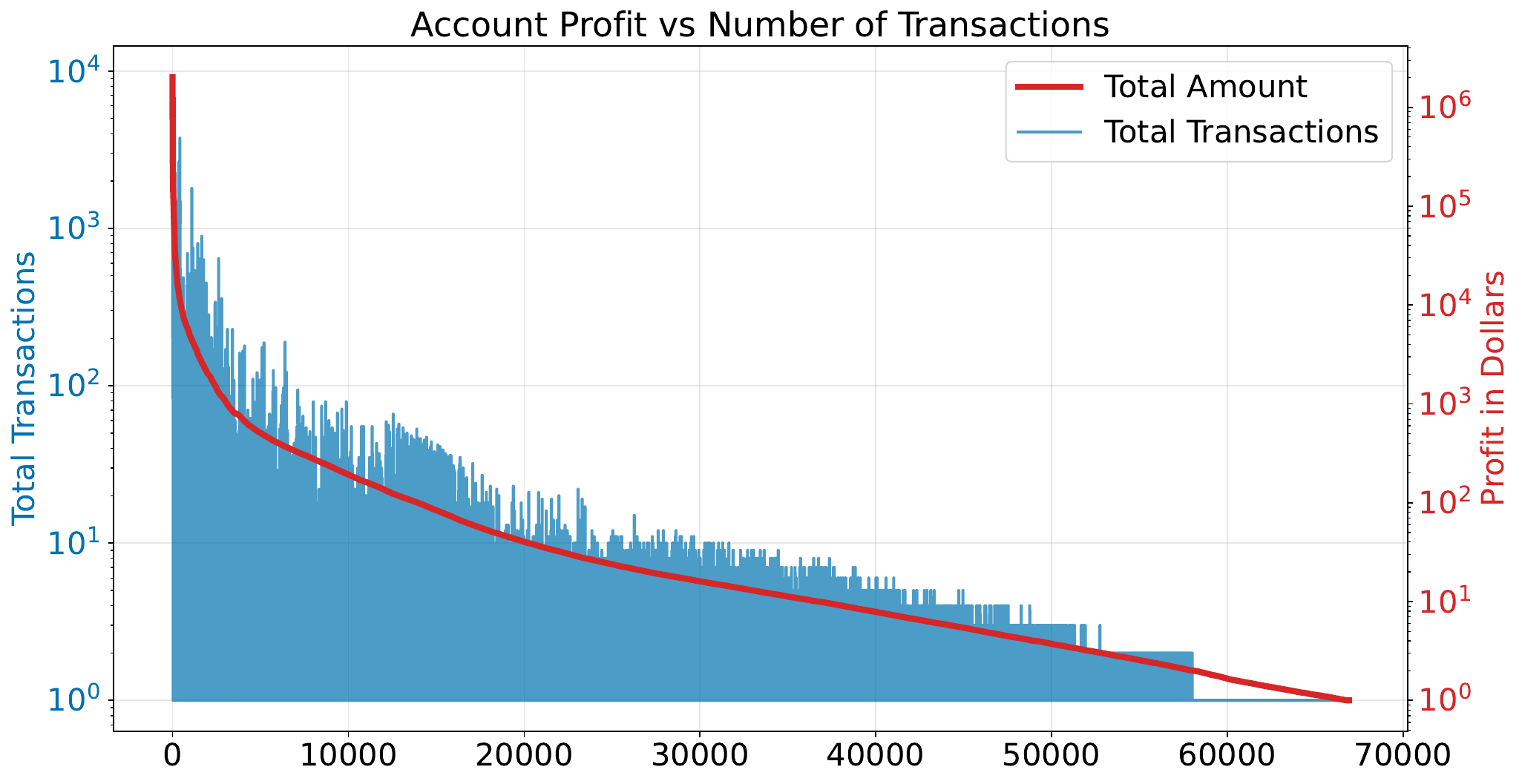}
    \caption{Total number of bids across accounts plotted alongside their aggregate profits in dollars. Both y-axes are represented on a logarithmic scale.}
    \label{fig:arb_strategies}
\end{figure}

\subsection{Arbitrageur Strategies}
Lastly, we briefly explore the strategies of top arbitrageurs. For all arbitrage we detect, we sum the total profit made by each account and the number of executed bids that went into this profit. We plot this in Figure~\ref{fig:arb_strategies}. The combined amount extracted from all strategies presented totals 
\$39{,}587{,}585.02, assuming an $\varepsilon = \$1$ profit per trade. We see some very big players with bot-like behaviour in the number of bids they participate in.
The user with the highest profits generated a total of \$2,009,631.76 through these strategies. The top 10 users by profit can be found in Table~\ref{tab:top_accounts}. 
A deeper study of these strategies is left for future work.

\begin{table}[t]
\centering
\begin{tabularx}{\linewidth}{r X r r}
\toprule
\# & Account (trimmed) & Amount (\$) & Transactions \\
\midrule
1  & 0xd218e474776403a3301422... & 2,009,631.76  & 4,049   \\
2  & 0x63d43bbb87f85af03b8f2f... & 1,273,058.68  & 2,215   \\
3  & 0x9d84ce0306f8551e02efef... & 1,092,616.17  & 4,294   \\
4  & 0x44c1dfe43260c94ed4f1d0... &   768,565.50  &   211   \\
5  & 0x59ee6c6a56d7b00223f0c3... &   749,795.99  & 3,468   \\
6  & 0xd42f6a1634a3707e27cbae... &   537,959.59  & 4,533   \\
7  & 0x4a64afa45a44a01890c216... &   476,766.58  & 3,341   \\
8  & 0xb7d54bf1d0a362beb916d9... &   468,391.71  & 2,287   \\
9  & 0x53d2d3c78597a78402d4db... &   424,505.34  &   200   \\
10 & 0x3cf3e8d5427aed066a7a59... &   383,569.94  & 2,720   \\
\bottomrule
\end{tabularx}
\caption{Top 10 accounts ranked by total amount and number of successful opportunities. 
}
\label{tab:top_accounts}
\end{table}

\section{Concluding Discussion}
\label{sec:discussion}

Despite the relatively modest volume of arbitrage compared to other markets like decentralized exchanges, where transactions are atomic and risk-free, our findings contribute valuable insights into the dynamics of prediction markets.
While our analysis of Polymarket data revealed a limited number of dependent markets, our methodology remains applicable to future prediction markets, particularly as platforms evolve towards greater decentralization. As arbitrageurs develop more specialized strategies, akin to those observed in decentralized finance (DeFi) automated market makers, we anticipate an increase in dependent markets emerging as part of these strategies.
Some limitations in the reasoning of the LLM for certain types of markets (e.g., reasoning loops we encountered) underscore the need for future enhancements in capabilities to better identify and interpret such dependencies, as well as to be able to handle larger inputs (allowing for determining dependencies among larger sets of markets). 
Our focus was on identifying unequivocal arbitrage opportunities—situations where purchasing positions guarantees a profit. However, the LLM-generated outcome tables also provide insights into weaker dependencies. For instance, in scenarios where one market pertains to ``\textit{Team A wins the semifinal}'' and another to ``\textit{Team A wins the final},'' the outcome of the first influences the second, creating a temporal window for arbitrage based on logical dependencies. Studying strategies in this weaker space of dependency remains an interesting open problem.

\bibliography{references}

\appendix

\section{All Market Descriptors}

\label{appendix:descriptors}

\begin{longtable}{|l|p{6cm}|p{4cm}|}
\hline
\textbf{Descriptor} & \textbf{Explanation} & \textbf{Example} \\
\hline
\texttt{accepting\_order\_timestamp} & Timestamp when the market began accepting orders. & \texttt{None} \\
\hline
\texttt{accepting\_orders} & Whether the market is currently accepting new orders. & \texttt{False} \\
\hline
\texttt{active} & Whether the market is currently active. & \texttt{False} \\
\hline
\texttt{archived} & Whether the market has been archived. & \texttt{False} \\
\hline
\texttt{closed} & Whether the market is closed. & \texttt{False} \\
\hline
\texttt{condition\_id} & Unique hexadecimal ’0x64’ identifier for the condition associated with the market. & \texttt{0x849753c23a3...54b1bb} \\
\hline
\texttt{description} & Detailed description of the market and its resolution criteria. & This market will resolve to ``Yes'' if the Boston Celtics win the 2023-24 NBA Championship. Otherwise, ``No''. \\
\hline
\texttt{enable\_order\_book} & Indicates if the order book feature is enabled. & \texttt{False} \\
\hline
\texttt{end\_date\_iso} & ISO 8601 formatted string representing the market’s end date. YYYY-MM-DD & \texttt{2024-06-06 00:00:00+00:00} \\
\hline
\texttt{fpmm} & Fixed price market maker setting or parameter. & \texttt{None} \\
\hline
\texttt{game\_start\_time} & The start time of the game or event related to the market. & \texttt{NaT} \\
\hline
\texttt{icon} & URL to the icon image for the market. & \url{https://polymarket-upload.s3.us-east-2.amazonaws.com/celtics.png} \\
\hline
\texttt{image} & URL to a larger image representing the market. & \url{https://polymarket-upload.s3.us-east-2.amazonaws.com/celtics.png} \\
\hline
\texttt{is\_50\_50\_outcome} & Whether the market has a 50/50 outcome. & \texttt{False} \\
\hline
\texttt{maker\_base\_fee} & Base fee charged to makers on this market. & \texttt{0} \\
\hline
\texttt{market\_slug} & URL-friendly string identifier for the market. & \texttt{will-boston-...-champions} \\
\hline
\texttt{minimum\_order\_size} & Minimum allowed order size. & \texttt{5.0} \\
\hline
\texttt{minimum\_tick\_size} & Minimum tick size allowed. & \texttt{0.001} \\
\hline
\texttt{neg\_risk} & Whether negative risk is enabled. & \texttt{True} \\
\hline
\texttt{neg\_risk\_market\_id} & ID of market defining negative risk; NaN if single condition. &
\texttt{0xd523a3175e3...c85100} \\
\hline
\texttt{neg\_risk\_request\_id} & Hexadecimal ’0x64’ identifier for the condition defining the market; if NaN, the market
has only one condition. & 
\texttt{0xda2119a68f1...030d24} \\
\hline
\texttt{notifications\_enabled} & Whether notifications are enabled for the market. & \texttt{False} \\
\hline
\texttt{question} & Text of the question posed in the market. & Will Boston Celtics 2023-24 NBA Champions? \\
\hline
\texttt{question\_id} & Unique hexadecimal ’0x64’ identifier for the question in the system. & 
\texttt{0xd523a3175e3...c85100} \\
\hline
\texttt{rewards} & Rewards structure. & \texttt{\{'rates': [\{'asset\_address': '0x2791B...'\}, ...]\}} \\
\hline
\texttt{seconds\_delay} & Delay time in seconds before the market resolves. & \texttt{0} \\
\hline
\texttt{tags} & List of tags related to the market. & \texttt{[Basketball, Sports, NBA, All]} \\
\hline
\texttt{taker\_base\_fee} & Base fee charged to takers on this market. & \texttt{0} \\
\hline
\texttt{token\_outcome} & Possible outcomes represented by tokens. & \texttt{Yes, No} \\
\hline
\texttt{token\_price} & Prices of tokens. & \texttt{1, 0} \\
\hline
\texttt{token\_token\_id} & Unique IDs for the YES and NO tokens. & \texttt{3383546...998, 3658699...743} \\
\hline
\texttt{token\_winner} & Indicates which token is the winner. & \texttt{True, False} \\
\hline
\texttt{tokens} & List of tokens available in the market. Gives the token ID, outcome description, price and if it is the winner of the market. & \texttt{[\{'token\_id':'3383...998', 'outcome':'Yes', 'price':1, 'winner':True\}, \{'token\_id':'3658...743', 'outcome':'No', 'price':0, 'winner':False\}]} \\
\hline
\texttt{question\_vectorized} & Vectorized form of the question text. & will boston celtics 2023-24 nba champions? this market will resolve ... \\
\hline
\texttt{all-mpnet-base-v2} & Vector embedding for the question. & [-0.871, 0.421, 3.459, ...] \\
\hline
\texttt{topic\_Politics} & Topic indicator for Politics. & \texttt{-0.015985} \\
\hline
\texttt{topic\_Economy} & Topic indicator for Economy. & \texttt{0.044717} \\
\hline
\texttt{topic\_Technology} & Topic indicator for Technology. & \texttt{0.077288} \\
\hline
\texttt{topic\_Crypto} & Topic indicator for Crypto. & \texttt{0.081052} \\
\hline
\texttt{topic\_Twitter} & Topic indicator for Twitter. & \texttt{0.133843} \\
\hline
\texttt{topic\_Culture} & Topic indicator for Culture. & \texttt{0.065833} \\
\hline
\texttt{topic\_Sports} & Topic indicator for Sports. & \texttt{0.145043} \\
\hline
\texttt{assigned\_topic} & The assigned topic for the market. & Sports \\
\hline
\end{longtable}

\section{Prompt For Pair Detection}

\label{appendix:prompt}

\begin{lstlisting}[caption={Pseudocode Prompt for Generating Valid Combinations}, label={lst:prompt-sample}]
You are given a set of binary (True/False) questions. Your task is to determine all valid logical combinations of truth values these questions can take.

Rules:
- Each tuple represents a possible valid assignment of truth values.
- Each tuple must contain exactly {len(statements)} values, corresponding to the listed questions.
- The output must be a JSON array where each entry is a list of Boolean values.
- The output must be valid JSON and contain no additional text.

Questions:
    for idx, (_, statement) in enumerate(statements):
        prompt += f"- ({idx}) {statement}\n"

    prompt += """

    **Expected Output Format:**
    ```json
    {
      "valid_combinations": [
        [true, false, ...],
        [false, true, ...],
        ...]
    }
    ```
    Ensure the output is strictly formatted as JSON without any additional explanation or formatting artifacts.
\end{lstlisting}

\newpage
\section{Liquidity In Multiple Conditions Markets}
\label{appendix:liqui-4}


\begin{figure}[H]
    \centering
    \includegraphics[scale=0.5]{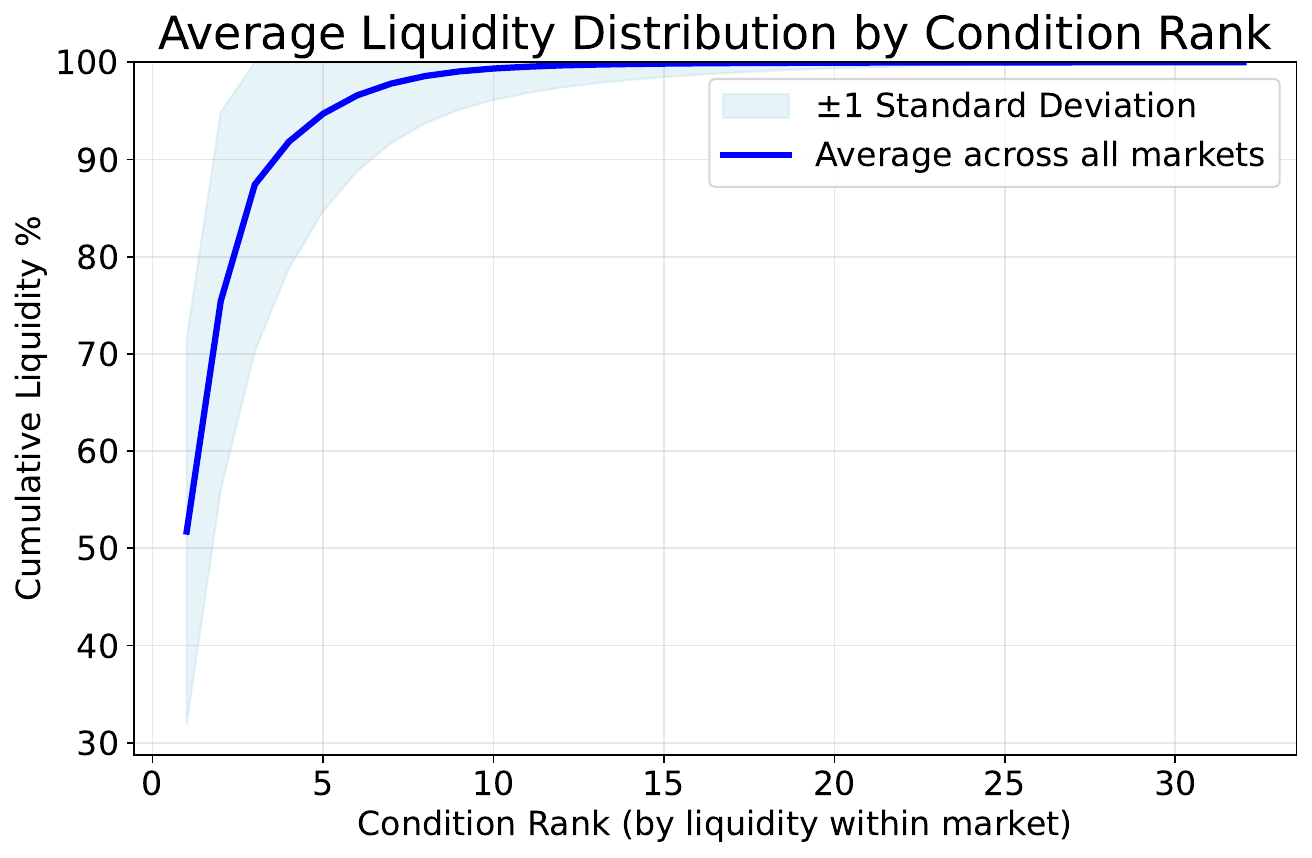}
    \caption{Average cumulative liquidity distribution by condition rank across all markets.
    The blue line represents the mean cumulative liquidity percentage as a function of condition rank, while the shaded area denotes $\pm 1$ standard deviation. Higher-ranked conditions capture the majority of liquidity, with the curve approaching 100\% within the first few ranks.}
    \label{fig:liqui-4}
\end{figure}

\section{Example of Weakly Dependent Markets}

\label{appendix:dependent}

\begin{table}[H]
\centering
\begin{tabular}{|c|p{6cm}|p{6cm}|}
\hline
\textbf{Date} & \textbf{Market A (ID: 0x326b...a800)} & \textbf{Market B (ID: 0x5a17...b100)} \\
\hline
2024-12-01 & 
(0) More than 25 named storms \newline
(1) Between 16 and 20 named storms \newline
(2) Less than 16 named storms \newline
(3) Between 21 and 25 named storms &
(4) More than 14 named storms \newline
(5) Less than 11 named storms \newline
(6) Between 11 and 14 named storms \\
\hline
\multicolumn{3}{|c|}{\textbf{Pairwise Dependency: Across markets A and B}} \\
\hline
\multicolumn{3}{|p{13cm}|}{
Dependencies exist between:
\begin{itemize}
    \item (0) and (4): Logical overlap in high storm count
    \item (2) and (5): Shared lower bound implications
    \item (1), (3), and (6): Middle range overlap with different granularities
\end{itemize}
} \\
\hline
\end{tabular}
\caption{The only market found by our LLM to have Pairwise Dependency in the set of markets not related to the U.S. election. While there are dependencies between the conditions in these two markets (i.e., knowing the resolution of one limits the space of possible outcomes in the other), it does not satisfy our strict dependency definition as the markets cannot be partitioned into two dependent subsets, as the range of numbers in each market strictly overlaps.}
\label{tab:dependent_no_ele}
\end{table}

\newpage

\section{Conflicting Markets for LLM Dependency}

\label{tab:invalid_markets}

\begin{longtable}{p{3cm}p{2cm}p{8cm}} \\
\toprule
Market ID & Markets Invalid & Questions \\
\midrule
\endfirsthead \\
\toprule
Market ID & Markets Invalid & Questions \\
\midrule
\endhead
\midrule
\multicolumn{3}{r}{Continued on next page} \\
\midrule
\endfoot
\bottomrule
\endlastfoot
0x2b3968...4cd600 & 256 & Will Vivek Ramaswamy be the next major GOP presidential race dropout? \newline ... \newline Will Donald Trump be the next major GOP presidential race dropout? \\
0x6f96e9...ae1b00 & 252 & Will Trump win Idaho by the largest margin? \newline ... \newline Will Trump win Wyoming by the largest margin? \\
0xa97980...79a800 & 239 & Will Kamala Harris win Rhode Island by the largest margin? \newline ... \newline Will Kamala Harris win Massachusetts by the largest margin? \\
0x58e978...528900 & 204 & Will 538 call 46 states correctly? \newline ... \newline Will 538 call 49 states correctly? \\
0x27a926...b3cd00 & 173 & Will North Carolina be the tipping point state? \newline ... \newline Will Minnesota be the tipping point state? \\
0x730481...857700 & 145 & Will RFK Jr. get the most votes of any 3rd party candidate? \newline ... \newline Will Chase Oliver get the most votes of any 3rd party candidate? \\
0x462064...753a00 & 115 & Will the Republican candidate win Iowa by 3.0-4.0\%? \newline ... \newline Will the Democratic candidate win Iowa by 1-2.0\%? \\
0xfa5d43...744700 & 100 & Will Elizabeth Warren win the popular vote in the 2024 Presidential Election? \newline ... \newline Will Vivek Ramaswamy win the popular vote in the 2024 Presidential Election? 
\end{longtable}

\captionsetup{type=table}
\caption{Markets that caused conflicts in the LLM's reasoning for pair dependencies. 
In the first market, the exclusivity condition is unclear since all candidates could drop out. 
In the second and third, the ``largest margin'' criterion is ambiguous because the opponent is not specified. In the fourth, it is unclear what ``X states'' refers to in the 538 context. 
In the fifth, exclusivity is uncertain because multiple states could serve as the tipping-point. The sixth, regarding most popular votes for other third-party candidates, is difficult to resolve due to its low-probability nature. 
In the seventh, exclusivity is unclear due to margin overlaps (e.g., a 4\% margin satisfies two conditions). In the last market, the popular vote question risks confusion.}

\newpage 

\section{Dependent Pair Markets}

\label{appendix:dependent_pairs}

\begin{longtable}{p{0.5cm}p{3.5cm}p{4cm}p{4cm}}
\label{tab:valid_arbitrage_pairs_dep} \\
\toprule
Pair \# & Markets & Questions 1 & Questions 2 \\
\midrule
\endfirsthead
\toprule
Pair \# & Markets & Question 1 & Question 2 \\
\midrule
\endhead
\midrule
\multicolumn{4}{r}{Continued on next page} \\
\midrule
\endfoot
\bottomrule
\endlastfoot
1 & ID1: 0x411a94...e89c3a00 \newline ID2: 0x4456a4...f8a81c00 & Will a Democrat win the popular vote and the Presidency? \newline ... \newline Will a Democrat win the popular vote and a Republican win the Presidency? & 2024 presidential election: GOP wins by 215+ \newline ... \newline 2024 presidential election: GOP wins by 1-4 \\
2 & ID1: 0x4456a4...f8a81c00 \newline ID2: 0xebbf62...edec5c00 & 2024 presidential election: GOP wins by 215+ \newline ... \newline 2024 presidential election: GOP wins by 1-4 & 2024 Balance of Power: R Prez R Senate R House \newline ... \newline 2024 Balance of Power: D Prez, R Senate, R House \\
3 & ID1: 0x8775b7...27b8bd00 \newline ID2: 0xebbf62...edec5c00 & Will Republicans have 56 or more seats in Senate after election? \newline ... \newline Will Republicans have 51 seats in Senate after election? & 2024 Balance of Power: R Prez R Senate R House \newline ... \newline 2024 Balance of Power: D Prez, R Senate, R House \\
4 & ID1: 0xe3b1bc...ec030f00 \newline ID2: 0xebbf62...edec5c00 & Will Kanye West win the 2024 US Presidential Election? \newline ... \newline Will any other Republican Politician win the 2024 US Presidential Election? & 2024 Balance of Power: R Prez R Senate R House \newline ... \newline 2024 Balance of Power: D Prez, R Senate, R House \\
5 & ID1: 0x1039dd...0c385f00 \newline ID2: 0x90d21a...2d933c00 & Will a Democrat win Georgia Presidential Election? \newline ... \newline Will a candidate from another party win Georgia Presidential Election? & Will the Democratic candidate win Georgia by 0\%-1.0\%? \newline ... \newline Will the Democratic candidate win Georgia by 3.0\%-4.0\%? \\
6 & ID1: 0x49e5aa...fd9c9200 \newline ID2: 0xa4805a...7ee80c00 & Will a candidate from another party win North Carolina Presidential Election? \newline ... \newline Will a Democrat win North Carolina Presidential Election? & Will the Democratic candidate win North Carolina by 1\%-2.0\%? \newline ... \newline Will the Democratic candidate win North Carolina by 3\%-4.0\%? \\
7 & ID1: 0x773a23...a642bb00 \newline ID2: 0x43eaa3...236a1a00 & Will a candidate from another party win Wisconsin Presidential Election? \newline ... \newline Will a Democrat win Wisconsin Presidential Election? & Will the Democratic candidate win Wisconsin by 4.0\% or more? \newline ... \newline Will the Democratic candidate win Wisconsin by 3\%-4.0\%? \\
8 & ID1: 0x61cf17...0d8eb900 \newline ID2: 0x8fb66d...0ec39e00 & Will a Democrat win Arizona Presidential Election? \newline ... \newline Will a candidate from another party win Arizona Presidential Election? & Will the Republican candidate win Arizona by 2.0\%-3.0\%? \newline ... \newline Will the Democratic candidate win Arizona by 3.0\%-4.0\%? \\
9 & ID1: 0x9d110b...c4f63300 \newline ID2: 0xf69f11...01a68f00 & Will a candidate from another party win Michigan Presidential Election? \newline ... \newline Will a Republican win Michigan Presidential Election? & Will the Democratic candidate win Michigan by 3.0\%-4.0\%? \newline ... \newline Will the Republican candidate win Michigan by 4.0\% or more? \\
10 & ID1: 0xf487c5...09d03e00 \newline ID2: 0x5c64d1...c3c21600 & Will a Republican win Pennsylvania US Senate Election? \newline ... \newline Will a candidate from another party win Pennsylvania US Senate Election? & Will the Democratic candidate win Pennsylvania by 1.5\%-2.0\%? \newline ... \newline Will the Democratic candidate win Pennsylvania by 2.5\% or more? \\
11 & ID1: 0x367be8...5d378300 \newline ID2: 0xc71d77...19a15400 & Will a Democrat win Nevada Presidential Election? \newline ... \newline Will a Republican win Nevada Presidential Election? & Will the Democratic candidate win Nevada by 3.0\%-4.0\%? \newline ... \newline Will the Republican candidate win Nevada by 1.0\%-2.0\%? \\

\end{longtable}

\captionsetup{type=table}
\caption{Valid arbitrage market pairs with their first and last questions. 
The first pair compares the party winning the election and the popular vote, while the second market focuses on the presidential election margins for the GOP or Democrats. 
Pairs 2--4 relate the balance of power in the presidency, Senate, and House to questions that can create impossible combinations. 
The remaining markets concern which party wins in a given state and their margins of victory, overlapping with the earlier pairs but framed around state-level margins.}

\begin{longtable}{p{0.5cm}p{3.5cm}p{4cm}p{4cm}}
\label{tab:valid_arbitrage_pairs_valid} \\
\toprule
Pair \# & Markets & Question 1 & Question 2 \\
\midrule
\endfirsthead
\caption[]{Appendix: Valid Arbitrage Market Pairs with First and Last Questions} \\
\toprule
Pair \# & Markets & Question 1 & Question 2 \\
\midrule
\endhead
\midrule
\multicolumn{4}{r}{Continued on next page} \\
\midrule
\endfoot
\bottomrule
\endlastfoot
12 & ID1: 0x7487a3...0dedfa00 \newline ID2: 0xDNE32bc & Will a candidate from another party win New York Presidential Election? \newline ... \newline Will a Republican win New York Presidential Election? & Will New York move right in the 2024 U.S. Presidential Election? \\
13 & ID1: 0x411a94...e89c3a00 \newline ID2: 0xDNE3b53 & Will a Democrat win the popular vote and the Presidency? \newline ... \newline Will a Democrat win the popular vote and a Republican win the Presidency? & Winning candidate also wins popular vote? 
\end{longtable}

\captionsetup{type=table}
\caption{Valid arbitrage market pairs with their first and last questions. 
Pair 12 exists because New York has historically voted Democratic; this creates arbitrage between a market on Republicans winning and another on Democrats prevailing. 
Pair 13 concerns combinations of the popular vote and Electoral College outcomes, compared against a market asking whether the same candidate wins both. 
}


\section{Difficult Markets for the LLM}

\label{tab:llm_confusion_dep}

\begin{longtable}{p{0.5cm}p{3.5cm}p{4cm}p{4cm}}
\label{tab:invalid_arbitrage_pairs_llm} \\
\toprule
Pair \# & Markets & Question 1 & Question 2 \\
\midrule
\endfirsthead
\caption[]{Appendix: Valid Arbitrage Market Pairs with First and Last Questions} \\
\toprule
Pair \# & Markets & Question 1 & Question 2 \\
\midrule
\endhead
\midrule
\multicolumn{4}{r}{Continued on next page} \\
\midrule
\endfoot
\bottomrule
\endlastfoot
1 & ID1: 0x4b6d82...9ed65200 \newline ID2: 0xDNE192c & Will a Democrat win Montana Presidential Election? \newline Will a Republican win Montana Presidential Election? & Kamala flips a 2020 Trump state? \\
2 & ID1: 0x29d02f...563c1500 \newline ID2: 0xDNE19b5 & Will a candidate from another party win North Dakota Presidential Election? \newline ... \newline Will a Republican win North Dakota Presidential Election? & Kamala Harris wins a solid red state? \\
3 & ID1: 0xffcf78...898bbc00 \newline ID2: 0xDNE2681 & Will a Democrat win New York US Senate Election? \newline ... \newline Will a candidate from another party win New York US Senate Election? & NY-19 election: Riley (D) vs. Molinaro (R)
\end{longtable}

\captionsetup{type=table}
\caption{Non-arbitrage dependency patterns detected by the LLM. 
The first pair reflects a recurring pattern found in 28 pairs, involving a potential Trump or Kamala flip from 2020 and the opposing party winning the state. 
In our case study, ``Democrat'' does not always imply Kamala; while that is highly probable, real-world scenarios exist where it may not hold. 
The second pair is a similar case to the first, but in a solidly Republican state rather than a flip state. 
The third pair illustrates a weak dependency, where the LLM sometimes confuses Senate elections with individual Senator races. 
Many other questions follow this third pattern, showing some dependency but not the kind of arbitrage relationship under investigation.}

\section{Additional Analysis: Detecting Arbitrage Opportunities}

\subsection{Arbitrage Within Single Conditions}
\label{app:detect_cond}
    \vspace{-1em} 
    \begin{figure}[H]
      \centering
      \begin{subfigure}[b]{0.325\textwidth}
        \centering
        \includegraphics[width=\linewidth]{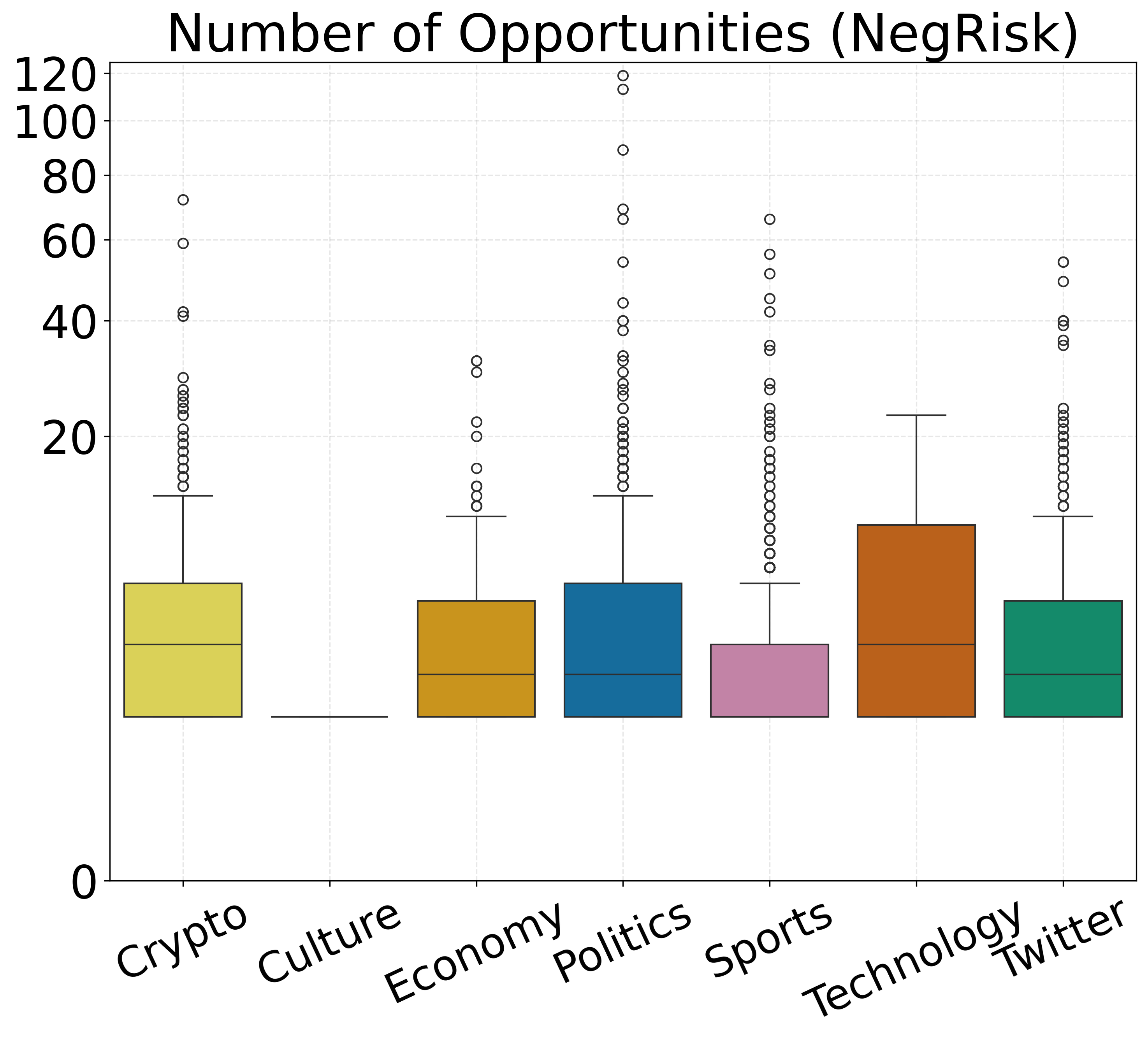}
      \end{subfigure}
       \hfill 
      \begin{subfigure}[b]{0.325\textwidth}
        \centering
        \includegraphics[width=\linewidth]{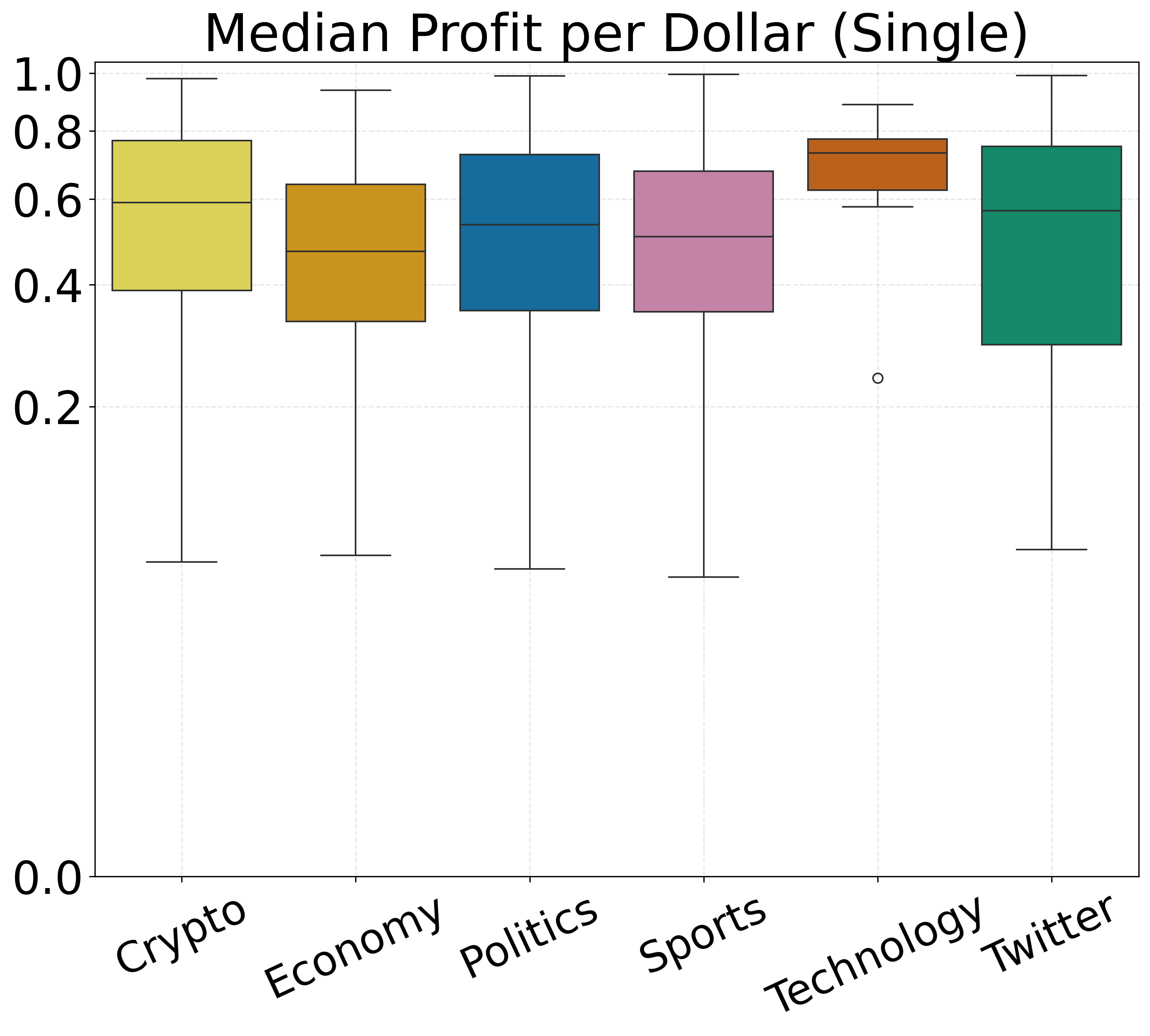}
      \end{subfigure}
      \hfill
      \begin{subfigure}[b]{0.325\textwidth}
        \centering
        \includegraphics[width=\linewidth]{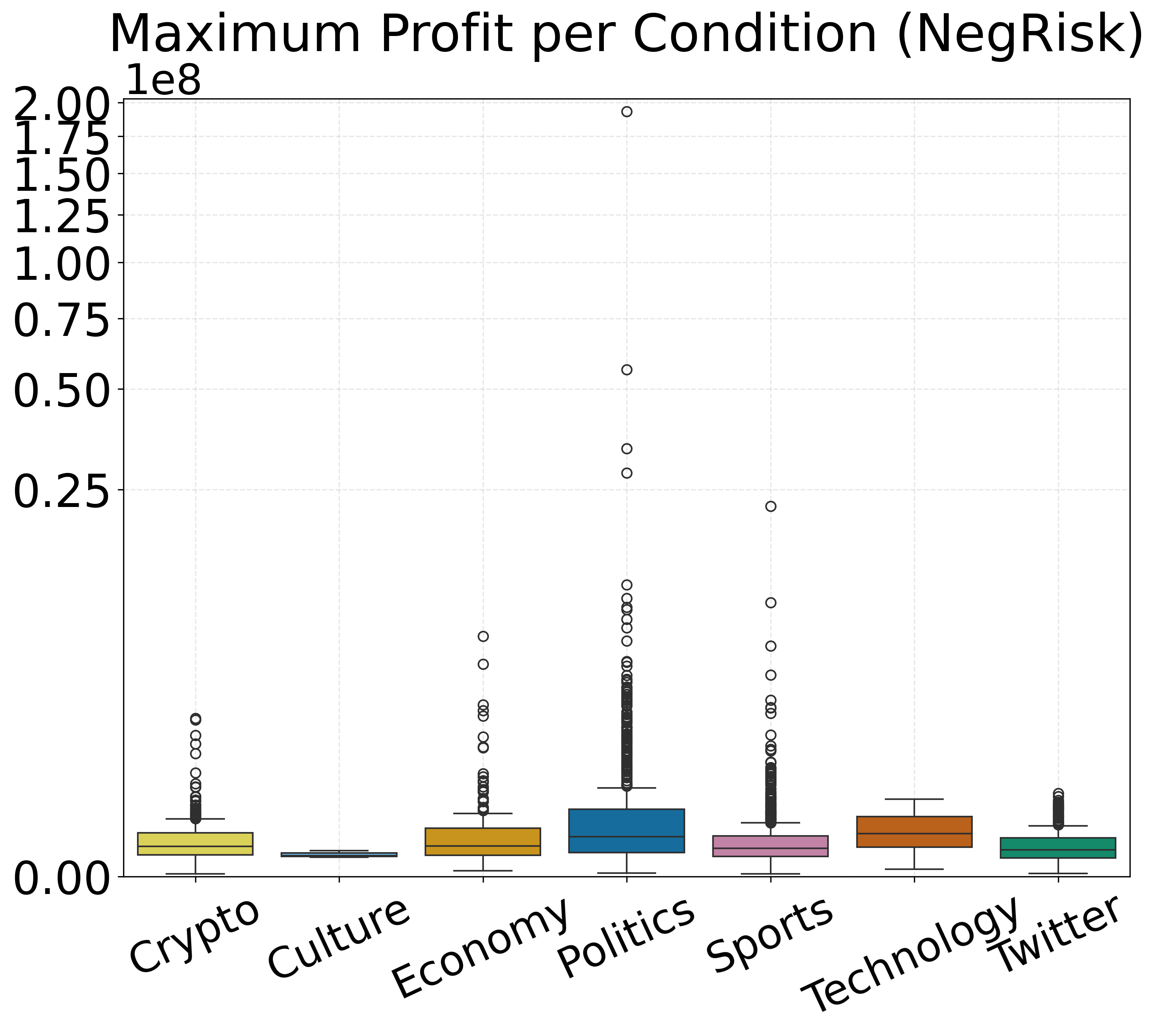}
      \end{subfigure}\\
      \centering
      \begin{subfigure}[b]{0.325\textwidth}
        \centering
        \includegraphics[width=\linewidth]{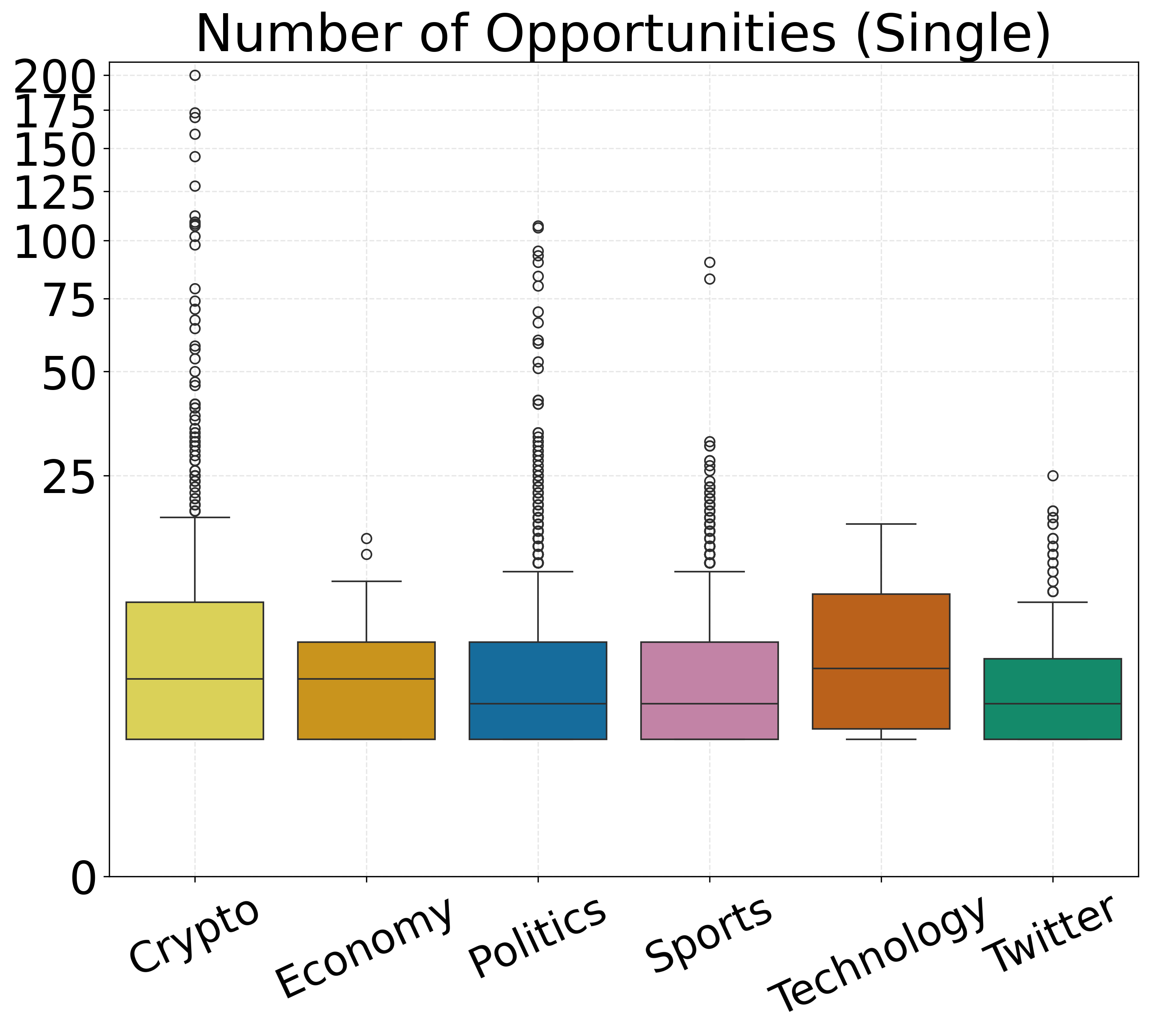}
        \vspace{-15pt} 
      \end{subfigure}
       \hfill 
      \begin{subfigure}[b]{0.325\textwidth}
        \centering
        \includegraphics[width=\linewidth]{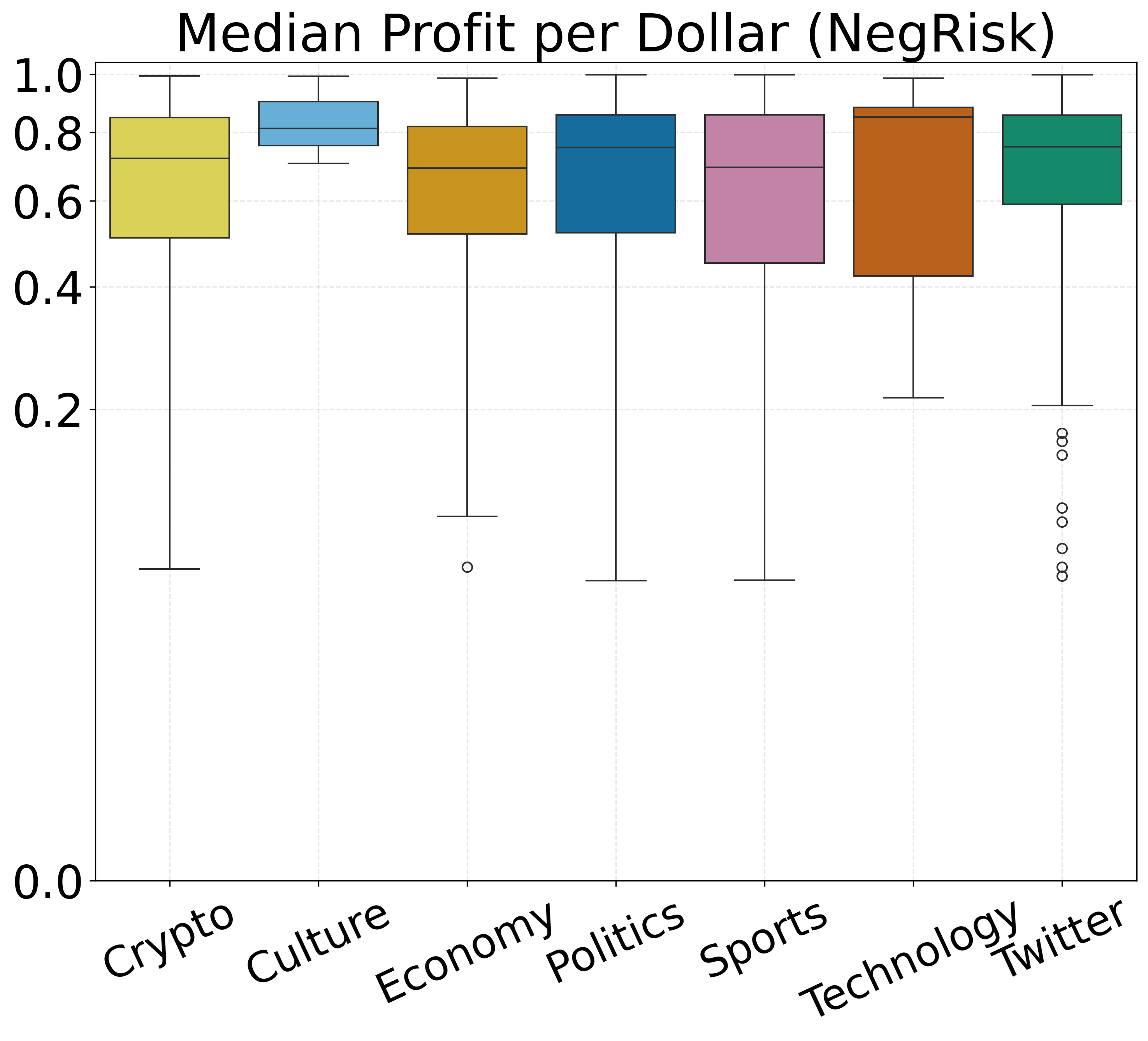}
        \vspace{-15pt} 
      \end{subfigure}
      \hfill
      \begin{subfigure}[b]{0.325\textwidth}
        \centering
        \includegraphics[width=\linewidth]{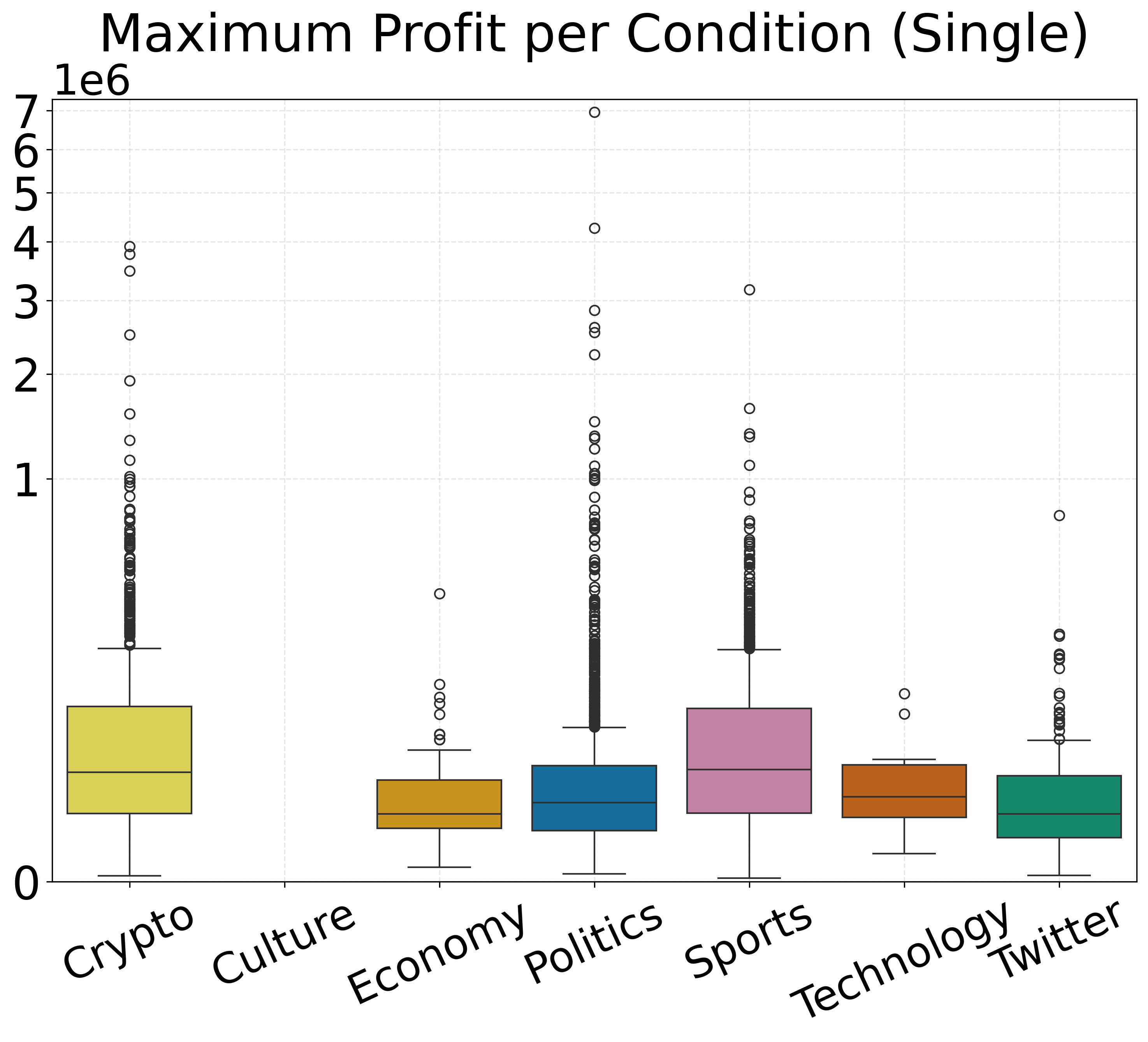}
        \vspace{-15pt} 
      \end{subfigure}
      \caption{Arbitrage opportunities detected within a single condition. The top is the arbitrage found in conditions of NegRisk (i.e., multi-condition) markets, and the bottom that of single-condition markets. We see that while single condition markets have generally more arbitrage opportunities per condition, those of the NegRisk markets are more lucrative. Over all types of markets, the median profit on the dollar is well above our 2 cent cap.} 
      \label{fig:cond_arb_box_splitup}
    \end{figure}

    \begin{figure}[H]
    \centering
    \resizebox{.85\linewidth}{!}{ 
        \begin{minipage}{\linewidth}
        \begin{subfigure}{\linewidth}
        \includegraphics[width=\linewidth]{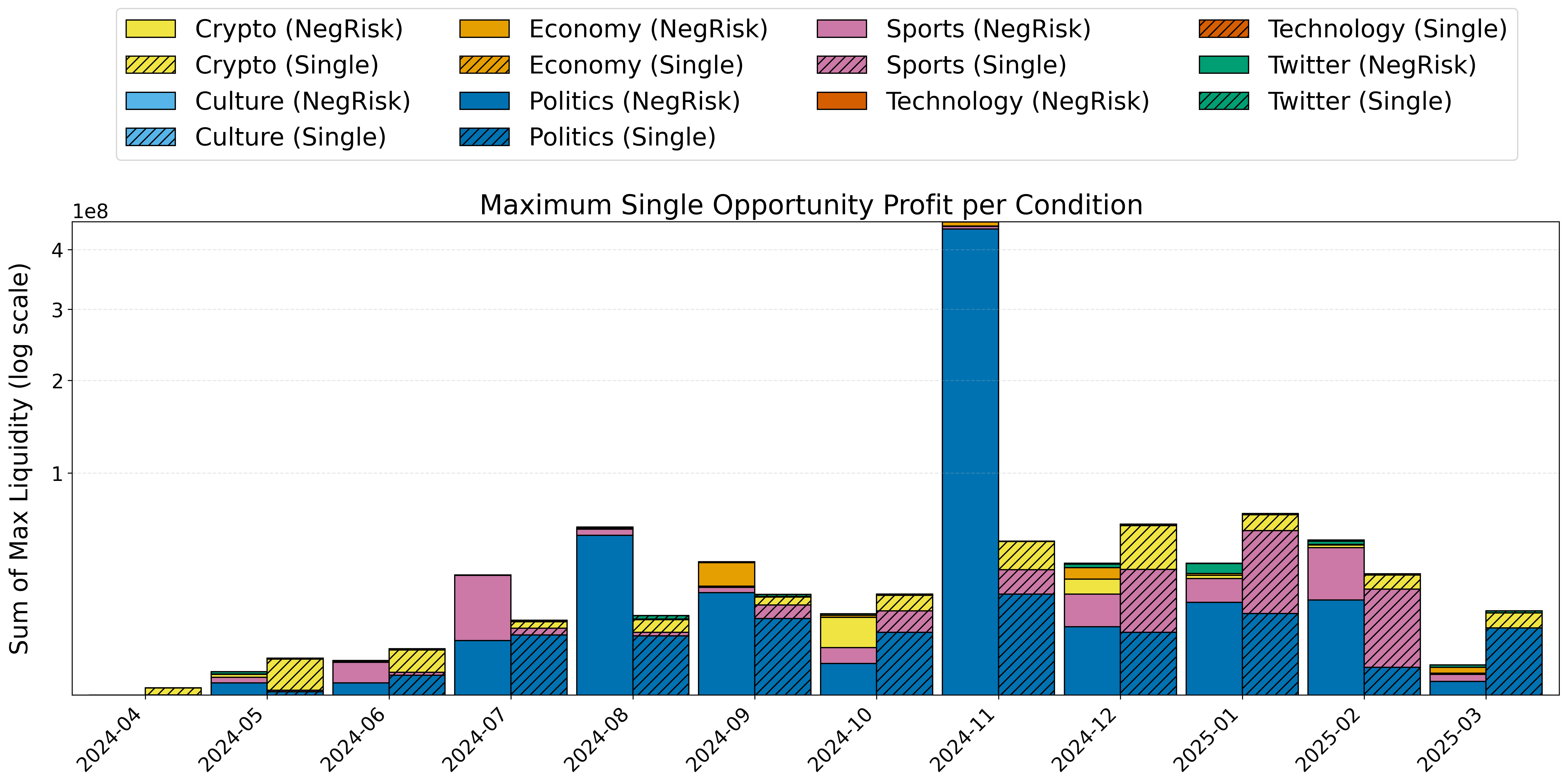}
    \end{subfigure}
    \begin{subfigure}{\linewidth}
        \includegraphics[width=\linewidth]{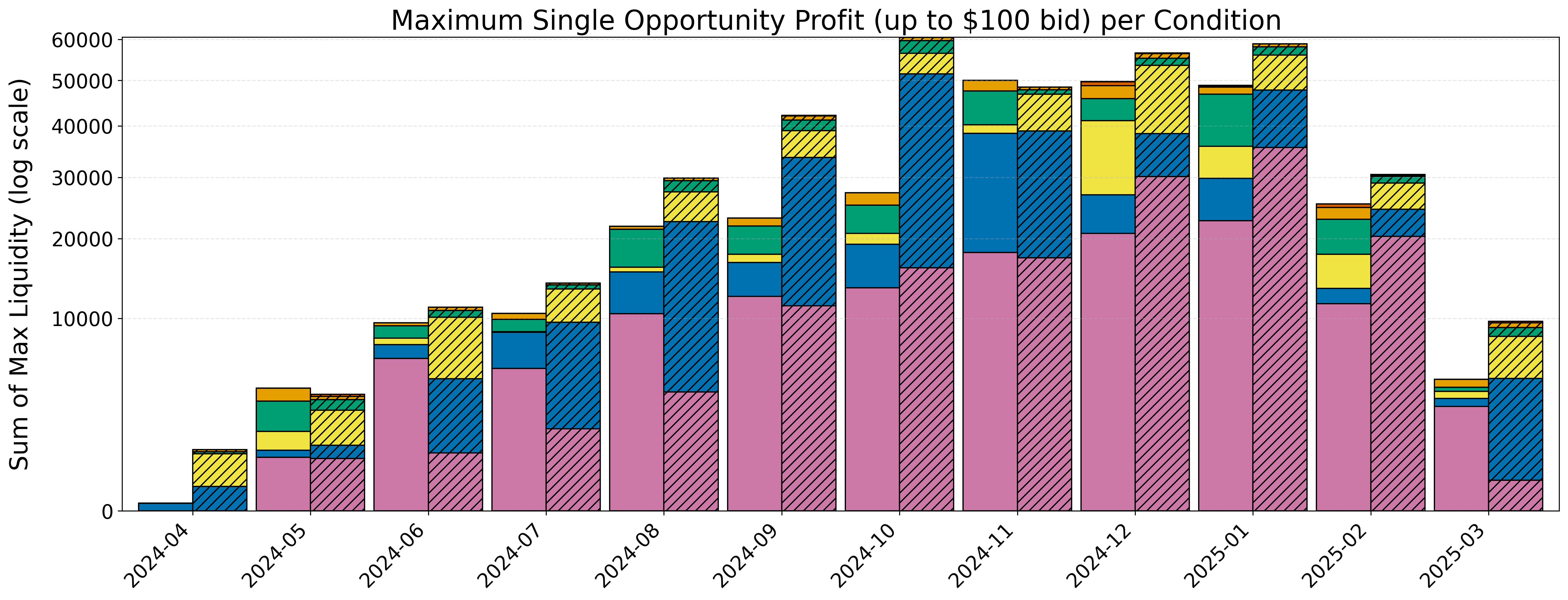}
    \end{subfigure}
    \end{minipage}
        }
    \caption{Here we explore the total arbitrage possible across each condition if an arbitrageur were to take advantage of the single most profitable opportunity in each condition at the maximum liquidity (top), and up to just \$100 of liquidity (bottom) \textbf{where the price of each conditional token is averaged over 100 blocks}. Compared to Figue~\ref{fig:con_arb_stack}, we see that averaging prices over a long period gets rid of some volatility where significant profit can be made (sum of profit is generally lower here). We do see higher profit in the lower plot, however, suggesting we longer averages leads to finding more markets where smaller value arbitrage is possible.}
    \label{fig:con_arb_stack_other}
    \end{figure}

    \begin{figure}[H]
        \centering
        \resizebox{\linewidth}{!}{ 
        \begin{minipage}{\linewidth}
        \begin{subfigure}{.49\linewidth}
            \includegraphics[width=\linewidth]{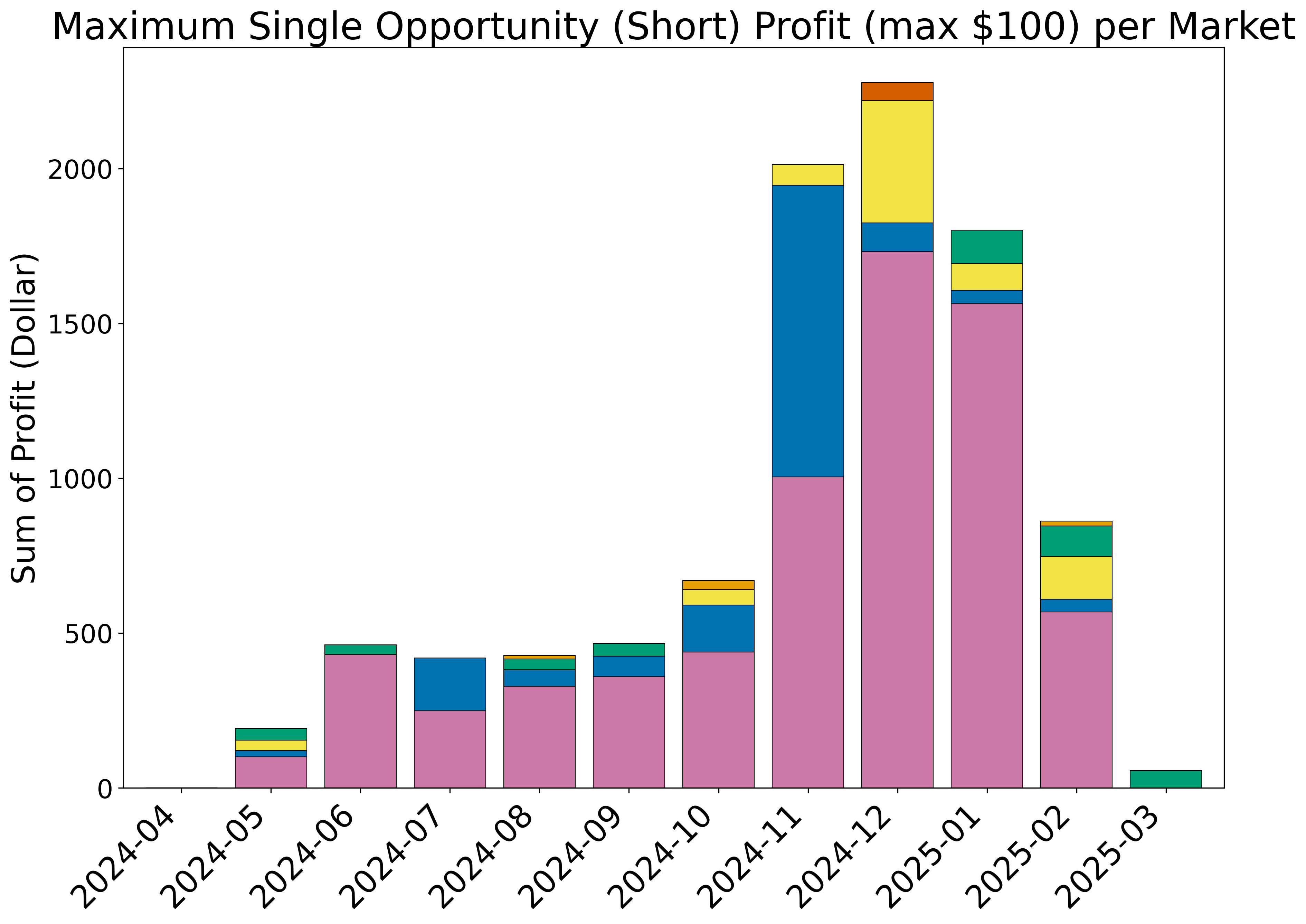}
        \end{subfigure}
        \hfill
        \begin{subfigure}{.49\linewidth}
            \includegraphics[width=\linewidth]{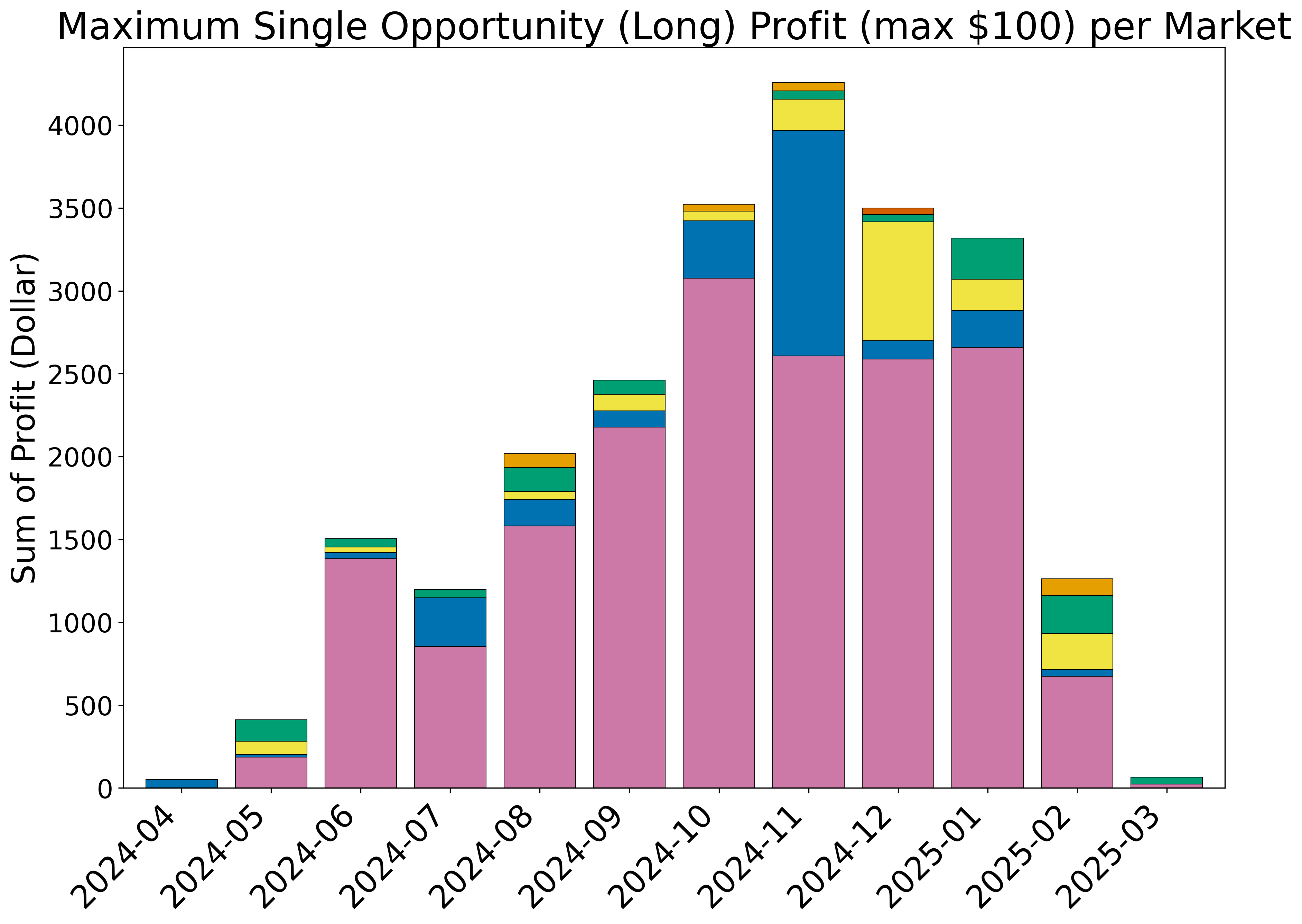}
        \end{subfigure}
        \end{minipage}
        }
        \caption{Here we explore the total arbitrage possible across each market if an arbitrageur were to take advantage of the most profitable opportunity up to just \$100 of liquidity. We see that overall shorting has more profit.}
    \end{figure}

\subsection{Arbitrage with Markets}

\label{app:detect_mark}
\begin{figure}[H]
  \centering
  \begin{subfigure}[b]{0.31\textwidth}
    \centering
    \includegraphics[width=\linewidth]{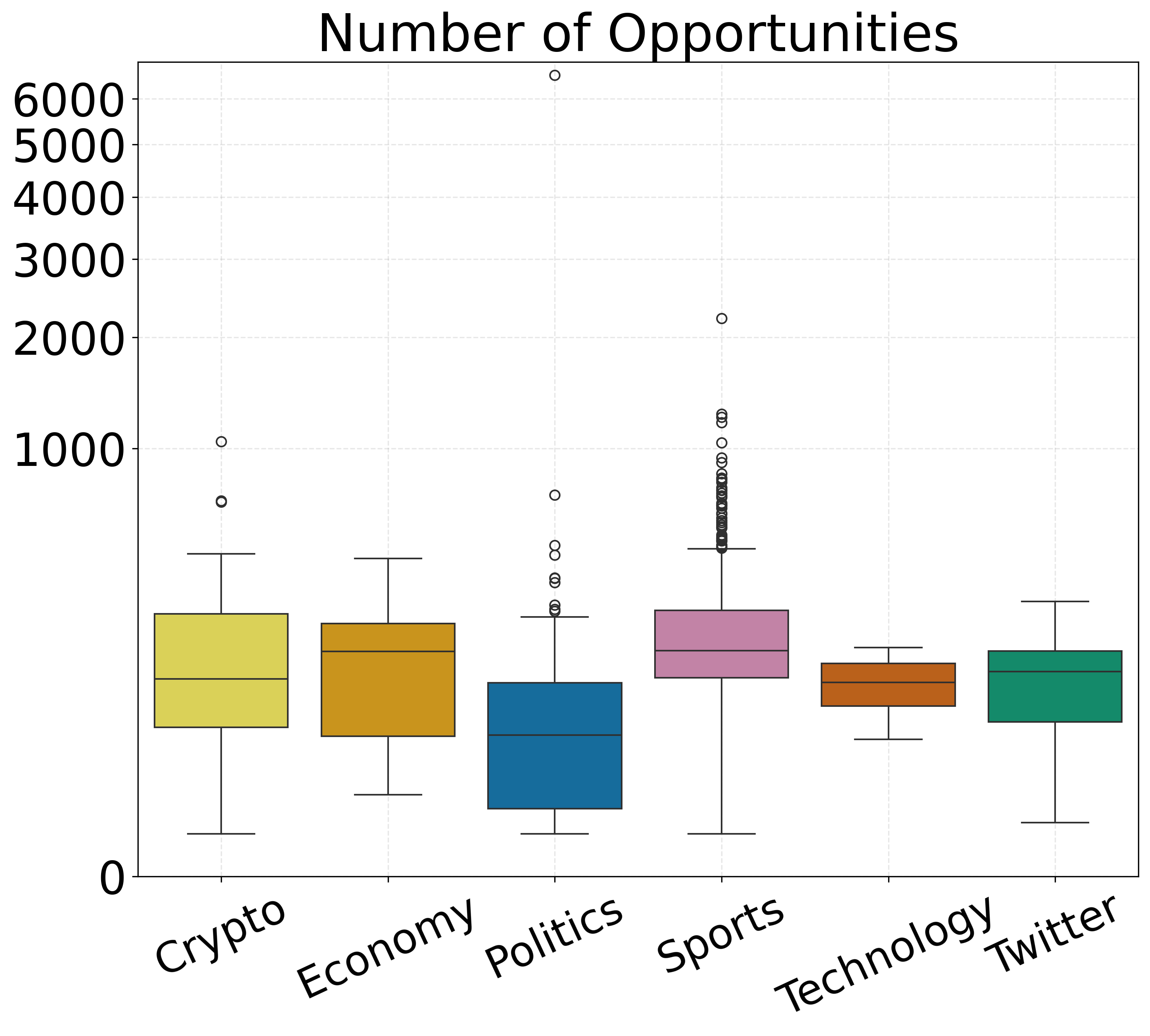}
    \vspace{-10pt} 
  \end{subfigure}
   \hfill 
  \begin{subfigure}[b]{0.31\textwidth}
    \centering
    \includegraphics[width=\linewidth]{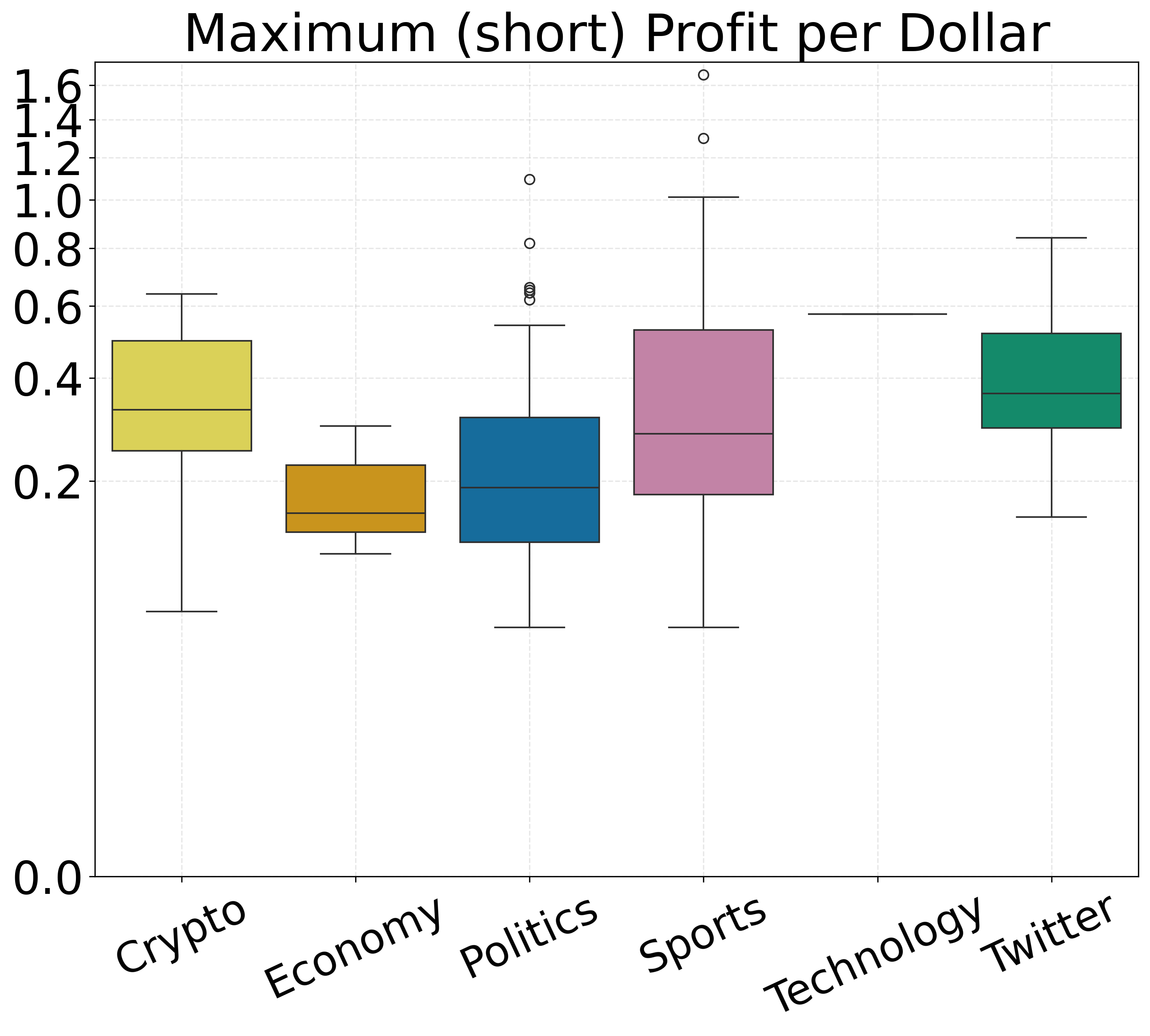}
    \vspace{-10pt} 
  \end{subfigure}
  \hfill
  \begin{subfigure}[b]{0.31\textwidth}
    \centering
    \includegraphics[width=\linewidth]{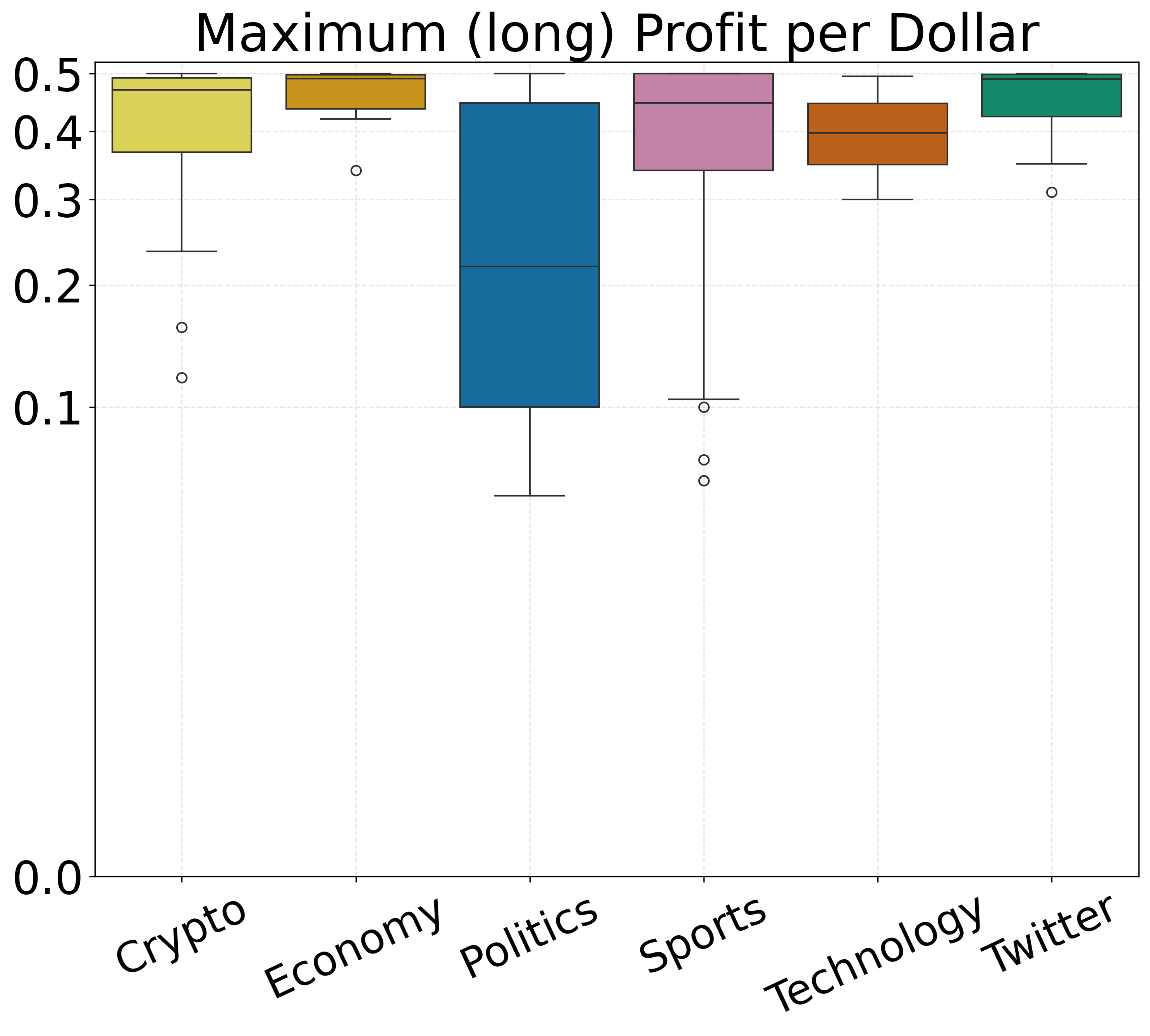}
    \vspace{-10pt} 
  \end{subfigure}\\
   \centering
  \begin{subfigure}[b]{0.31\textwidth}
    \centering
    \includegraphics[width=\linewidth]{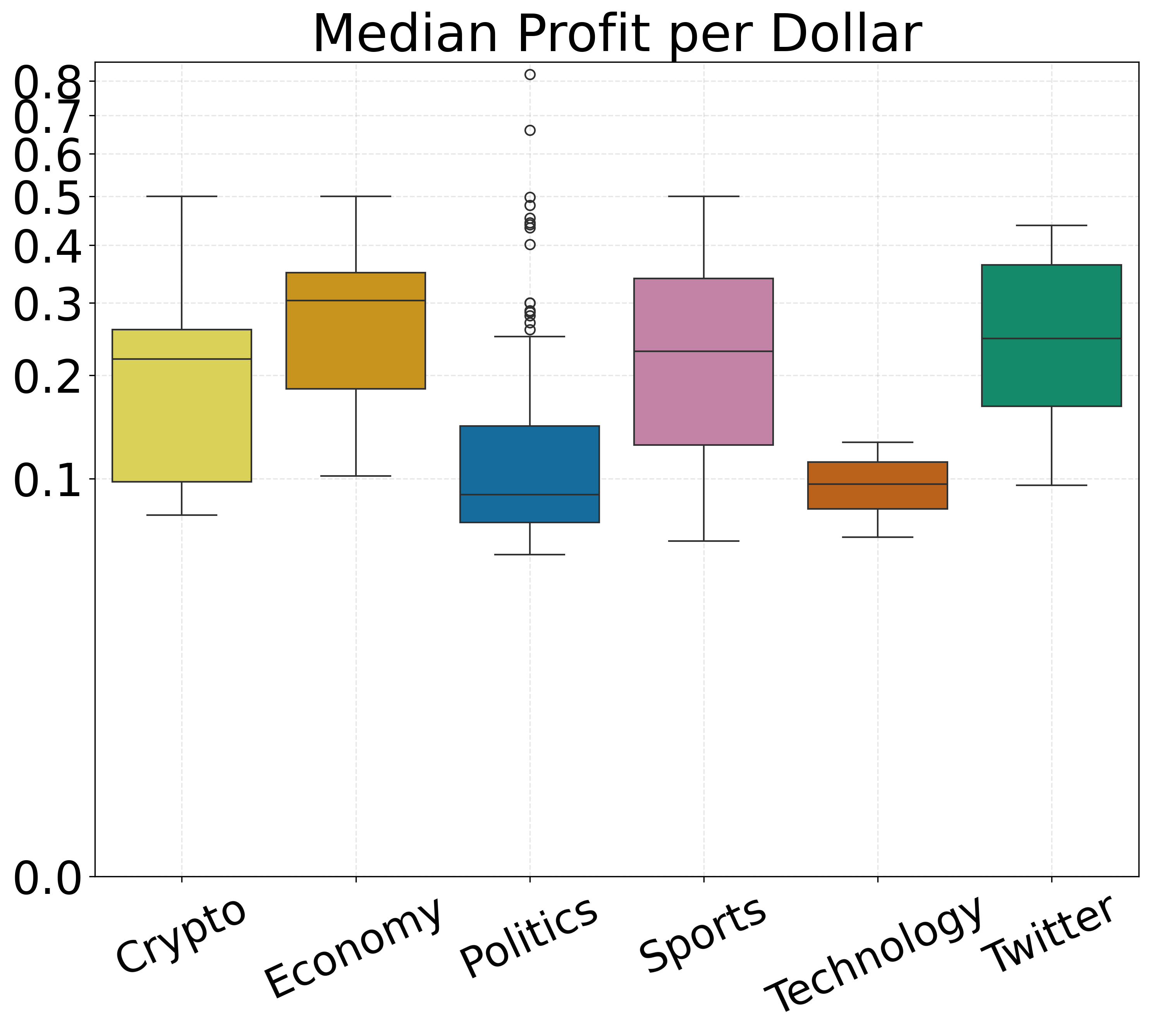}
    \vspace{-15pt} 
  \end{subfigure}
   \hfill 
  \begin{subfigure}[b]{0.31\textwidth}
    \centering
    \includegraphics[width=\linewidth]{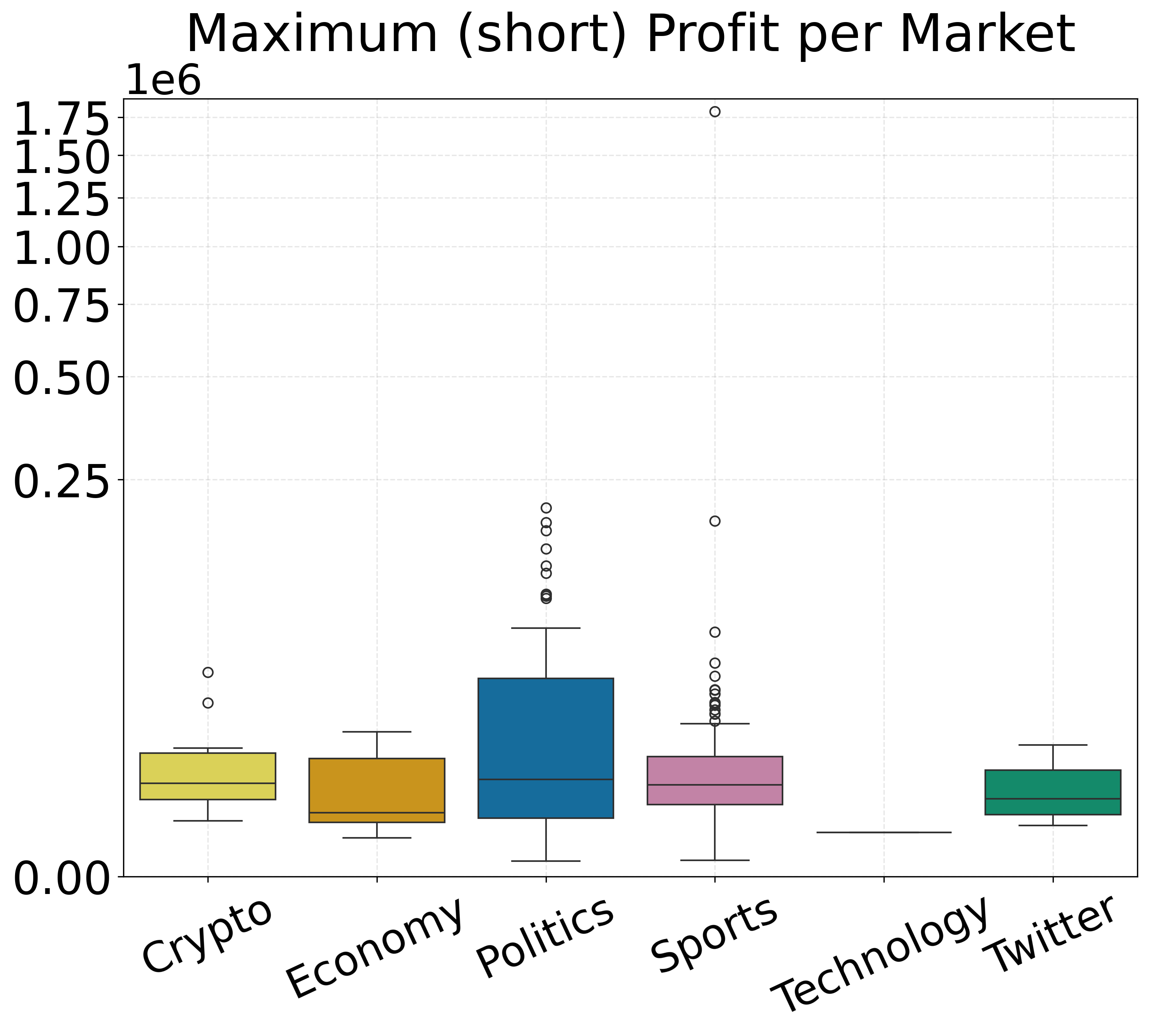}
    \vspace{-15pt} 
  \end{subfigure}
  \hfill
  \begin{subfigure}[b]{0.31\textwidth}
    \centering
    \includegraphics[width=\linewidth]{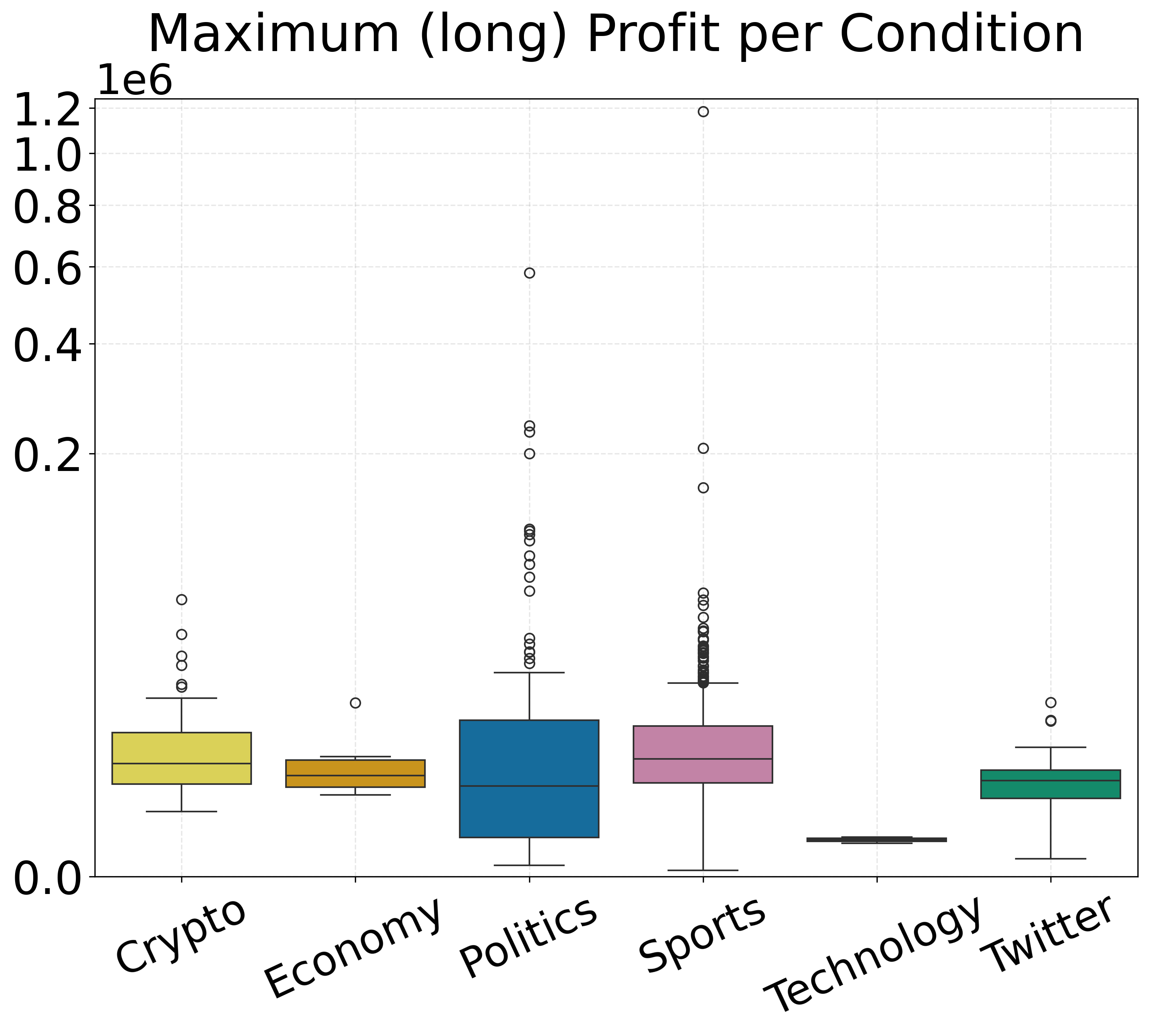}
    \vspace{-15pt} 
  \end{subfigure}
  \caption{ Properties of arbitrage opportunities detected within NegRisk markets.}
  \label{fig:market_arb_box}
\end{figure}

\begin{figure}[H]
    \centering
    \begin{subfigure}{.49\linewidth}
        \includegraphics[width=\linewidth]{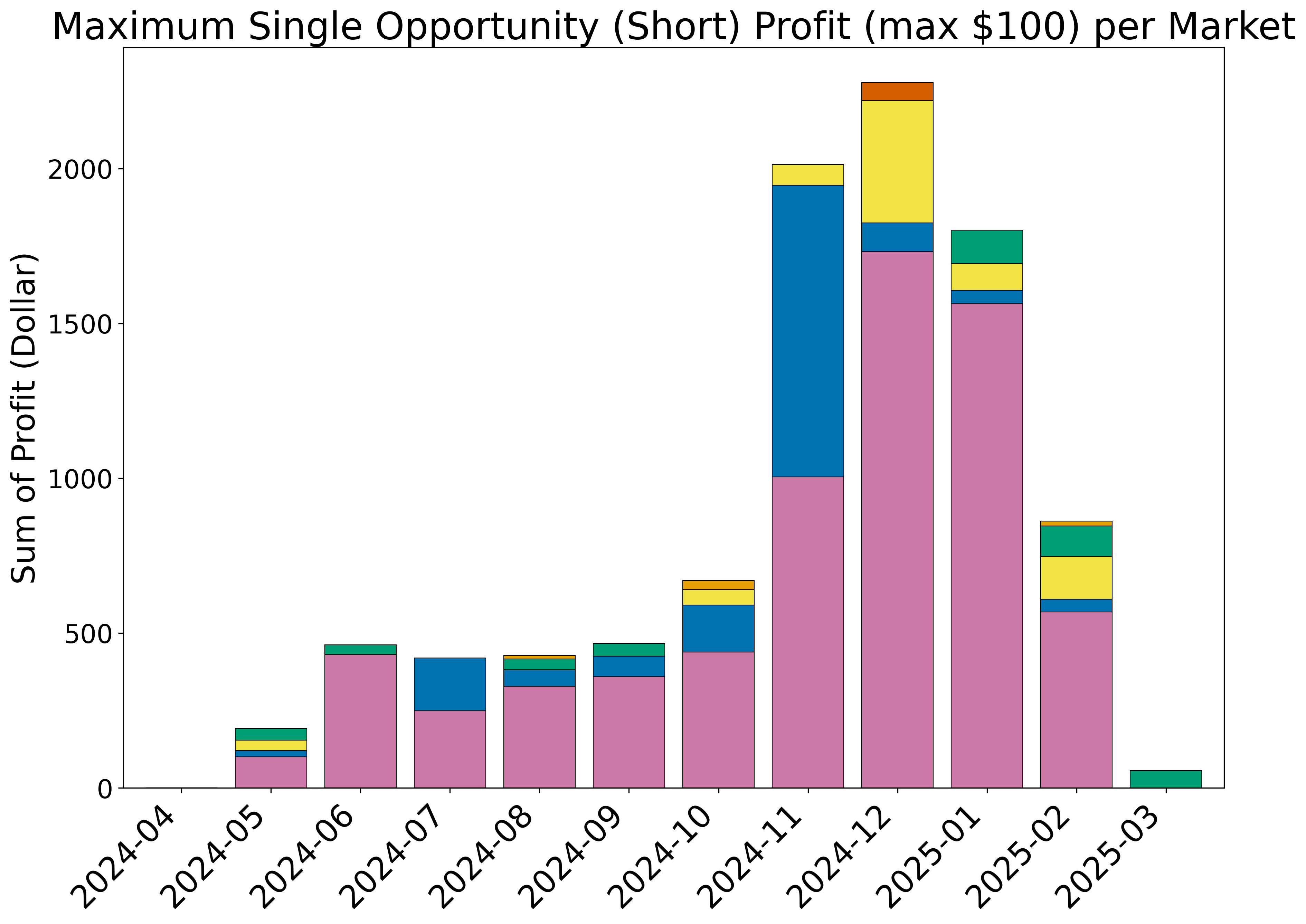}
    \end{subfigure}
    \hfill
    \begin{subfigure}{.49\linewidth}
        \includegraphics[width=\linewidth]{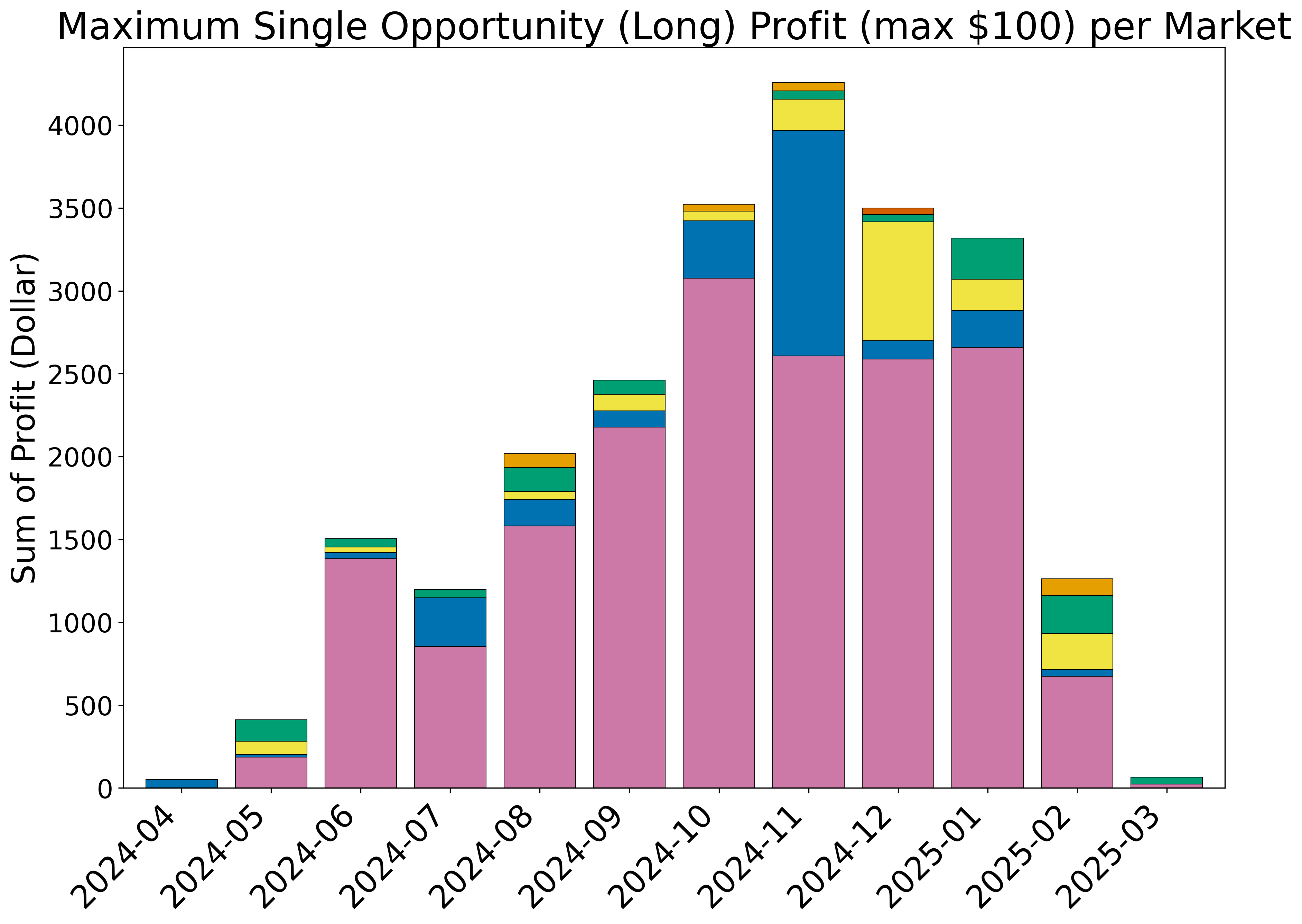}
    \end{subfigure}
    \caption{Here we explore the total arbitrage possible across each market if an arbitrageur were to take advantage of the single most profitable opportunity in each market up to \$100.}
\end{figure}

\section{Additional Analysis: Uncovering Arbitrageurs}

\subsection{Bids Statistics and Delta Measurement}

\label{appendix:bids}

\begin{figure}[H]
  \centering
  \begin{minipage}[t]{0.47\textwidth}
    \vspace{0pt}
    \centering
    \begin{tabular}{@{}lr@{}}
        \toprule
        Statistic & Value \\ \midrule
        \# of txs & 86,620,143 \\
        \hline
        Mean & 135.616 \$ \\
        Median & 8.289 \$  \\
        Minimum & 0.000001 \$  \\
        Maximum & 2,478,476.448 \$ \\
        Standard Deviation & 1,831.994 \$ \\
        25th Percentile & 0.999999 \$ \\
        75th Percentile & 46.437 \$ \\
        \bottomrule
    \end{tabular}
  \end{minipage}%
  \hfill
  \begin{minipage}[t]{0.47\textwidth}
    \vspace{0pt}
    \centering
    \includegraphics[width=\linewidth]{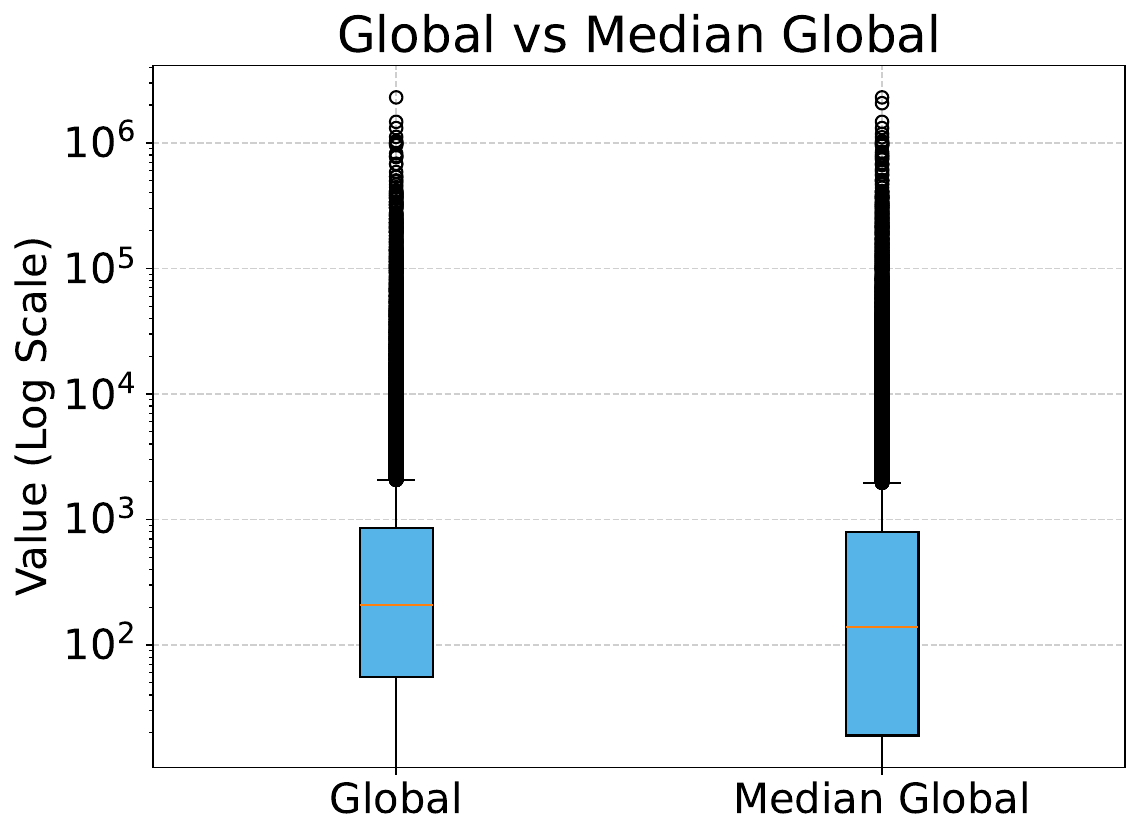}
  \end{minipage}
  \caption{Summary statistics of all bids (left) and the distribution of deltas (right). 
A delta is defined as the number of blocks between the placement and execution of an order 
for a given user $u$ and condition $c$ in a market, i.e., the interval between $t$ and $t'$. 
All deltas across users and markets are aggregated to compute the average and median, which are shown in the boxplot.}
\end{figure}

\subsection{Profit on the Dolar Different Strategies}

\begin{figure}[H]
    \centering
    \begin{minipage}{0.48\textwidth}
        \centering
        \includegraphics[width=\linewidth]{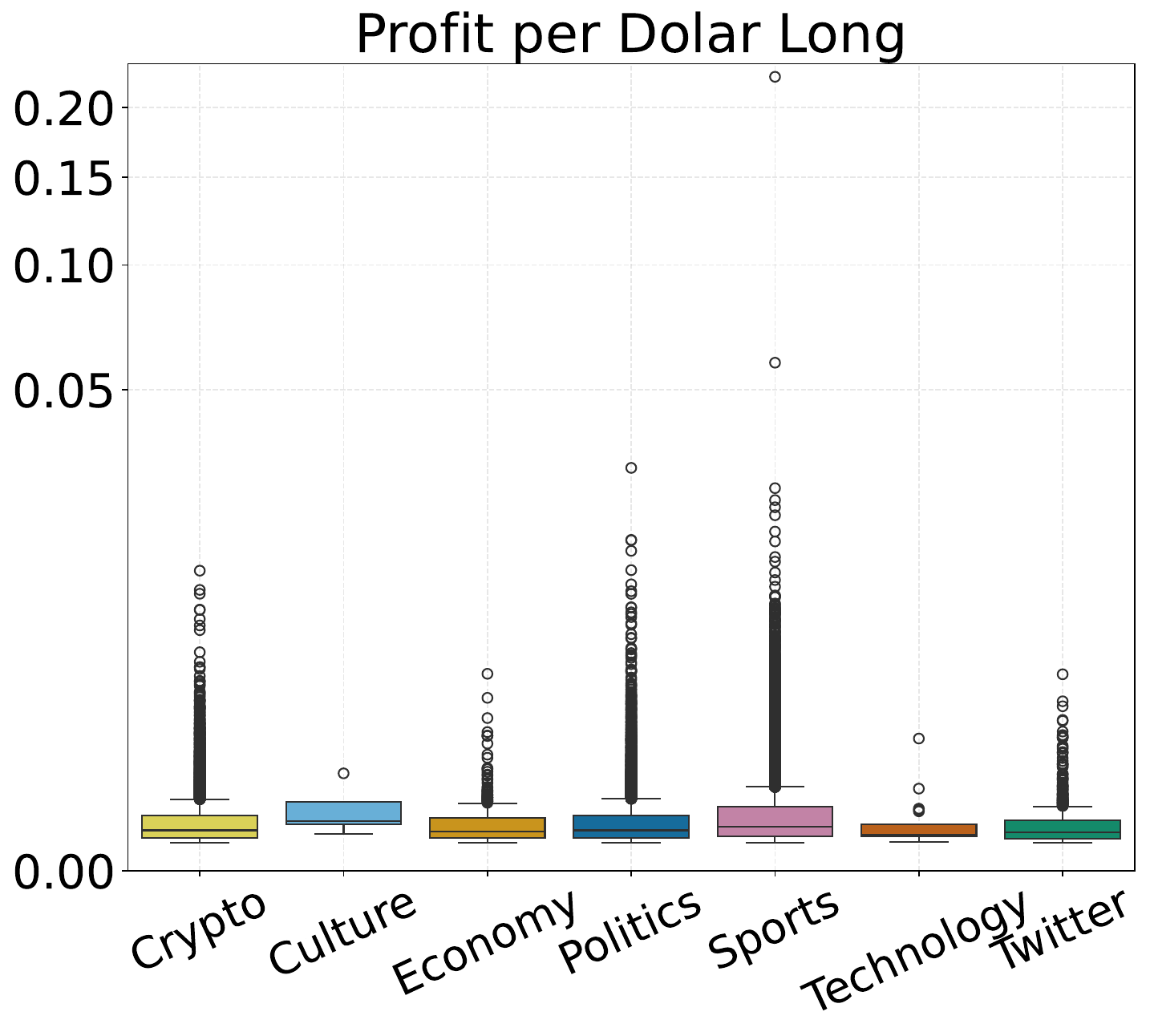}
    \end{minipage}
    \hfill
    \begin{minipage}{0.48\textwidth}
        \centering
        \includegraphics[width=\linewidth]{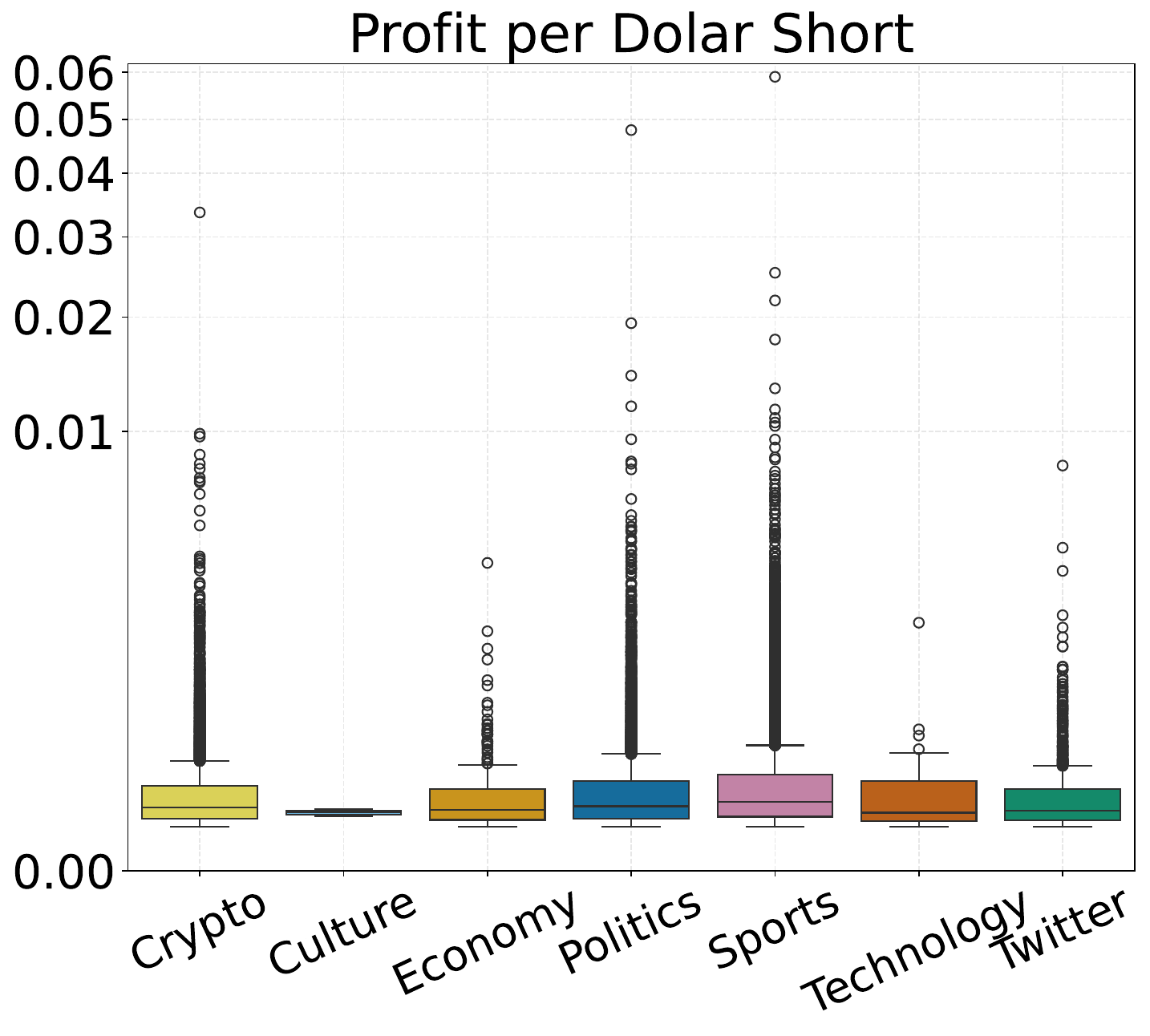}
    \end{minipage}
    \caption{Comparison of the dollar profit per trade when applying long or short strategies 
under single-condition scenarios. The plots separate cases where the total profit sum is 
less than 1 versus greater than 1.}
\end{figure}

\begin{figure}[H]
    \centering
    \begin{minipage}{0.48\textwidth}
        \centering
        \includegraphics[width=\linewidth]{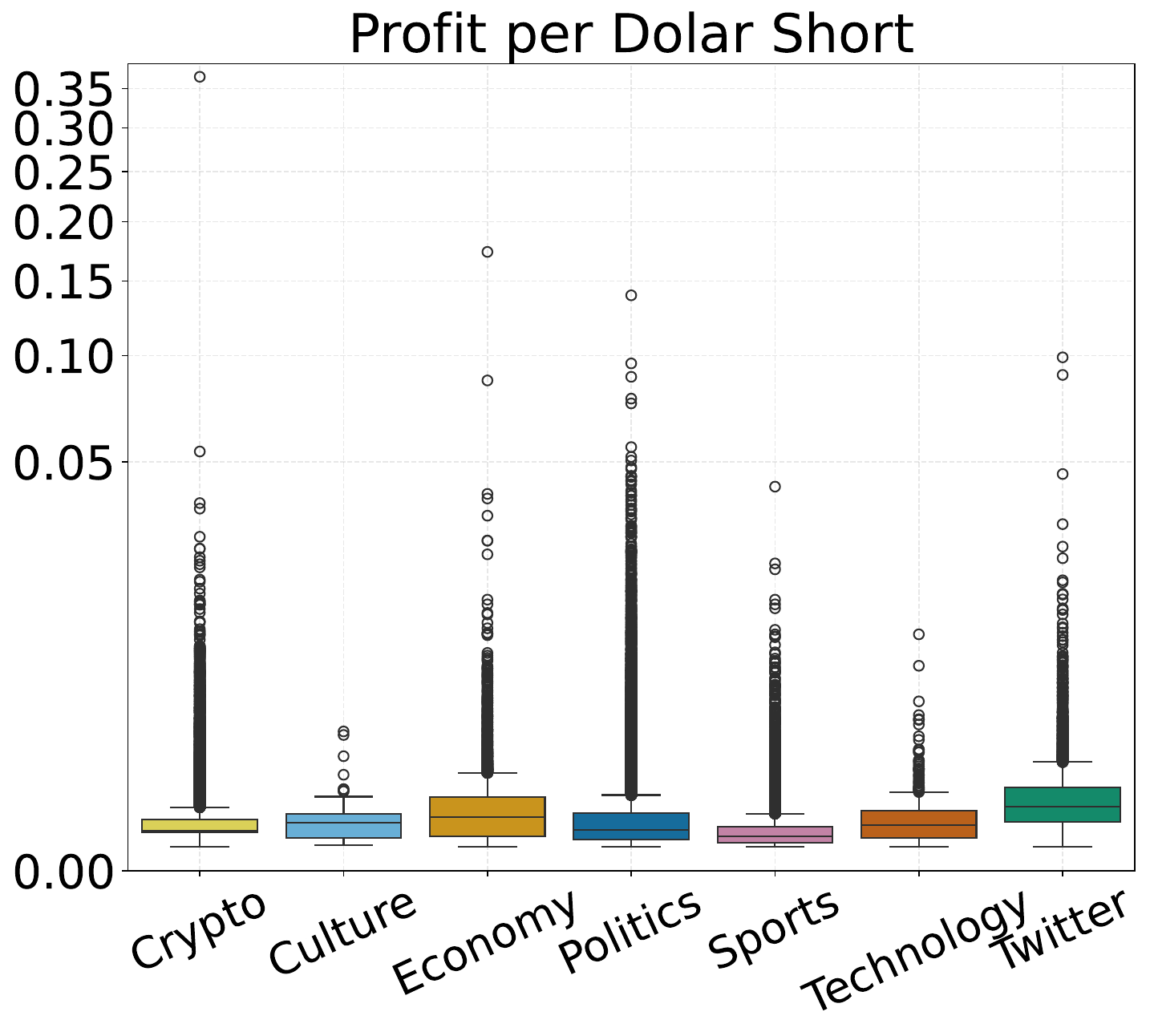}
    \end{minipage}
    \hfill
    \begin{minipage}{0.48\textwidth}
        \centering
        \includegraphics[width=\linewidth]{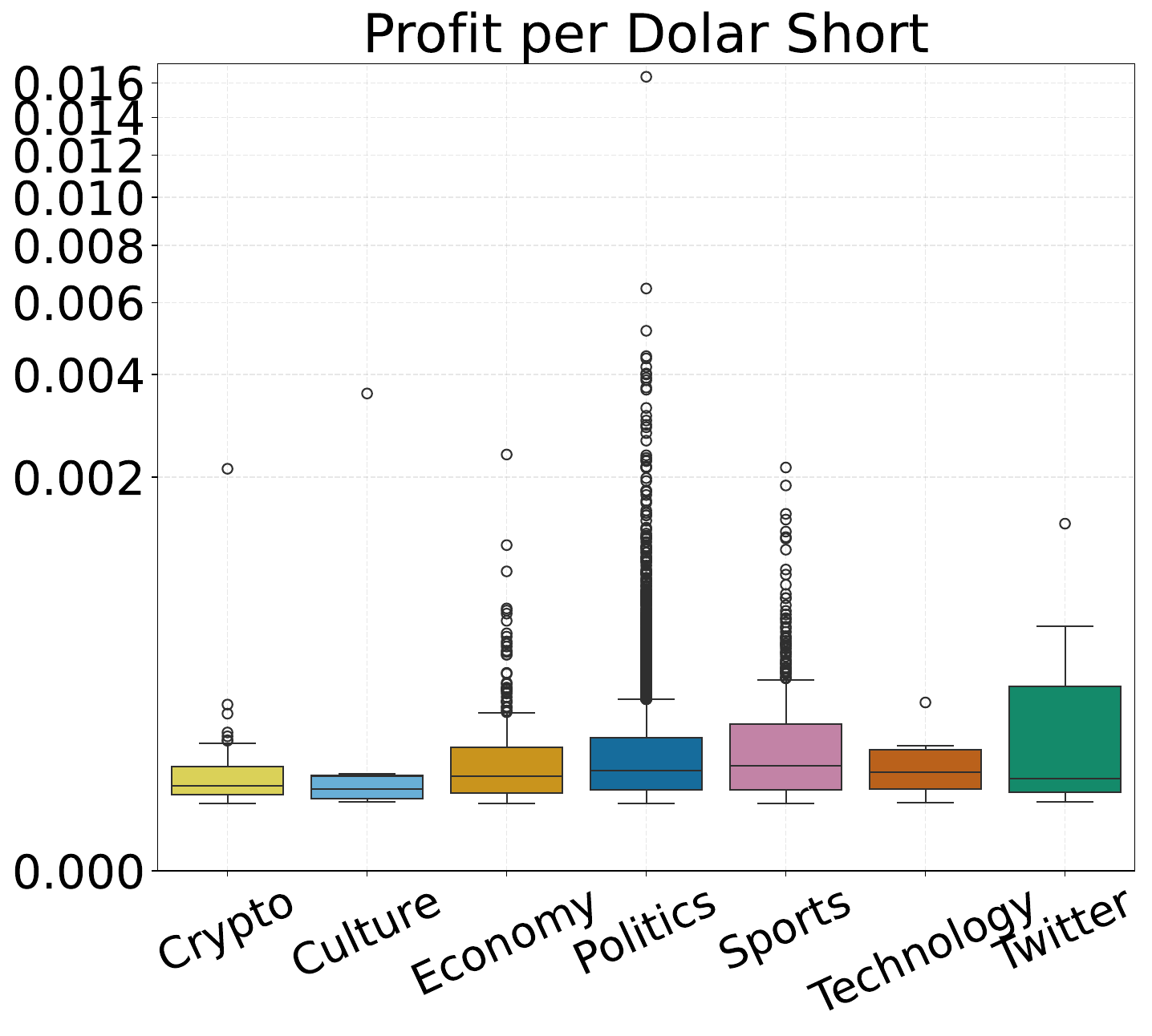}
    \end{minipage}
    \caption{Comparison of the dollar profit per trade when applying long and short strategies 
in markets with multiple conditions. The plots distinguish scenarios where the total profit 
sum is less than 1 for YES rebalancing or less than \(n-1\) for NO rebalancing versus cases 
exceeding those thresholds.}
\end{figure}

\end{document}